\documentclass{article}

\usepackage[preprint]{neurips_2026}
\usepackage[utf8]{inputenc} 
\usepackage[T1]{fontenc}    
\usepackage{hyperref}       
\usepackage{url}            
\usepackage{booktabs}       
\usepackage{amsfonts}       
\usepackage{nicefrac}       
\usepackage{microtype}      
\usepackage{xcolor}         
\usepackage{listings}
\usepackage{graphicx}

\usepackage{amssymb}
\usepackage{caption}        
\usepackage{subcaption}     
\usepackage{wrapfig}
\usepackage{xspace}
\usepackage{pifont}

\newcommand{\repoa}{BDE}
\newcommand{\repob}{BMQ}
\newcommand{\Tool}{Spec-Agent}

\newcommand{\ymark}{\textcolor{green!60!black}{\ding{51}}\xspace}
\newcommand{\nmark}{\textcolor{red!80!black}{\ding{55}}\xspace}
\newcommand{\no}{\nmark}
\newcommand{\yes}{\ymark}

\definecolor{dkgreen}{rgb}{0,0.6,0}
\definecolor{gray}{rgb}{0.5,0.5,0.5}
\definecolor{mauve}{rgb}{0.58,0,0.82}
\definecolor{dblue}{rgb}{0,71,150}

\definecolor{logicFOSL}{rgb}{0.10,0.30,0.75}
\definecolor{logicPSL}{rgb}{0.00,0.55,0.55}
\definecolor{logicFOL}{rgb}{0.85,0.45,0.00}
\definecolor{logicPL}{rgb}{0.75,0.10,0.20}
\newcommand{\logicFOSL}{\textbf{First Order Separation Logic}}
\newcommand{\logicPSL}{\textbf{Propositional Separation Logic}}
\newcommand{\logicFOL}{\textbf{First Order Logic}}
\newcommand{\logicPL}{\textbf{Propositional Logic}}

\usepackage{tcolorbox}
\tcbuselibrary{breakable}
\newcommand{\specbreaks}{%
  \sloppy%
  \emergencystretch=3em%
  \hbadness=10000%
  \tolerance=9999%
}
\newtcolorbox{blockFOSL}{breakable, colback=logicFOSL!5!white, colframe=logicFOSL!75!black, boxrule=0.3mm, arc=1.5mm, boxsep=1.5mm, before upper=\specbreaks}
\newtcolorbox{blockPSL}{breakable, colback=logicPSL!7!white, colframe=logicPSL!70!black, boxrule=0.3mm, arc=1.5mm, boxsep=1.5mm, before upper=\specbreaks}
\newtcolorbox{blockFOL}{breakable, colback=logicFOL!7!white, colframe=logicFOL!75!black, boxrule=0.3mm, arc=1.5mm, boxsep=1.5mm, before upper=\specbreaks}
\newtcolorbox{blockPL}{breakable, colback=logicPL!5!white, colframe=logicPL!70!black, boxrule=0.3mm, arc=1.5mm, boxsep=1.5mm, before upper=\specbreaks}

\newcommand{\bigast}{\mathop{\scalebox{2.5}{\raisebox{-0.2ex}{$\ast$}}}}%

\title{Agentic Separation Logic Specification Synthesis}

\author{
Tarun Suresh, David Korczynski and Julien Vanegue \vspace{2mm} \\
Bloomberg, NY, USA \\
\texttt{\{tsuresh10,dkorczynski,jvanegue\}@bloomberg.net}
}

\begin{document}

\maketitle

\begin{abstract}
Specification synthesis, the task of automatically inferring formal specifications from program implementations and natural language, is important for refactoring, transpilation, optimization, and verification, yet remains an open challenge for large C++ repositories. Existing LLM-based approaches fail to simultaneously scale to such repositories, produce specifications expressive enough to capture systems-code features such as dynamic memory and heap-allocated data structures, and systematically validate those specifications to rule out incorrect candidates. We present \Tool{}, an agentic system for synthesizing expressive, well-validated specifications across large C++ codebases. \Tool{} targets a \emph{ladder} of specification languages: propositional logic, first-order logic, propositional separation logic, and first-order separation logic. For each function, \Tool{} uses static analysis and runtime heap tracing to select the appropriate target specification language, generalizes existing functional tests into fuzz harnesses, and iteratively refines LLM-generated candidates via counterexample-guided feedback. We evaluate \Tool{} on open source C++ codebases comprising millions of lines of code. \Tool{} synthesizes valid specifications for \textbf{85\%} of target functions, with no false positives observed under fuzzing and expert validation, outperforming Claude Code Opus~4.6 at \textbf{10$\times$} lower token cost.
\end{abstract}

\section{Introduction}

A major problem arising from the success of LLM-based automatic programming~\cite{swebench} is the absence of correctness guarantees for the generated software. Software correctness requires that the code contains no unwanted bugs or security vulnerabilities that could compromise safety~\cite{shadowscode, secweakness}. High-profile incidents of compromised systems underscore the critical need for verifying the correctness of human-written code, let alone LLM-generated code~\cite{liu2023is}. It is therefore of foremost importance that the next generation of program synthesis agents be equipped with sufficient guardrails.
Unfortunately, software verification is undecidable in general~\cite{turing1936computable}, with many interesting program properties provably unprovable. While LLMs approximate such undecidable tasks with impressive success on benchmarks, their accuracy drops sharply on real-world repositories~\cite{sultan2026llms}.

In this paper, we address the problem of \emph{code contract} synthesis at scale. Code contracts~\cite{fahndrich2010static} capture the intent of code without contingent implementation details and serve as formal documentation by characterizing a function's \emph{preconditions} and \emph{postconditions}: they relate inputs to outputs via conditional logic and capture iterative behaviors in logical form. Code contract synthesis enables several important applications: (1) translators can use contracts to migrate legacy code from one language to another (e.g., C/C++ to Rust), (2) security tools can use contracts to guide bug finding, and (3) developers can use contracts as unambiguous documentation of function intent.

Prior work has shown the promise of LLMs for automated code contract synthesis~\cite{WenCSXQHLCT24,MaL0XB25,SunACTBDQL25,EndresFCL24,ugare2025fun2spec}, but no existing system meets the demands of real-world systems software in a single pipeline. An ideal contract synthesis system must: (1) scale to repositories spanning millions of lines of code, (2) produce preconditions and postconditions expressive enough to capture rich program features such as dynamic memory and heap-allocated data structures, and (3) systematically validate synthesized candidates to rule out incorrect ones.

\Tool{} addresses these gaps by combining two techniques that, to our knowledge, have not previously been brought together for specification synthesis. First, we adopt \textbf{separation logic}~\cite{o2019separation}, a program logic widely used in verifiers, to express not only the logical intent of a contract but also the resource-aware constraints it imposes on memory, a prerequisite for reasoning about the heap-manipulating code that pervades systems software. Second, we employ \textbf{fuzz testing} to drive contract validation: by generalizing each project's existing unit tests into fuzz harnesses, we obtain far broader execution coverage than unit tests alone, and use that coverage to stress-test every candidate contract before accepting it.

While fuzz testing is traditionally a bug-finding technique, we repurpose it as a strong \emph{pseudo-oracle} for specification inference: any input that violates a candidate contract constitutes direct evidence that the contract is wrong, and that input is fed back to the LLM as a counterexample in the next refinement attempt. This is particularly valuable for C++, where no well-established full-program verifier exists and prior verification attempts have proven prohibitively costly~\cite{vanegue2013towards}. To our knowledge, \Tool{} is the first system to couple fuzz testing with specification inference, and the first to do so at the scale of millions of lines of code. We have applied our system to large, widely-used open source C++ libraries.

\begin{figure}
    \centering
    \includegraphics[width=0.8\textwidth]{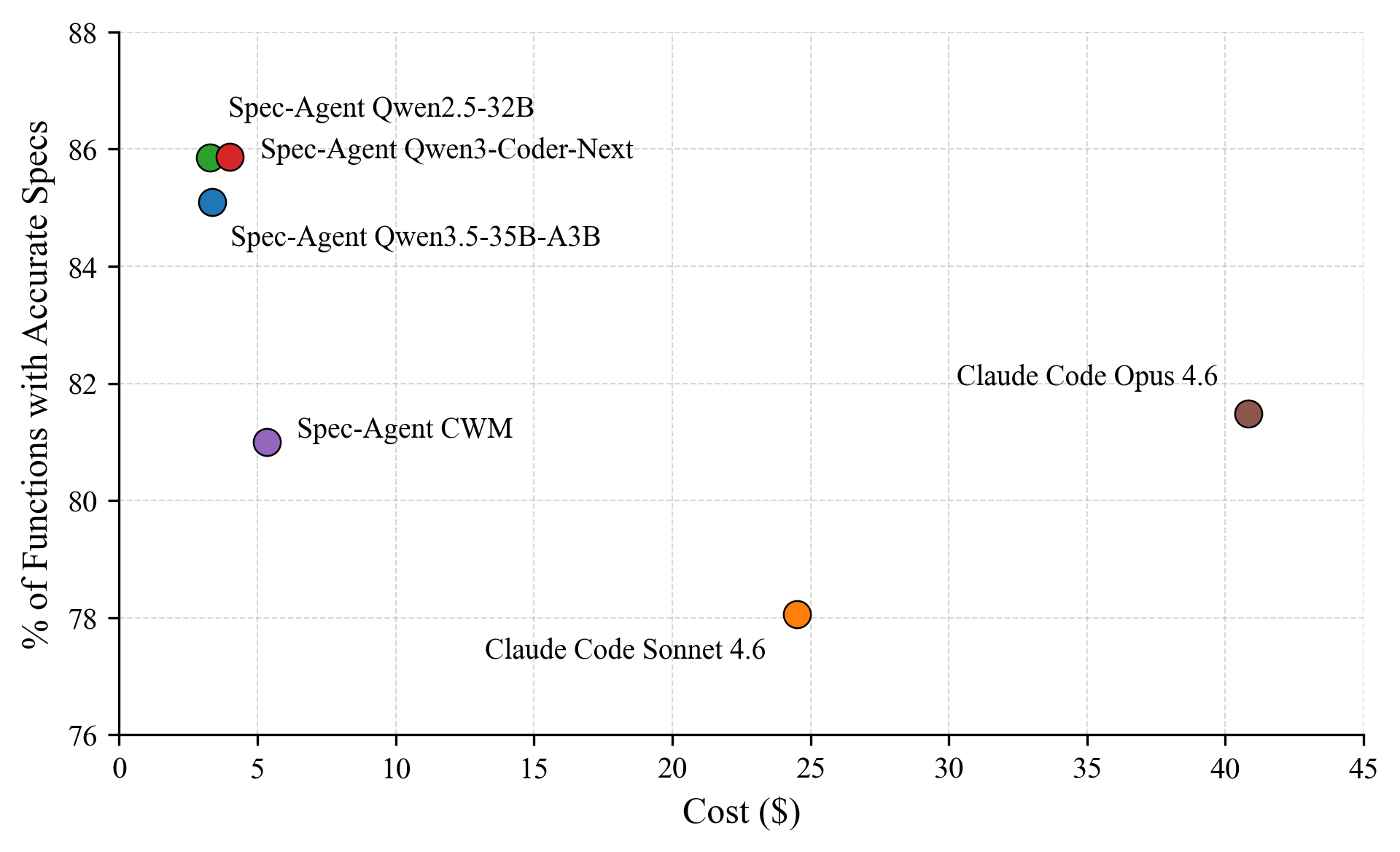}
    \caption{Specification validity and cost comparison of \Tool{} versus Claude Code.}
    \label{fig:cost}
\end{figure}

In summary, our work makes three contributions:

\begin{itemize}
    \item \Tool{}, an end-to-end agentic system that combines counterexample-guided contract inference with automatically generated fuzz harnesses for systematic validation of synthesized specifications.
    \item An adaptive specification-language selector spanning four program logics of increasing expressivity (propositional, first-order, propositional separation, and first-order separation logic), enabling \Tool{} to reason about conditional behavior, iterative loops, and heap structure within a single pipeline.
    \item An experimental study on widely-used projects totaling over four million lines of C++ code, in which \Tool{} synthesizes valid contracts for up to \textbf{85\%} of target functions with \textbf{no} false positives observed under fuzzing and expert validation, outperforming Claude Code Opus~4.6 at \textbf{10$\times$} lower token cost (Figure~\ref{fig:cost}).
\end{itemize}

The rest of the paper is organized as follows. Section~\ref{sec:background} recalls the program logics and specification languages used by \Tool{}. Section~\ref{sec:design} presents \Tool{}'s design. Section~\ref{sec:eval} reports our experimental study, including quantitative results and qualitative analysis of the inferred contracts. Section~\ref{sec:related} surveys related work, and Section~\ref{sec:concl} concludes.

\section{Background}
\label{sec:background}

We provide necessary background on program logics and specification languages. Figure~\ref{fig:speclang} shows the formal grammars of the specification languages supported by \Tool{}.

\paragraph{\textbf{Program Verification.}} The gold standard of software correctness is program verification, which guarantees that a program always implements the developer's intent. Pioneered by Floyd~\cite{floyd} and Hoare~\cite{hoare}, axiomatic verification provides the ultimate criterion for determining whether software satisfies its specification. In Hoare logic, we write $\{P\}\; c\; \{Q\}$ where $P$ is the precondition and $Q$ is the postcondition of code fragment $c$: any state satisfying $P$ leads to a state satisfying $Q$ after executing $c$, provided $c$ terminates (partial correctness). The role of a prover such as Frama-C~\cite{framac} is to discharge these proof obligations. Creating a specification is thus the first requirement for verification.

\paragraph{\textbf{Software Specification.}} From propositional logic and its familiar operators conjunction ($\wedge$), disjunction ($\vee$), and negation ($\lnot$), one can already model simple code contracts, including those with conditional code where guards are encoded as disjunctive implications, e.g., \texttt{if}~$p$~\texttt{then}~$q_1$~\texttt{else}~$q_2$ becomes $(p \Rightarrow q_1) \vee (\lnot p \Rightarrow q_2)$. Quantifiers are needed for code with loops, where quantifying over objects in a container ensures that they all satisfy desired constraints. Consider the program on the right of Figure \ref{fig:ex} that returns \texttt{true} if some element of a list satisfies predicate $P$, or \texttt{false} otherwise:

\begin{figure}[ht]
\centering

\begin{minipage}[t]{0.48\textwidth}
 \begin{lstlisting}[language=C++,
    basicstyle=\ttfamily\small,
    breaklines=true,
    mathescape=true
    ]
  pre: $x \mapsto v_1 \star y \mapsto v_2$
  swap(int *x, int *y) {
    int z = *x;
    *x = *y;
    *y = z;
  }
  post: $x \mapsto v_2 \star y \mapsto v_1$
\end{lstlisting}
\end{minipage}
\hfill
\begin{minipage}[t]{0.48\textwidth}
 \begin{lstlisting}[language=C++,
    basicstyle=\ttfamily\small,
    breaklines=true,
    mathescape=true
    ]
 lookup(std::list<int> &lst) {
   std::list<int>::iterator it = lst.begin();
   while (it != lst.end()) {
    if (P(it)) return true; it++;
   }
   return false;
  }
\end{lstlisting}
\end{minipage}

\caption{Left: A memory-manipulating program with separation logic annotations. Right: A function with a loop described with first-order specifications. More examples are given in Appendix \ref{sec:longex}.}
\label{fig:ex}
\end{figure}

The specification for the \texttt{lookup} function on the right side of Figure~\ref{fig:ex} becomes $\mathit{post}\colon (\forall x \in \mathit{lst}.\; \lnot P(x) \Rightarrow \mathit{ret} = \texttt{false}) \vee (\exists x \in \mathit{lst}.\; P(x) \Rightarrow \mathit{ret} = \texttt{true})$, with the precondition being simply \texttt{true} since no special conditions on the input are needed for the function to execute correctly.

\paragraph{\textbf{Separation Logic.}} Verifying programs with pointers, dynamic memory allocation, and heap-allocated data structures requires separation logic~\cite{o2019separation}, a modular program logic designed for low-level memory-manipulating programs. Separation logic introduces two key constructs: (1) a \emph{maps-to} operator $x \mapsto n$, denoting that memory cell $x$ contains value $n$, and (2) a \emph{separating conjunction} $\star$, where $p \star q$ asserts that the memory footprints of $p$ and $q$ are strictly disjoint. This is essential for describing heap data structures such as linked lists, trees, and graphs that pervade low-level systems code. It is also needed for simpler examples like \texttt{swap} on the left of Figure~\ref{fig:ex}: swapping a variable with itself is ill-defined, and the specification must explicitly state that $x$ and $y$ refer to distinct memory locations. Separation logic is at the core of state-of-the-art program verifier \texttt{infer}~\cite{infer}, enabling compositional reasoning for very large programs, a key enabler for the scalability of our approach.

\begin{figure}[!h]
\centering
\begin{tabular}{l|ll}
Logic                           &  Grammar & \\
\hline
Propositional Logic             & \texttt{p  } \hspace{1mm} := & $\texttt{true}$ $\mid$ $\texttt{false}$ $\mid$ $p$ $\wedge$ $p$ $\mid$ $p$ $\vee$ $p$ $\mid$ $\lnot p$ \\
First Order Logic               & \texttt{fp } := &  $p$ $\mid$ $\exists x.fp(x)$ $\mid$ $\forall x.fp(x)$\\
Propositional Separation Logic  & \texttt{sp } := &  $p$ $\mid$ $sp$ $\star$ $sp$ $\mid$ $x$ $\mapsto$ $n$ \\
First Order Separation Logic    & \texttt{fsp} := &  $sp$ $\mid$ $\exists x.fsp(x)$ $\mid$ $\forall x.fsp(x)$ \\
\hline
\end{tabular}
\caption{Program specification languages supported by \Tool{}.}
\label{fig:speclang}
\end{figure}

\paragraph{\textbf{Specification Inference.}} We address the dual problem of program verification: given a code fragment $c$, what are the appropriate $P$ and $Q$ for any safe execution? This is the specification inference problem, recovering the weakest precondition (the most permissive condition on inputs) and the strongest postcondition (the most informative condition on outputs given the inputs). Inferring preconditions and postconditions using LLM-assisted tools can enable program translation, full program verification, and serve as formal documentation.

\paragraph{\textbf{Fuzz Testing.}} Fuzzing is a technique for dynamically exploring code execution paths~\cite{AnEmpiricalStudyOfReliabilityOfUnixUtils, fuzzingbook2024}. Coverage-guided fuzzing, exemplified by engines such as libFuzzer~\cite{libfuzzer} (part of the LLVM toolchain), instruments the target software with a coverage-feedback mechanism and employs a genetic algorithm to discover unique inputs that trigger distinct code paths. Fuzzing is most frequently used for security testing, often combined with sanitizers~\cite{DBLP:conf/usenix/SerebryanyBPV12}, to identify inputs that trigger vulnerabilities. LLMs have shown promise for generating fuzz harnesses~\cite{Yang2025HarnessAgentSA}; in our work, we leverage coding agents to generate fuzz harnesses for their control-flow exploration capability, using them to stress-test generated contracts far beyond the coverage of unit tests alone.

\section{\Tool{} Design}
\label{sec:design}

\begin{figure}
    \centering
    \includegraphics[width=\textwidth]{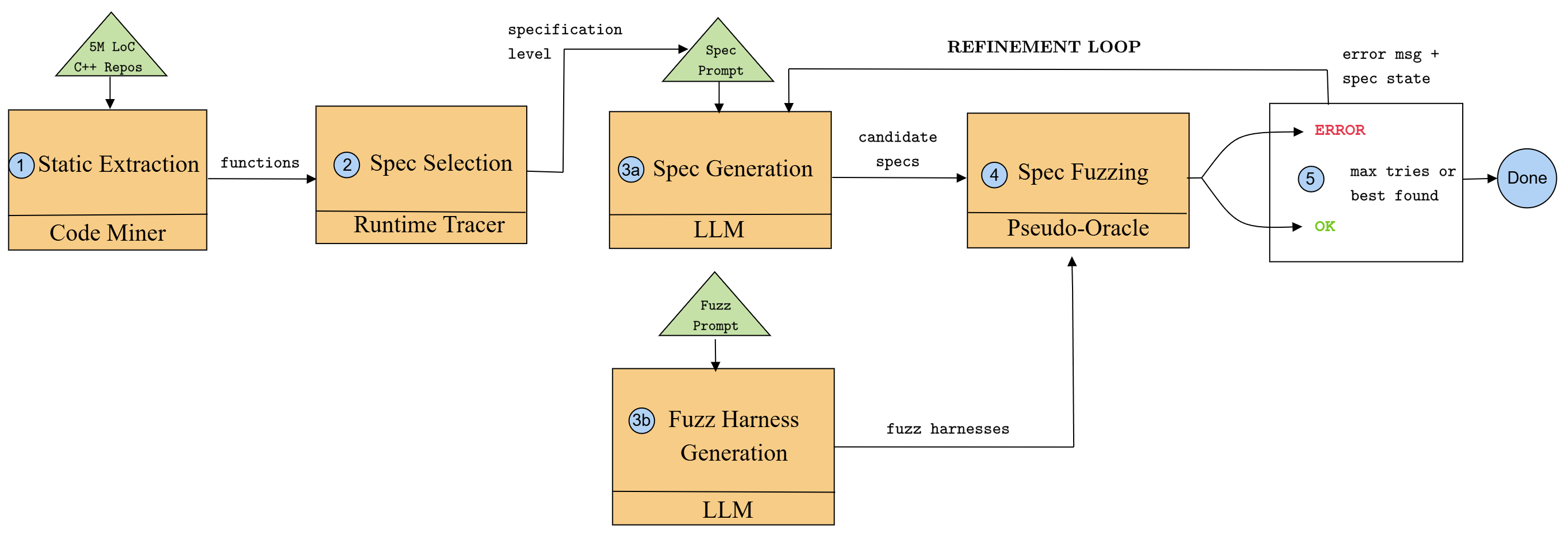}
    \caption{Inside \Tool{}: Automated fuzz-guided refinement of candidate specifications. The refinement loop iterates until the best possible specification is inferred or the maximum number of retries is reached.}
    \label{fig:class2spec}
\end{figure}

We describe the architecture and workflow of \Tool{} for inferring expressive specifications at the scale of millions of lines of low-level code (Figure~\ref{fig:class2spec}).

\paragraph{\textbf{Code Mining.}} The first step extracts per-function artifacts from the target repositories. For each function $f$, \Tool{} recovers its source code, in-tree documentation comments, and existing unit tests, then computes a set of program features used by all subsequent stages. Static features, presence of loops, conditionals, and induction-variable usage, are computed via Tree-sitter~\cite{treesitter} parsing of $f$. A dynamic feature, whether $f$ accesses the heap, is determined by running the existing tests of $f$ under heap tracing and analyzing the resulting heap profile.

\paragraph{\textbf{Fuzz Harness Generation.}} Fuzz harness generation consists of generalizing existing functional tests into fuzz harnesses. Functional tests target specific hard-coded inputs, while fuzz harnesses are input generators that exercise the function over a wide variety of executions. To produce a harness for function $f$, \Tool{} prompts a coding agent with $f$'s signature, source, and existing unit tests. The agent extracts shared test fixtures and setup logic, lifts the hard-coded inputs into fuzzer-controlled parameters, and emits a harness that reuses the original setup while invoking $f$ on fuzzer-generated inputs. This system constitutes a \emph{pseudo-oracle}: it can detect when a candidate specification is wrong (via counterexample), but cannot assert correctness, which would require a full C++ program verifier.

\paragraph{\textbf{Specification Language Selection.}}\label{par:specsel} \Tool{} supports four specification languages: propositional logic (Prop), first-order logic (FOL), propositional separation logic (Prop~SL), and first-order separation logic (FOSL). From the code-mining features, this stage selects a target language $L$ for $f$. Whenever $f$ contains a loop, first-order logic is required to model predicates over entire collections (e.g., elements of a list or container). Separation logic is selected when $f$ uses dynamic memory allocation. When both heap reasoning and quantification over collections are required, $L$ is set to FOSL.

\paragraph{\textbf{Specification Generation.}}  We instruct a large language model to generate a candidate specification $\langle P, Q \rangle$ for $f$ based on the selected specification language, the function's source code, and any documentation comments present in the project. Note that unit tests are not used during the specification generation phase, they are reserved for validation. This step is agnostic to the specific LLM used and treats it as a black-box generator. The prompt includes specific instructions for using the chosen specification language, up to 10 manually-written correct examples in that language, and the formal grammar of the selected logic.

\paragraph{\textbf{Specification Fuzz Testing.}} Each candidate is parsed against the grammar of $L$, compiled to an executable runtime assertion, and stress-tested under libFuzzer. Boolean, arithmetic, and implication operators map to their C++ counterparts. Bounded quantifiers compile to bounded loops over a range $[a,b)$, where the bounds are LLM-supplied expressions over the function's parameters and standard container/iterator interfaces (e.g., \texttt{0} and \texttt{v.size()} for a container, \texttt{begin} and \texttt{end} for an iterator pair). Separation-logic operators are realized over the observed heap state at runtime: the points-to atom $x \mapsto n$ asserts that the cell at address $x$ holds value $n$, and the separating conjunction $p \star q$ tracks each side's heap footprint and asserts that the two address sets are disjoint. Separating quantifiers extend $\star$ index-wise across the iteration range. The assertion is exercised by libFuzzer through harnesses lifted from the project's existing unit tests, ensuring the assertion is checked over a broad fuzzer-driven input distribution rather than a handful of hand-written cases. If any input violates the assertion, the candidate is rejected and the failing input together with the violated assertion is recorded as feedback for the next refinement attempt.

\begin{figure}
    \centering
    \includegraphics[width=\textwidth]{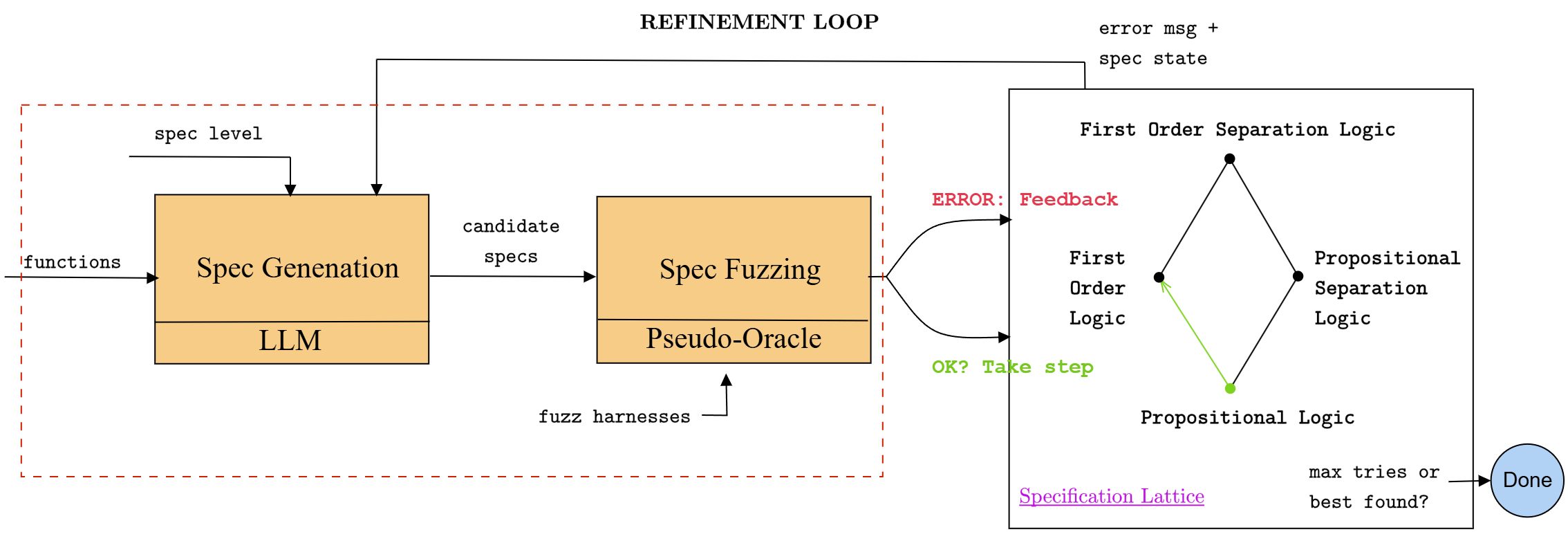}
    \caption{The specification lattice: \Tool{} maintains an order-theoretic lattice of specification state to track progress in synthesizing the most expressive specification, not only a correct one. Boxes in the dashed red rectangle summarize the full architecture view from Figure~\ref{fig:class2spec}.}
    \label{fig:speclattice}
\end{figure}

\paragraph{\textbf{Specification Refinement.}} Acceptance requires both that the fuzzer find no counterexample \emph{and} that the candidate be expressed in the targeted logic $L$. The four languages form a partial order $\sqsubseteq$ depicted in Figure~\ref{fig:speclattice}: Prop~$\sqsubseteq$~FOL, Prop~$\sqsubseteq$~Prop~SL, FOL~$\sqsubseteq$~FOSL, and Prop~SL~$\sqsubseteq$~FOSL, with Prop as the bottom element and FOSL as the top. A candidate sits at the lattice level its operators require, and \Tool{} accepts it if and only if $L \sqsubseteq \ell(\mathit{cand})$, that is, the candidate reaches at least the target's expressivity. This flexibility accounts for the heuristic nature of the selector: a function with a loop may not actually need a quantifier in its contract, and a function without an explicit loop may still benefit from a quantified summary. If $\ell(\mathit{cand}) \sqsubset L$ (e.g., a candidate that omits $\mapsto$ for a heap-touching function targeted at FOSL), the candidate is rejected with a structural diagnostic fed back to the LLM in the next attempt. The loop terminates when the candidate is both fuzz-validated and at the target lattice level, or when the retry budget is exhausted.

\section{Evaluation}
\label{sec:eval}

\paragraph{Experimental Setup.} All experiments run on a single machine with an Intel Xeon Silver 4410Y CPU and four NVIDIA H100 GPUs. \Tool{} is built on top of~\citep{ugare2025fun2spec}, using vLLM~\cite{vllm} for model serving, libFuzzer for fuzzing, and Clang~\cite{clang} with Tree-sitter for C++ parsing. We evaluate \Tool{} with four open-weight base models spanning general-purpose and code-specialized families at varying scales: Qwen3.5-35B-A3B~\cite{qwen3.5}, Qwen2.5-32B-Instruct~\cite{qwen2.5}, CWM~\cite{cwm}, and Qwen3-Coder-Next~\cite{qwen_qwen3_coder_next_tech_report}, each decoding greedily with a maximum of 400 tokens per refinement attempt. As a baseline, we run Claude Code~\cite{anthropic2024claude_code} with Claude Sonnet~4.6 and Claude Opus~4.6 on the same function prompts, unit tests, and fuzz harnesses with identical time and parameter budgets. Claude Code can spawn subagents that synthesize and refine candidate contracts under a budget of 6M weighted tokens per repository, larger than the budget consumed by any model with \Tool{}. We use Claude Code as the baseline because, as Table~\ref{tab:related-work-comparison} shows, no prior specification-synthesis system simultaneously targets large C++ codebases, both pre- and postconditions, and separation logic. To make dollar costs comparable, we report Anthropic's published pricing for Claude and Together AI's pricing~\cite{togetherai} for the open-weight models as a commercial reference point.

Following~\citep{ugare2025fun2spec}, we evaluate on two large open source C++ repositories: \repoa{}~\cite{BDE}, a modular C++ library suite of foundational data-structure and utility components used by thousands of developers, and \texttt{BlazingMQ}~\cite{BlazingMQ} (\repob{}), a high-performance, fault-tolerant message-queue library used by thousands of low-latency applications. Both ship with well-documented interfaces and strong test suites. We automatically extract the public functions that carry documentation and existing unit tests, yielding 651 functions for \repoa{} and 508 for \repob{}. To complement the automated pseudo-oracle, two experts in separation logic and C++ manually reviewed every generated fuzz harness and a random sample of 100 synthesized specifications per repository; both experts were in agreement. This sample is fully detailed in Appendix \ref{CPPEVAL}.

\paragraph{Accuracy.}
\label{sec:accuracy}

Table~\ref{tab:model_metrics} compares \Tool{} against Claude Code Sonnet~4.6 and Opus~4.6 on \repoa{} and \repob{}. For each configuration, we report: \emph{Test Valid}, the fraction of functions whose synthesized contract passes every instrumented test execution; \emph{Test Invalid}, the fraction for which the fuzzing pseudo-oracle produces a counterexample; \emph{Compilation Error}, the fraction whose assertion-instrumented build fails; \emph{Trivial}, the fraction of valid contracts that reduce to \texttt{true}; and \emph{Avg.~Atoms}, the mean number of logical atoms per synthesized contract.

\begin{table*}[ht]
    \centering
    \caption{Model Performance with \Tool{} on large C++ repositories}
    \vspace{-.07in}
    \setlength{\tabcolsep}{4pt} 
    \begin{tabular}{
    p{0.8cm}
    p{1.8cm}
    p{3.1cm}
    c c c c c
    }
    \toprule
    \textbf{Repo.} & \textbf{Method} & \textbf{Model} & \textbf{Test} & \textbf{Test} & \textbf{Compile} & \textbf{Trivial} & \textbf{Avg.} \\
    & & & \textbf{Valid (\%)} & \textbf{Invalid (\%)} & \textbf{Error (\%)} & \textbf{(\%)} & \textbf{Atoms} \\
    \midrule
    \repoa{} & Claude Code & Sonnet 4.6 & 78.06 & 4.52 & 17.42 & 10.89 & 2.24 \\
    \repoa{} & Claude Code & Opus 4.6 & 81.48 & 4.71 & 13.80 & 17.34 & 2.34 \\
    \repoa{} & \Tool{} & CWM & 81.00 & 0.00 & 19.00 & 8.30 & 2.78 \\
    \repoa{} & \Tool{} & Qwen3.5-35B-A3B & 85.10 & 0.00 & 14.90 & 10.29 & 3.21 \\
    \repoa{} & \Tool{} & Qwen2.5-32B & 85.85 & 0.96 & 13.18 & 5.63 & 3.86 \\
    \textbf{\repoa{}} & \textbf{\Tool{}} & \textbf{Qwen3-Coder-Next} & \textbf{85.87} & \textbf{0.00} & \textbf{14.13} & \textbf{5.84} & \textbf{3.35} \\
    \midrule
    \repob{} & Claude Code & Sonnet 4.6 & 62.64 & 2.16 & 35.19 & 14.18 & 2.13 \\
    \repob{} & Claude Code & Opus 4.6 & 66.69 & 2.19 & 31.12 & 11.92 & 2.46 \\
    \repob{} & \Tool{} & CWM & 69.77 & 0.00 & 30.23 & 8.94 & 3.07 \\
    \repob{} & \Tool{} & Qwen2.5-32B & 72.41 & 0.38 & 27.20 & 6.13 & 3.97 \\
    \repob{} & \Tool{} & Qwen3.5-35B-A3B & 76.97 & 0.20 & 22.84 & 17.13 & 4.05 \\
    \textbf{\repob{}} & \textbf{\Tool{}} & \textbf{Qwen3-Coder-Next} & \textbf{77.73} & \textbf{0.00} & \textbf{22.27} & \textbf{16.94} & \textbf{4.15} \\
    \bottomrule
    \end{tabular}
    \label{tab:model_metrics}
    \end{table*}

 On \repoa{}, the best \Tool{} configuration (Qwen3-Coder-Next) synthesizes valid specifications for 85.87\% of functions, compared to 81.48\% for the strongest Claude Code baseline (Claude Opus~4.6) and 78.06\% for Claude Sonnet~4.6. On \repob{}, the best \Tool{} configuration (also Qwen3-Coder-Next) reaches 77.73\%, compared to 66.69\% for Claude Opus~4.6 and 62.64\% for Claude Sonnet~4.6. Furthermore, on \repoa{} and \repob{}, the best \Tool{} configurations average 3.35 and 4.15 atoms per contract, respectively, compared to Claude Opus~4.6's 2.34 and 2.46. We corroborate the expressivity claim with a language-level breakdown. Table~\ref{tab:correct_specs} reports the number of functions with valid specifications for \Tool{} and Claude Code across both repositories at each specification language. \Tool{} synthesizes substantially more valid contracts using first-order quantifiers and separation-logic operators, while Claude Code is more heavily skewed toward propositional logic.

\begin{table*}[ht]
    \centering
    \caption{Number of correctly synthesized specifications per specification language}
    \vspace{-.07in}
    \setlength{\tabcolsep}{4pt}
    \begin{tabular}{
    p{0.8cm}
    p{1.8cm}
    p{3.1cm}
    c c c c
    }
    \toprule
    \textbf{Repo.} & \textbf{Method} & \textbf{Model Name} & \textbf{Prop} & \textbf{FOL} & \textbf{Prop SL} & \textbf{FOSL} \\
    \midrule
    \repoa{} & \Tool{} & Qwen3.5-35B-A3B & 289 & 200 & 52 & 13 \\
    \repoa{} & Claude Code & Claude Opus 4.6 & 369 & 118 & 42 & 1 \\
    \midrule
    \repob{} & \Tool{} & Qwen3.5-35B-A3B & 186 & 173 & 26 & 4 \\
    \repob{} & Claude Code & Claude Opus 4.6 & 236 & 71 & 29 & 3 \\
    \bottomrule
    \end{tabular}
    \label{tab:correct_specs}
    \end{table*}

\begin{figure*}[t]
    \centering
    \begin{subfigure}[t]{0.48\textwidth}
        \centering
        \includegraphics[width=\linewidth]{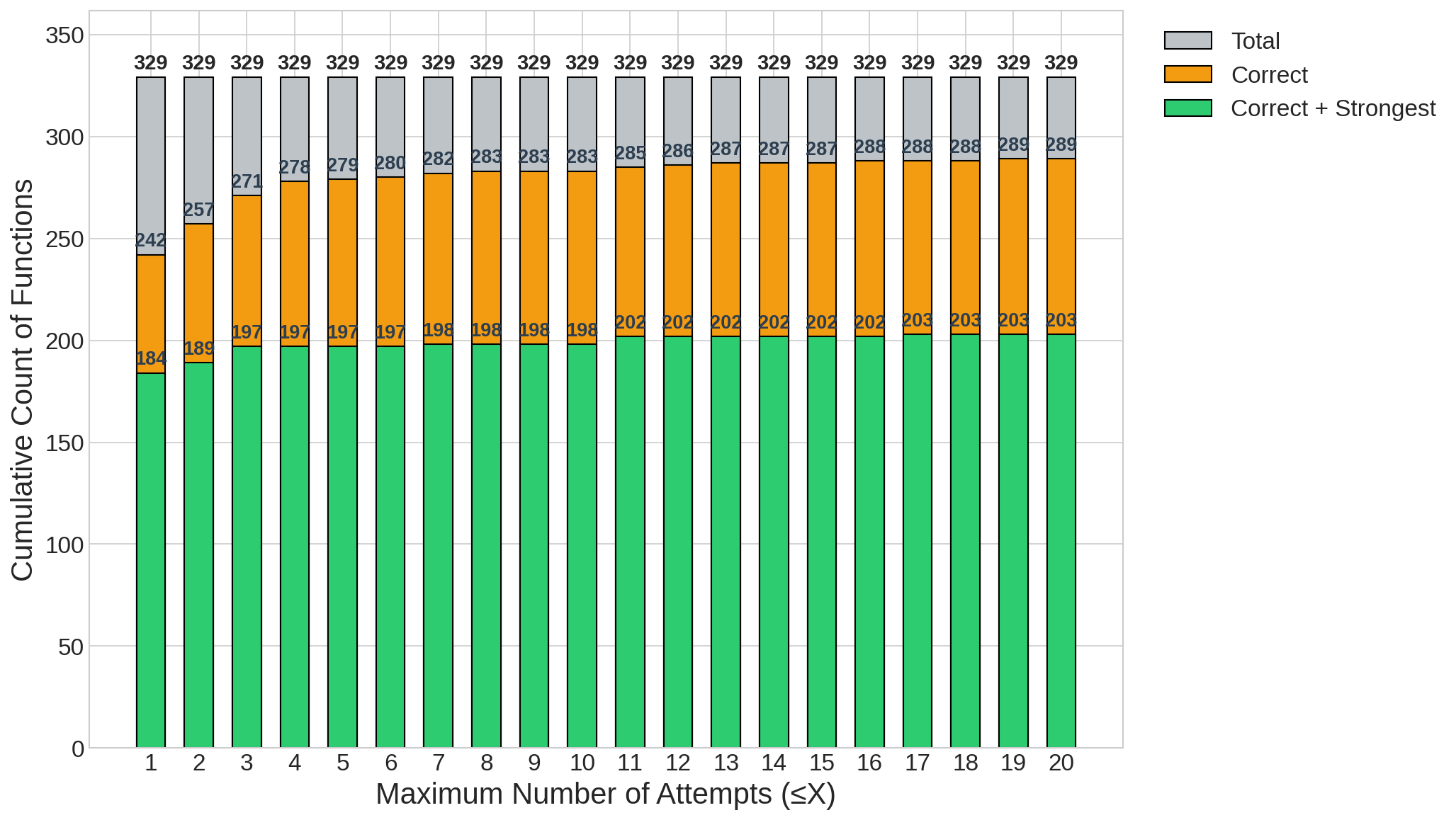}
        \caption{Propositional logic (Prop).}
        \label{fig:stacked-prop}
    \end{subfigure}
    \hfill
    \begin{subfigure}[t]{0.48\textwidth}
        \centering
        \includegraphics[width=\linewidth]{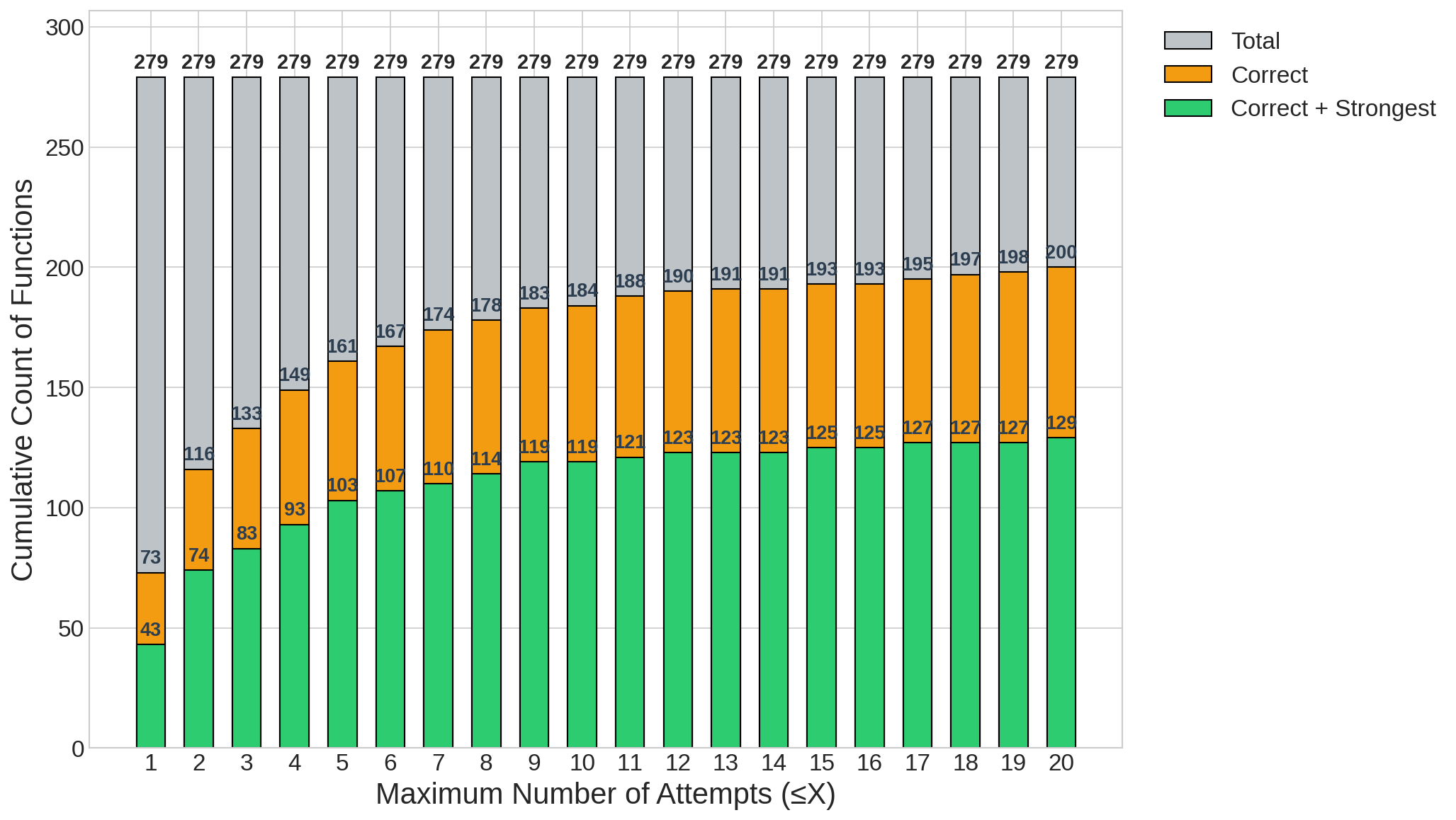}
        \caption{First-order logic (FOL).}
        \label{fig:stacked-fol}
    \end{subfigure}

    \vspace{0.6em}

    \begin{subfigure}[t]{0.48\textwidth}
        \centering
        \includegraphics[width=\linewidth]{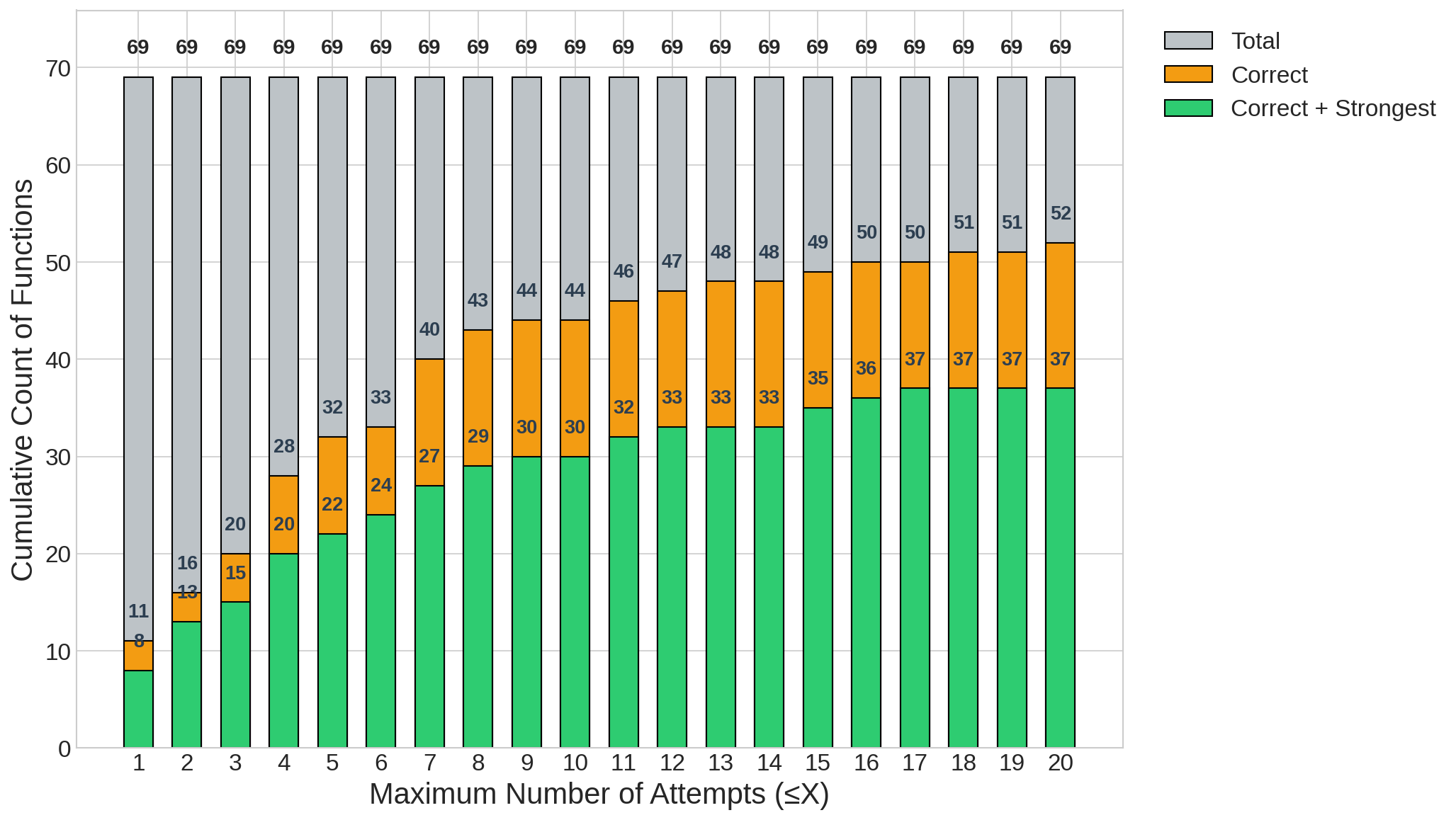}
        \caption{Propositional separation logic (Prop SL).}
        \label{fig:stacked-prop-sl}
    \end{subfigure}
    \hfill
    \begin{subfigure}[t]{0.48\textwidth}
        \centering
        \includegraphics[width=\linewidth]{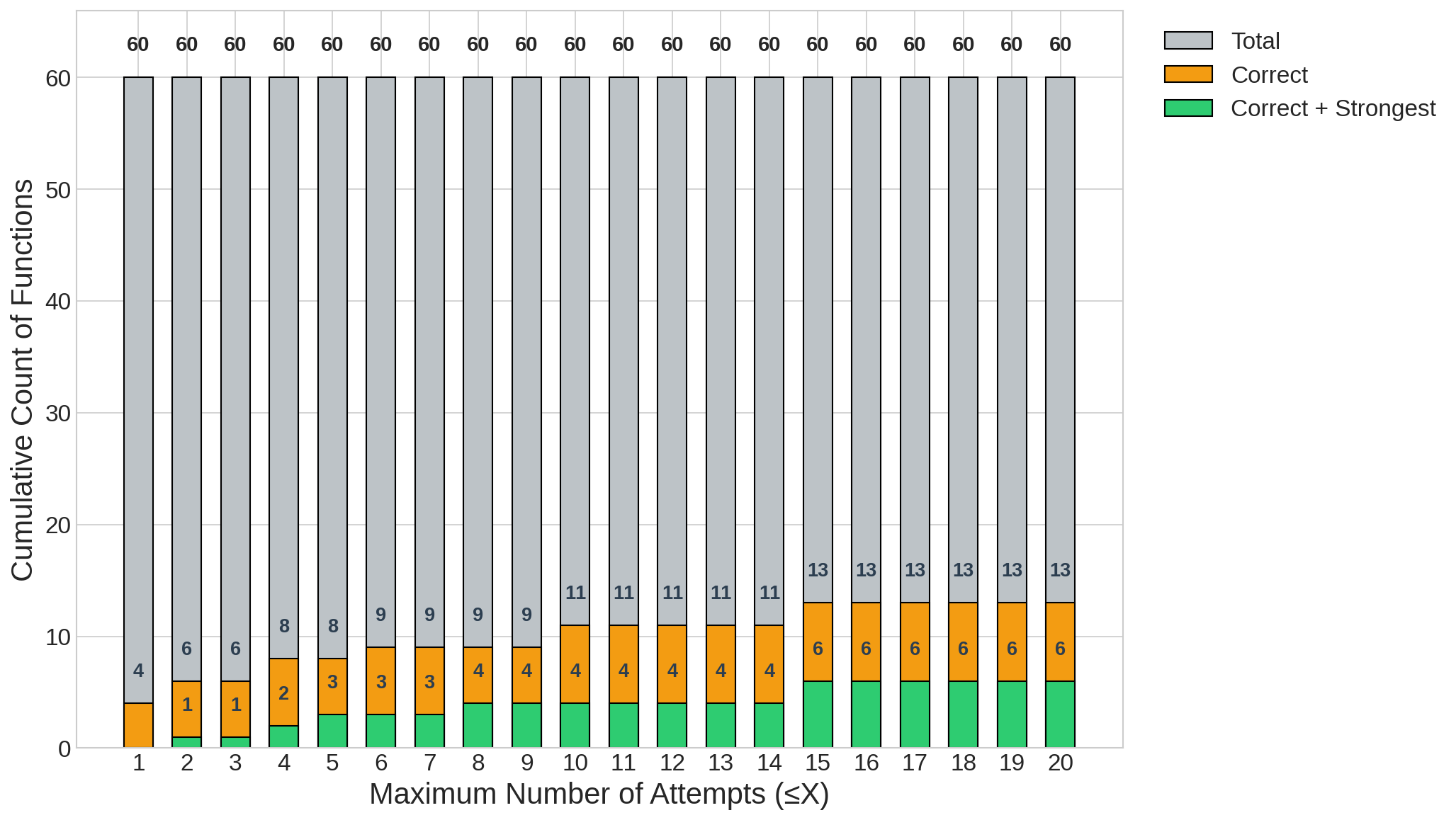}
        \caption{First-order separation logic (FOSL).}
        \label{fig:stacked-fosl}
    \end{subfigure}
    \caption{Number of \repoa{} functions with correctly synthesized specifications by number of refinement attempts, for each target specification language. Results shown for \Tool{} with Qwen3.5-35B-A3B.}
    \label{fig:stacked-attempts}
\end{figure*}

Figure~\ref{fig:stacked-attempts} characterizes how the number of valid synthesized specifications scales with the number of retry attempts on \repoa{} for \Tool{} with Qwen3.5-35B-A3B. The curves grow rapidly through the first five attempts, continue to climb with diminishing returns, and effectively plateau by 20 attempts.

We attribute these improvements to \Tool{}'s more structured refinement loop compared to Claude Code's. Claude Code delegates to subagents that can explore, manage their own context, and write and execute code. This exploration consumes substantially more tokens per step than a focused refinement pass, which, under a fixed compute budget, reduces the number of effective refinement attempts. In contrast, \Tool{}'s loop is deterministic: static and dynamic analyses select the target specification language before synthesis; each attempt makes a single LLM call with the candidate, the prompt, and any prior counterexample; the fuzzer runs after every attempt; and acceptance requires both that the fuzzer report no counterexample and that the candidate be expressed in the targeted logic.

\paragraph{\textbf{Qualitative Analysis.}} Figure~\ref{fig:stacked-attempts} distinguishes between \textbf{correct} contracts (yellow bars) and \textbf{strongest} contracts (green bars). Both categories of inferred contracts are correct in that they describe real program behaviors. The strongest contracts describe all possible behaviors of the function, constituting the majority of contracts inferred by \Tool{}. \Tool{} achieves this through its specification selection step (Section~\ref{sec:design}), which ensures that conditional code is represented as disjunctive implications, loops are expressed using quantifiers, and memory operations are captured with separation logic operators. A contract is strongest when it characterizes all expected executions as well as all error paths without exception. While \Tool{}'s performance is strong across most logics, it struggles to discover the strongest contract for the most complex functions where FOSL is selected as the target, as shown in Figure~\ref{fig:stacked-attempts}d. This may indicate a ceiling in current techniques for these most expressive specification languages, suggesting that new algorithmic insights may be required.

\paragraph{\textbf{Impact of Fuzzing on Contract Quality.}}
\label{sec:fuzz-impact}

\begin{wraptable}{r}{0.5\textwidth}
\centering
\caption{Fuzz testing results: \emph{Passed} = no counterexample, \emph{Violated} = counterexample found, \emph{Timeout} = fuzz binary timed out.}
\label{tab:fuzz_results}
\begin{tabular}{lrrr}
\toprule
Dataset & Passed & Violated & Timeout \\
\midrule
BDE & 98.3\% & 0.4\% & 1.2\% \\
BMQ & 92.0\% & 8.0\% & 0.0\% \\
\bottomrule
\end{tabular}
\end{wraptable}

A central design choice in \Tool{} is to generalize the repositories' existing unit tests into fuzz harnesses for more systematic validation of synthesized specifications. Every candidate entering the fuzzing stage has \emph{already} passed the project's existing unit tests under our assertion instrumentation. Despite this prior filter, Table~\ref{tab:fuzz_results} shows the fuzzer still rejects 0.4\% of \repoa{} survivors and 8.0\% of \repob{} survivors, demonstrating the value of fuzzing for improving specification validity beyond fixed test suites, especially for projects with weaker unit-test coverage like \repob{}. Each rejected contract corresponds to a counterexample that the project's unit tests alone would have missed, preventing an unsound contract from being accepted.

\begin{figure}[h]
    \centering
    \begin{subfigure}[t]{0.48\textwidth}
        \centering
        \includegraphics[width=\linewidth]{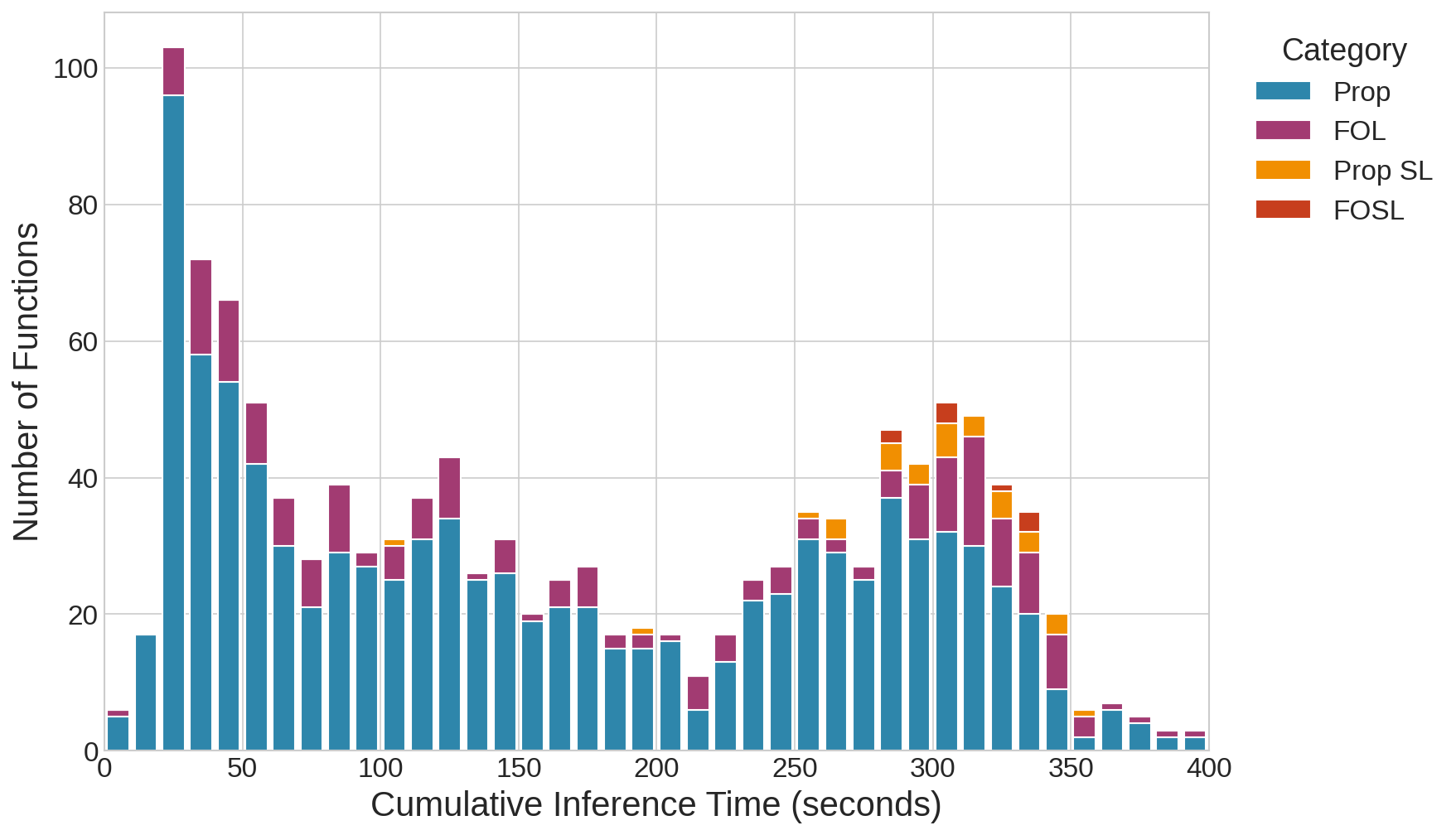}
        \caption{Cumulative inference time.}
        \label{fig:inference-time}
    \end{subfigure}
    \hfill
    \begin{subfigure}[t]{0.48\textwidth}
        \centering
        \includegraphics[width=\linewidth]{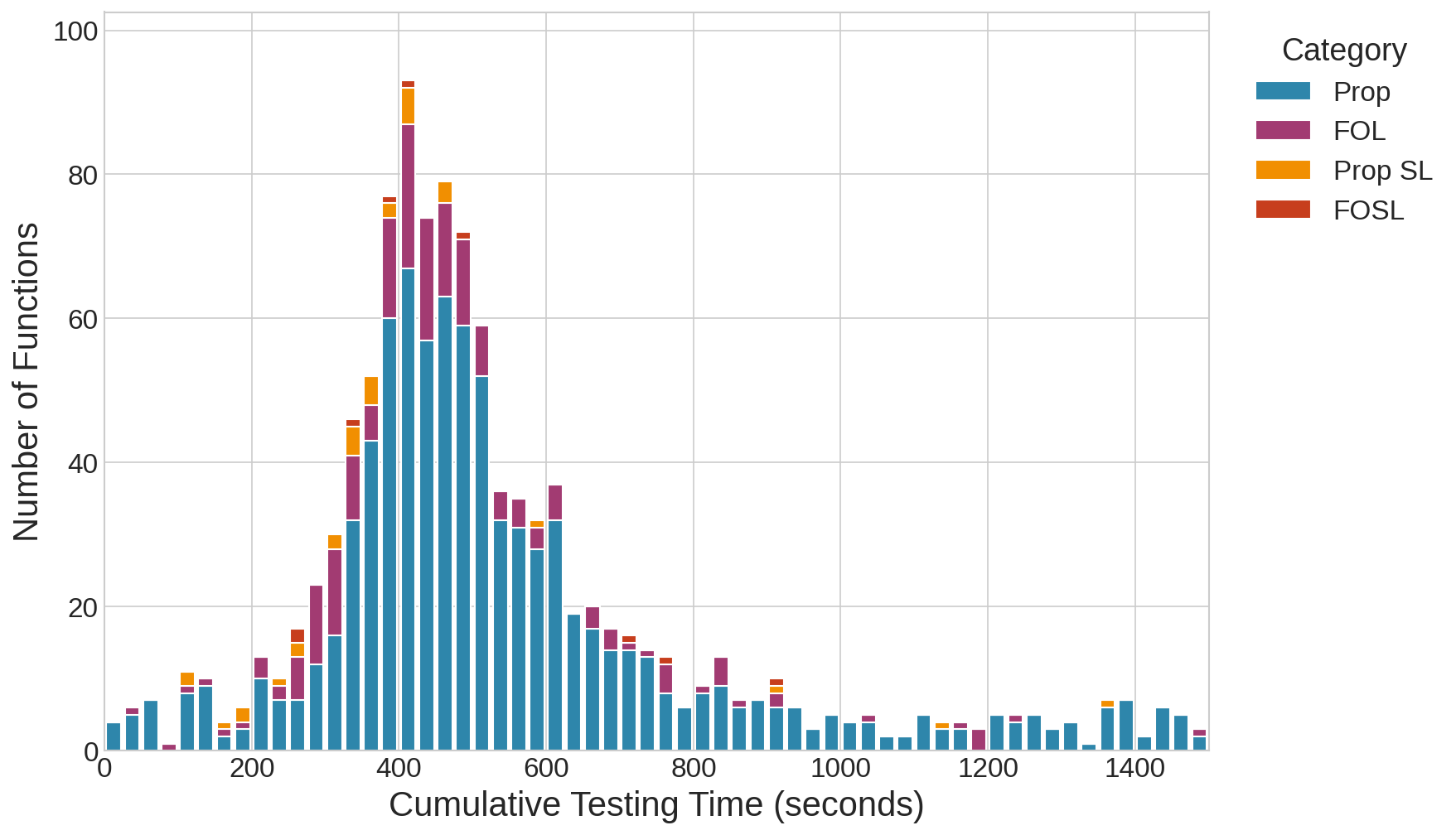}
        \caption{Cumulative testing time.}
        \label{fig:testing-time}
    \end{subfigure}
    \caption{Distribution of cumulative inference and testing time for each \repoa{} function with an accepted synthesized specification, for \Tool{} with Qwen3.5-35B-A3B.}
    \label{fig:inference-testing-time}
\end{figure}

Figure~\ref{fig:inference-testing-time} shows the runtime distribution of contract inference (via LLM) and contract testing (via fuzz testing). The choice of target logic does not noticeably influence runtime, as all specification languages are evenly distributed across the timing spectrum. Additionally, the overall runtime is dominated by testing rather than inference, with cumulative per-function means of 170 seconds for inference and 400 seconds for testing. Additional fuzzing-induced coverage measurements for BDE and BMQ are given in Appendix \ref{app:additional-stats}.

\section{Related Work}
\label{sec:related}

Earlier work in automated specification synthesis includes Daikon~\cite{ErnstPGMPTX07}, which infers likely pre- and postconditions from execution traces against a fixed invariant grammar, and SLING~\cite{LeZN19}, which extends this trace-based approach to separation logic for heap-manipulating C programs.

\begin{table*}[h]
    \centering
    \small
    \setlength{\tabcolsep}{5pt}
    \renewcommand{\arraystretch}{1.15}
    \caption{Comparison with prior work. \yes denotes the work has this capability, and \no denotes that the capability is not targeted. “Test Gen. / Verification” denotes either dynamically generated or fuzzed tests beyond fixed unit tests, or verifier-based validation of synthesized specifications. }
    \label{tab:related-work-comparison}
    \begin{tabular}{lcccccc}
    \toprule
    \textbf{Work} &
    \shortstack{\textbf{Large C++}\\\textbf{Projects}} &
    \shortstack{\textbf{Pre +}\\\textbf{Post}} &
    \shortstack{\textbf{Separation}\\\textbf{Logic}} &
    \shortstack{\textbf{Adaptive Spec}\\\textbf{Lang. Selection}} &
    \shortstack{\textbf{Test Gen. /}\\\textbf{Verification}} &
    \shortstack{\textbf{Counterex.}\\\textbf{Repair}} \\
    \midrule

    Daikon~\cite{ErnstPGMPTX07}               & \no  & \yes      & \no  & \no & \no       & \no       \\
    SLING~\cite{LeZN19}                & \no  & \yes      & \yes & \no & \no       & \no       \\
    NL2Postcond~\cite{EndresFCL24}          & \no  & \no       & \no  & \no & \no       & \no       \\
    LLM-SE~\cite{LiuW0CY24}      & \no  & \no       & \yes & \no & \yes      & \no       \\
    AutoSpec~\cite{WenCSXQHLCT24}             & \no  & \yes      & \no  & \no & \yes      & \no       \\
    ClassInvGen~\cite{SunACTBDQL25}          & \no  & \yes      & \no  & \no & \yes      & \yes      \\
    SpecGen~\cite{MaL0XB25}              & \no  & \yes      & \no  & \no & \yes      & \no       \\
    Fun2spec~\cite{ugare2025fun2spec}             & \yes & \no       & \no  & \no & \no       & \yes       \\
    \midrule
    \textbf{\Tool{}} (our work)  & \yes & \yes      & \yes & \yes & \yes     & \yes      \\
    \bottomrule
    \end{tabular}
    \end{table*}

Recent efforts have shifted toward LLM-based specification synthesis, which can leverage natural-language documentation, produce specifications outside any fixed grammar, and incorporate validator feedback into iterative refinement. AutoSpec~\cite{WenCSXQHLCT24} synthesizes pre- and postconditions for C and validates them with the Frama-C theorem prover; SpecGen~\cite{MaL0XB25} synthesizes specifications for Java using OpenJML and a mutation-and-select stage; ClassInvGen~\cite{SunACTBDQL25} synthesizes C++ class invariants using LLM-generated test sequences as the pseudo-oracle; LLM-SE~\cite{LiuW0CY24} synthesizes separation-logic loop invariants from a fine-tuned LLM with symbolic-execution validation; and NL2Postcond~\cite{EndresFCL24} translates natural-language docstrings into postconditions for Python and Java. Much of this prior work focuses on small benchmark programs rather than real-world codebases. Recently, Fun2Spec~\cite{ugare2025fun2spec} scales LLM postcondition synthesis to C++ codebases with millions of lines of code, but synthesizes only postconditions, uses propositional and first-order logic without separation logic, and validates against fixed unit tests only. As Table~\ref{tab:related-work-comparison} summarizes, \Tool{} is the first system that simultaneously scales to large C++ repositories and targets pre- and postcondition synthesis, separation logic, adaptive specification-language selection, fuzzing as a non-fixed pseudo-oracle, and counterexample-guided refinement in a unified pipeline.

\section{Limitations}
\label{sec:lim}

\paragraph{Fuzz Testing vs.\ Full Verification.} While our specifications are broadly fuzz-tested with high execution coverage and thousands of test inputs per specification (generalizing from narrow functional tests), \Tool{} does not employ full C++ verification using a logic prover. The most mature code contract tools do not support C++, and fuzz testing constitutes a strong, scalable alternative that supersedes functional testing, the typical approach used to validate inferred specifications in prior work. Implementing a solver for full C++ program verification is out of scope for this paper.

\paragraph{Strongest Specifications.} Our method produced no incorrect specifications under our fuzzing and expert validation regime, with fuzz testing shown to be a cost-effective proxy to full verification in practice. While we instruct the LLM to select logical operators based on the program's syntactic structure, LLMs are not always capable of generating the strongest possible specification capturing the most general behavior of the program (without contingent implementation details). As a result, some inferred contracts may represent only a subset of all possible program behaviors. This subset satisfies all functional tests and fuzz tests employed by \Tool{} for validation and therefore constitutes a useful approximation. To our knowledge, no existing algorithm can decide whether a given specification for a function is the strongest possible one.

\section{Conclusion}
\label{sec:concl}

\Tool{} is a new tool for agentic code specification synthesis, tested at scale on millions of lines of C++ code. \Tool{} synthesizes code contracts in a variety of expressive specification languages for memory-intensive code with loops and conditional reasoning, providing higher accuracy than frontier models at a fraction of the cost. \Tool{} employs fuzz testing to validate candidate contracts, providing broad code coverage for the synthesized specifications. Automated specification synthesis is key to generating formal code documentation, translating legacy code to modern programming languages, and identifying security vulnerabilities. We expect such capability to become essential as agentic software vulnerability discovery pipelines scale, making automated specification synthesis a necessary countermeasure.

\newpage

\bibliography{bibliography}

\newpage
\appendix

\section{Additional Repository and Fuzzing Statistics}
\label{app:additional-stats}

This appendix includes a distributional supplement to the main evaluation showing the per-harness branch coverage of the generated fuzz harnesses. The $X$ axis is given in logarithmic scale while the $Y$ axis is given in linear scale.

\begin{figure}[h]
    \centering
    \begin{subfigure}[t]{0.48\textwidth}
        \centering
        \includegraphics[width=\linewidth]{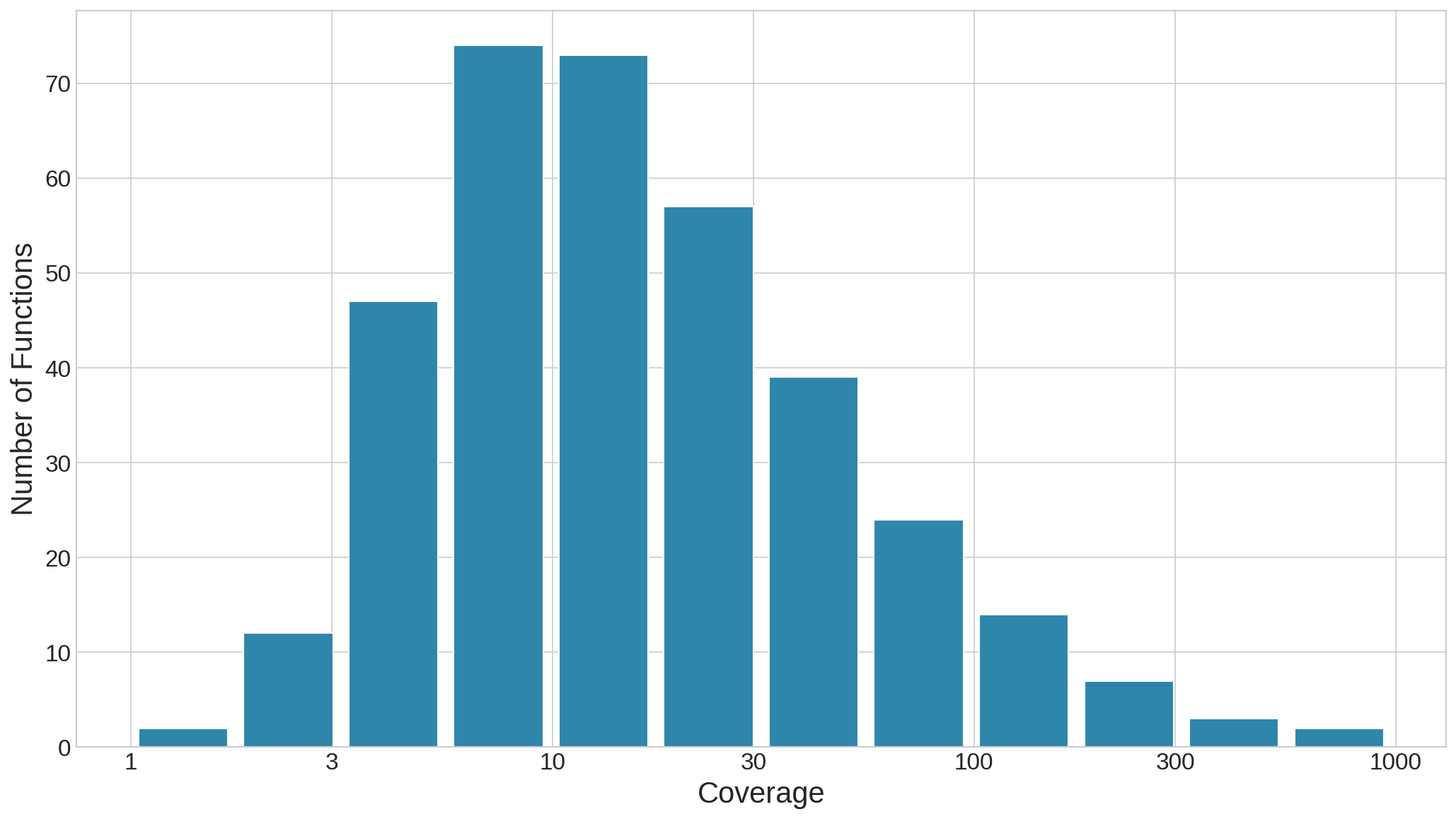}
        \caption{\repoa{}.}
        \label{fig:fuzz-cov-bde}
    \end{subfigure}
    \hfill
    \begin{subfigure}[t]{0.48\textwidth}
        \centering
        \includegraphics[width=\linewidth]{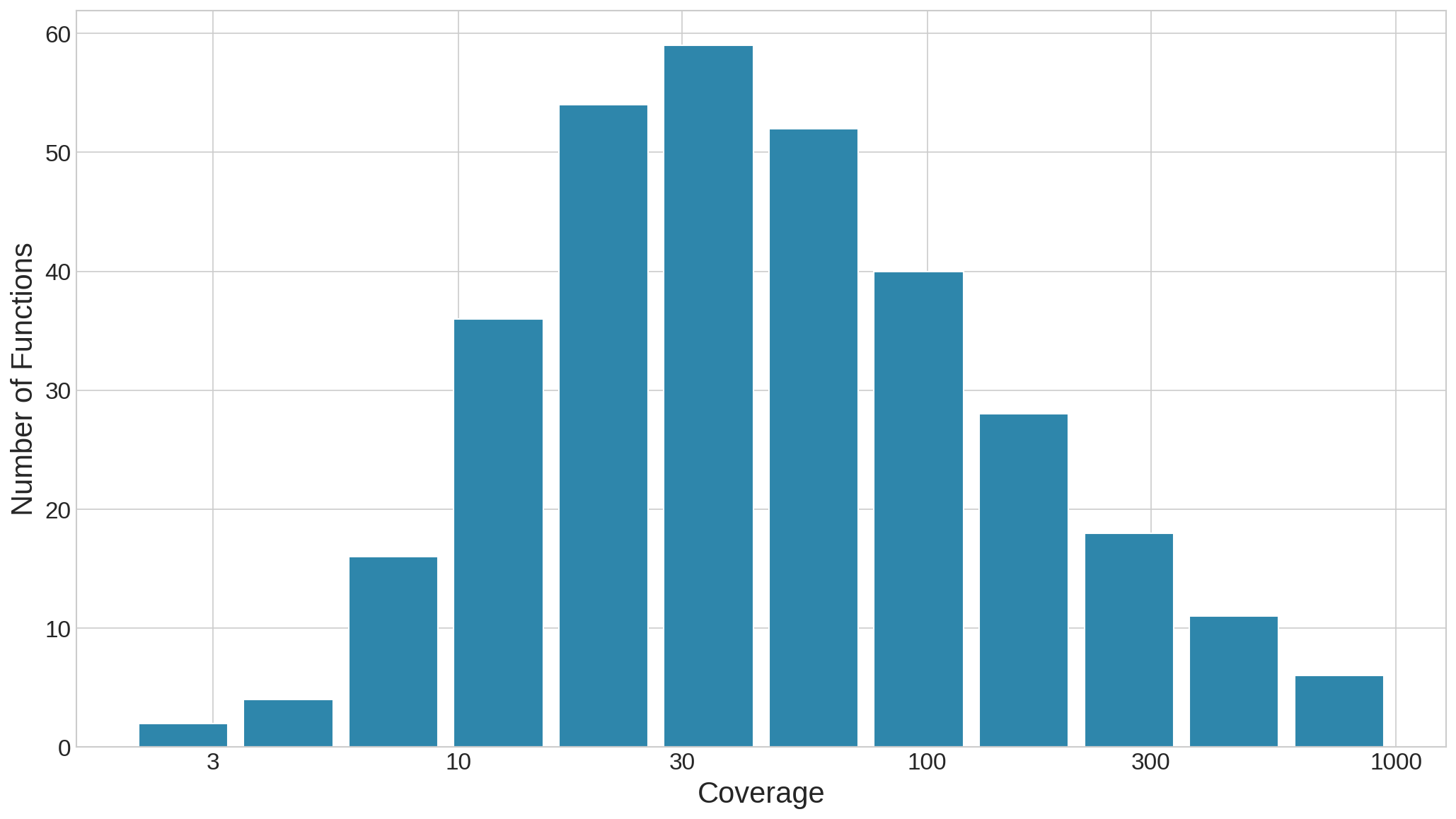}
        \caption{\repob{}.}
        \label{fig:fuzz-cov-bmq}
    \end{subfigure}
    \caption{Coverage distributions for \Tool{}'s LLM-generated fuzz harnesses on \repoa{} and \repob{}.}
    \label{fig:fuzz-coverage}
\end{figure}

We can witness that our fuzz testing harnesses achieve strong coverage for both projects, the number of reached basic blocks being normally distributed.

\begin{figure}[h]
    \centering
    \includegraphics[width=\linewidth]{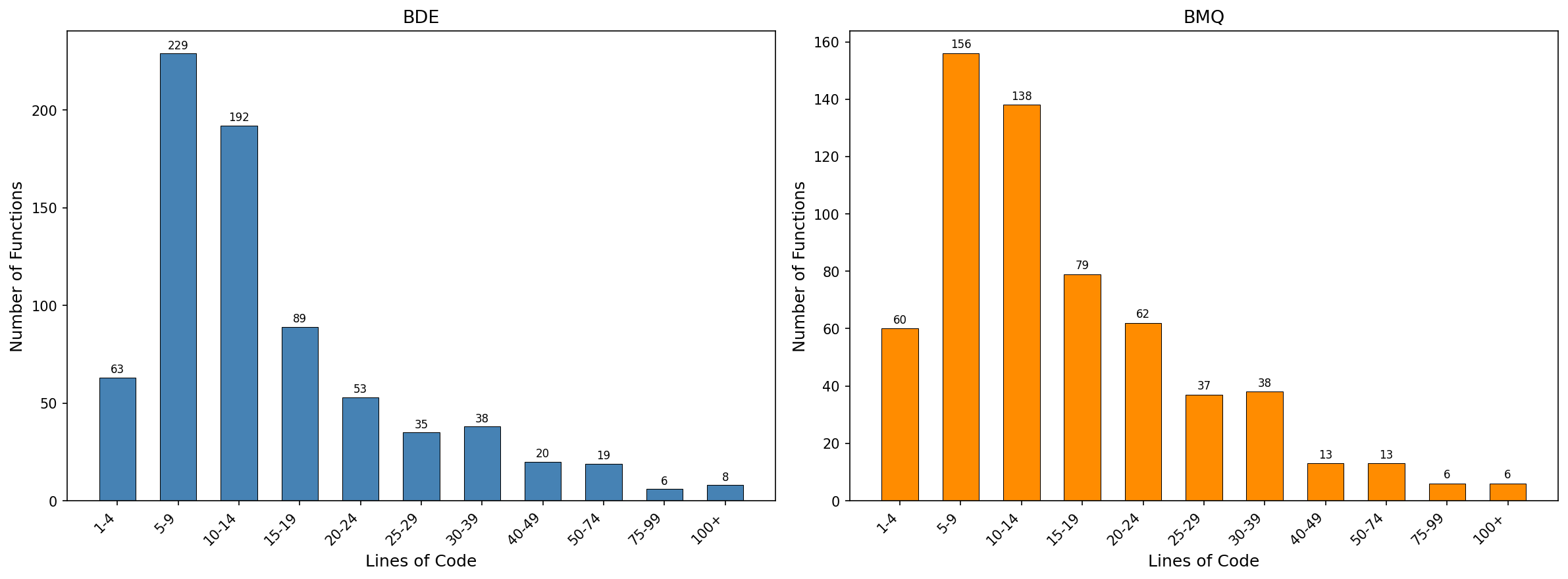}
    \caption{Distribution of lines of code per function for the evaluated functions in \repoa{} (left) and \repob{} (right).}
    \label{fig:loc-distribution}
\end{figure}

We also report the distribution of lines of code per function across the evaluated functions in \repoa{} and \repob{} in Figure~\ref{fig:loc-distribution}. Both repositories exhibit a similar long-tailed distribution: most functions are short (5--14 lines), while a non-trivial fraction span 20 or more lines, with a small number of functions exceeding 100 lines.

\newpage

\section{Cost Breakdown on \repoa{}}
\label{sec:cost}

\begin{table*}[h]
    \centering
    \caption{Per-model token usage and total dollar cost on \repoa{}. Token counts are in millions of tokens (Mtok).}
    \vspace{-.07in}
    \setlength{\tabcolsep}{4pt}
    \small
    \begin{tabular}{l l c c c}
    \toprule
    \textbf{Method} & \textbf{Model} & \textbf{Input (Mtok)} & \textbf{Output (Mtok)} & \textbf{Total Cost (\$)} \\
    \midrule
    \Tool{}     & Qwen2.5-32B        & 3.0  & 0.7 & 3.5 \\
    \Tool{}     & Qwen3-Coder-Next   & 3.2  & 1.0 & 4.0 \\
    \Tool{}     & Qwen3.5-35B-A3B    & 2.6  & 0.6 & 3.0 \\
    \Tool{}     & CWM                & 3.5  & 1.2 & 5.5 \\
    \midrule
    Claude Code & Sonnet 4.6         & 3.1  & 2.0 & 24.8 \\
    Claude Code & Opus 4.6           & 3.3  & 2.2 & 41.0 \\
    \bottomrule
    \end{tabular}
    \label{tab:cost-breakdown}
\end{table*}

To accompany the cost-versus-accuracy comparison in Figure~\ref{fig:cost}, Table~\ref{tab:cost-breakdown} reports the full breakdown of input tokens, output tokens, per-token pricing, and total dollar spend for every model we evaluate, on \repoa{}.  For \Tool{}, the reported tokens sum across every LLM call made by the system, fuzz harness generation, specification generation, and specification refinement. For Claude Code, the reported tokens aggregate over all top-level and subagent calls performed within the per-repository budget.

\newpage
\section{End-to-End Inferred Code Contracts on Large C++ Projects}
\label{sec:longex}

We illustrate two complete examples from BDE and BMQ, for which we infer pre and post conditions. We use the following notations throughout the contracts to encode specifications: \\

\begin{tabular}{r c l}
$ \texttt{FORALL}(a,i,b, P)$     & $\Leftrightarrow$ & $\bigwedge_{i=a}^{b} P(i)$ \\ 
\vspace{1mm}
$ \texttt{SEPFORALL}(a,i,b, P)$  & $\Leftrightarrow$ & $\bigast_{i=a}^{b} P(i)$ \\ 
\vspace{1mm}
$ \texttt{EXISTS}(a,i,b,P)$      & $\Leftrightarrow$ & $\exists i .a \leq i \wedge i < b \wedge P(i)$ when $\star$ or $\mapsto$ may not be in $P$\\
\vspace{1mm}
$ \texttt{SEPEXISTS}(a,i,b,P)$   & $\Leftrightarrow$ & $ \exists i .a \leq i \wedge i < b \wedge P(i)$ when $\star$ or $\mapsto$ may be in $P$ \\
\end{tabular}

\begin{figure*}[h]
\begin{tabular}{|l|}
\hline
 \begin{lstlisting}[language=C++,
    basicstyle=\ttfamily\small,
    breaklines=true,
    mathescape=true
    ]
// requires: true
// ensures:
// (__out == true ==>
//  EXIST(trans.begin(), trans.end(), it, descr == it->descr())) &&
// (__out == false ==>
// FORALL(trans.begin(), trans.end(), it, descr != it->descr()))
static bool containsDescriptor(
const bsl::vector<baltzo::ZoneinfoTransition>& trans,
const baltzo::LocalTimeDescriptor& descr) {
    auto it  = trans.begin();
    auto end = trans.end();
    for (; it != end; ++it)
        if (descr == it->descriptor())
            return true;
    return false;
}
\end{lstlisting} \\
\hline
\end{tabular}
\caption{First-order logic contract capturing a loop invariant directly in the postcondition (ensures), allowing any input in the precondition (requires).}
\end{figure*}
\begin{figure*}[h]
\begin{tabular}{|l|}
\hline
\begin{lstlisting}[language=C++,
basicstyle=\ttfamily\small,
breaklines=true,
label=lst:strongest,
mathescape=true
]
// requires:
// SEP_FORALL(0, rhs.d_len, i, (rhs.d_head_p + i)->d_value_p $\mapsto$ _)
// ensures:
// (__out == *this) && (__out.d_length == rhs.d_length) &&
// SEP_FORALL(0, rhs.d_length, i,
//       __out.d_head_p->d_value_p + i $\mapsto$ rhs.d_head_p->d_value_p + i)
AttributeContainerList&
AttributeContainerList::operator=(const AttributeContainerList& rhs) {
    if (&rhs != this) {
        // Append the 'rhs' elements to this list.
        Node **prevNextAddr = &d_head_p;
        node = new (*d_allocator.mechanism()) Node();
        for (iterator it = rhs.begin(); it != rhs.end(); ++it) {
            node->d_value_p = *it;
            node->d_next_p = 0;
            node->d_prevNextAddr_p = prevNextAddr;
            *prevNextAddr = node;
            prevNextAddr = &node->d_next_p;
            ++d_length;
        }
    }
    return *this;
}
\end{lstlisting} \\
\hline
\end{tabular}
\caption{A first-order separation logic contract capturing deep memory constraints between C++ class attributes using the separation logic \emph{points-to} ($\mapsto$) relation and separating conjunctions between subformulae (implicit in SEP\_FORALL), guaranteeing non-aliasing of memory cells.}
\end{figure*}

\section{Case Study}
\label{CPPEVAL}
We list 100 examples of rich C++ specifications synthesized by \Tool{} on \texttt{BDE}~\cite{BDE} and \texttt{BlazingMQ}~\cite{BlazingMQ}. The specifications are expressed in various program logics, including propositional logic (red box), first-order logic (yellow box), separation logic (green box), and first-order separation logic (blue box). The specification language is chosen as described in section \ref{sec:design} paragraph \ref{par:specsel}.

\begin{enumerate}
    \item \textbf{Filename:} balcl\_\allowbreak{}\allowbreak{}option\allowbreak{}type.\allowbreak{}c\allowbreak{}pp, \textbf{Function:} Option\allowbreak{}Type::\allowbreak{}\allowbreak{}print, \textbf{Logic:} \logicFOSL, \href{https://github.com/bloomberg/bde/blob/main/groups/bal/balcl/balcl_optiontype.cpp#L68}{\textcolor{blue!60!black}{[code]}}
    \begin{blockFOSL}
    \noindent \textbf{requires:} true \\
    \noindent \textbf{ensures:} \_\allowbreak{}\_\allowbreak{}out == stream \&\& (SEPFORALL(0,\allowbreak{} toAscii(value).\allowbreak{}size(),\allowbreak{} i,\allowbreak{} stream + i $\mapsto$ toAscii(value)[i]))
    \end{blockFOSL}
    \item \textbf{Filename:} baljsn\allowbreak{}\_\allowbreak{}encod\allowbreak{}er\_\allowbreak{}tes\allowbreak{}ttypes\allowbreak{}.\allowbreak{}cpp, \textbf{Function:} Encode\allowbreak{}rTestC\allowbreak{}hoiceW\allowbreak{}ithAll\allowbreak{}Catego\allowbreak{}riesEn\allowbreak{}umerat\allowbreak{}ion::\allowbreak{}f\allowbreak{}romStr\allowbreak{}ing, \textbf{Logic:} \logicFOSL, \href{https://github.com/bloomberg/bde/blob/main/groups/bal/baljsn/baljsn_encoder_testtypes.cpp#L477}{\textcolor{blue!60!black}{[code]}}
    \begin{blockFOSL}
    \noindent \textbf{requires:} result != 0 \&\& string != 0 \&\& string\allowbreak{}Length $>$= 0 \\
    \noindent \textbf{ensures:} (\_\allowbreak{}\_\allowbreak{}out == 0 ==$>$ SEPEXISTS(0,\allowbreak{} 2,\allowbreak{} i,\allowbreak{} (string\allowbreak{}Length == Encode\allowbreak{}rTestC\allowbreak{}hoiceW\allowbreak{}ithAll\allowbreak{}Catego\allowbreak{}riesEn\allowbreak{}umerat\allowbreak{}ion::\allowbreak{}E\allowbreak{}NUMERA\allowbreak{}TOR\_\allowbreak{}IN\allowbreak{}FO\_\allowbreak{}ARR\allowbreak{}AY[i].\allowbreak{}d\_\allowbreak{}nam\allowbreak{}eLengt\allowbreak{}h \&\& 0 == bsl::\allowbreak{}m\allowbreak{}emcmp(Encode\allowbreak{}rTestC\allowbreak{}hoiceW\allowbreak{}ithAll\allowbreak{}Catego\allowbreak{}riesEn\allowbreak{}umerat\allowbreak{}ion::\allowbreak{}E\allowbreak{}NUMERA\allowbreak{}TOR\_\allowbreak{}IN\allowbreak{}FO\_\allowbreak{}ARR\allowbreak{}AY[i].\allowbreak{}d\_\allowbreak{}nam\allowbreak{}e\_\allowbreak{}p,\allowbreak{} string,\allowbreak{} string\allowbreak{}Length) \&\& (*result == static\allowbreak{}\_\allowbreak{}cast$<$\allowbreak{}Encode\allowbreak{}rTestC\allowbreak{}hoiceW\allowbreak{}ithAll\allowbreak{}Catego\allowbreak{}riesEn\allowbreak{}umerat\allowbreak{}ion::\allowbreak{}V\allowbreak{}alue$>$(Encode\allowbreak{}rTestC\allowbreak{}hoiceW\allowbreak{}ithAll\allowbreak{}Catego\allowbreak{}riesEn\allowbreak{}umerat\allowbreak{}ion::\allowbreak{}E\allowbreak{}NUMERA\allowbreak{}TOR\_\allowbreak{}IN\allowbreak{}FO\_\allowbreak{}ARR\allowbreak{}AY[i].\allowbreak{}d\_\allowbreak{}value))))) \&\& (\_\allowbreak{}\_\allowbreak{}out == -1 ==$>$ SEPFORALL(0,\allowbreak{} 2,\allowbreak{} i,\allowbreak{} !(string\allowbreak{}Length == Encode\allowbreak{}rTestC\allowbreak{}hoiceW\allowbreak{}ithAll\allowbreak{}Catego\allowbreak{}riesEn\allowbreak{}umerat\allowbreak{}ion::\allowbreak{}E\allowbreak{}NUMERA\allowbreak{}TOR\_\allowbreak{}IN\allowbreak{}FO\_\allowbreak{}ARR\allowbreak{}AY[i].\allowbreak{}d\_\allowbreak{}nam\allowbreak{}eLengt\allowbreak{}h \&\& 0 == bsl::\allowbreak{}m\allowbreak{}emcmp(Encode\allowbreak{}rTestC\allowbreak{}hoiceW\allowbreak{}ithAll\allowbreak{}Catego\allowbreak{}riesEn\allowbreak{}umerat\allowbreak{}ion::\allowbreak{}E\allowbreak{}NUMERA\allowbreak{}TOR\_\allowbreak{}IN\allowbreak{}FO\_\allowbreak{}ARR\allowbreak{}AY[i].\allowbreak{}d\_\allowbreak{}nam\allowbreak{}e\_\allowbreak{}p,\allowbreak{} string,\allowbreak{} string\allowbreak{}Length))))
    \end{blockFOSL}
    \item \textbf{Filename:} ball\_\allowbreak{}a\allowbreak{}ttribu\allowbreak{}tecont\allowbreak{}ainerl\allowbreak{}ist.\allowbreak{}cp\allowbreak{}p, \textbf{Function:} Attrib\allowbreak{}uteCon\allowbreak{}tainer\allowbreak{}List::\allowbreak{}\allowbreak{}operat\allowbreak{}or=, \textbf{Logic:} \logicFOSL, \href{https://github.com/bloomberg/bde/blob/main/groups/bal/ball/ball_attributecontainerlist.cpp#L47}{\textcolor{blue!60!black}{[code]}}
    \begin{blockFOSL}
    \noindent \textbf{requires:} SEPFORALL(rhs.\allowbreak{}begin(),\allowbreak{} rhs.\allowbreak{}end(),\allowbreak{} it,\allowbreak{} *it $\mapsto$ \_\allowbreak{}) \\
    \noindent \textbf{ensures:} \_\allowbreak{}\_\allowbreak{}out == *this \&\& (\_\allowbreak{}\_\allowbreak{}out == *this ==$>$ SEPFORALL(rhs.\allowbreak{}begin(),\allowbreak{} rhs.\allowbreak{}end(),\allowbreak{} it,\allowbreak{} EXISTS(\_\allowbreak{}\_\allowbreak{}out.\allowbreak{}\allowbreak{}begin(),\allowbreak{} \_\allowbreak{}\_\allowbreak{}out.\allowbreak{}end(),\allowbreak{} jt,\allowbreak{} *jt == *it)))
    \end{blockFOSL}
    \item \textbf{Filename:} ball\_\allowbreak{}l\allowbreak{}og.\allowbreak{}cpp, \textbf{Function:} Log::\allowbreak{}f\allowbreak{}ormat, \textbf{Logic:} \logicFOSL, \href{https://github.com/bloomberg/bde/blob/main/groups/bal/ball/ball_log.cpp#L40}{\textcolor{blue!60!black}{[code]}}
    \begin{blockFOSL}
    \noindent \textbf{requires:} format != 0 \&\& buffer != 0 \&\& SEPFORALL(0,\allowbreak{} numBytes,\allowbreak{} i,\allowbreak{} buffer + i $\mapsto$ \_\allowbreak{}) \\
    \noindent \textbf{ensures:} (\_\allowbreak{}\_\allowbreak{}out == -1) $||$ (\_\allowbreak{}\_\allowbreak{}out != -1 \&\& SEPFORALL(0,\allowbreak{} \_\allowbreak{}\_\allowbreak{}out,\allowbreak{} i,\allowbreak{} buffer + i $\mapsto$ \_\allowbreak{}))
    \end{blockFOSL}
    \item \textbf{Filename:} ball\_\allowbreak{}u\allowbreak{}serfie\allowbreak{}lds.\allowbreak{}cp\allowbreak{}p, \textbf{Function:} UserFi\allowbreak{}elds::\allowbreak{}\allowbreak{}print, \textbf{Logic:} \logicFOSL, \href{https://github.com/bloomberg/bde/blob/main/groups/bal/ball/ball_userfields.cpp#L19}{\textcolor{blue!60!black}{[code]}}
    \begin{blockFOSL}
    \noindent \textbf{requires:} true \\
    \noindent \textbf{ensures:} (stream\allowbreak{}.\allowbreak{}bad() ==$>$ \_\allowbreak{}\_\allowbreak{}out == stream) \&\& (stream\allowbreak{}.\allowbreak{}good() ==$>$ SEPFORALL(0,\allowbreak{} length(),\allowbreak{} i,\allowbreak{} EXISTS(0,\allowbreak{} \_\allowbreak{}\_\allowbreak{}out.\allowbreak{}\allowbreak{}size(),\allowbreak{} j,\allowbreak{} (\_\allowbreak{}\_\allowbreak{}out + j) $\mapsto$ value(i))))
    \end{blockFOSL}
    \item \textbf{Filename:} balm\_\allowbreak{}m\allowbreak{}etricd\allowbreak{}escrip\allowbreak{}tion.\allowbreak{}c\allowbreak{}pp, \textbf{Function:} Metric\allowbreak{}Descri\allowbreak{}ption:\allowbreak{}:print, \textbf{Logic:} \logicFOSL, \href{https://github.com/bloomberg/bde/blob/main/groups/bal/balm/balm_metricdescription.cpp#L20}{\textcolor{blue!60!black}{[code]}}
    \begin{blockFOSL}
    \noindent \textbf{requires:} true \\
    \noindent \textbf{ensures:} \_\allowbreak{}\_\allowbreak{}out == stream \&\& (SEPFORALL(0,\allowbreak{} d\_\allowbreak{}cate\allowbreak{}gory\_\allowbreak{}p\allowbreak{}-$>$name().\allowbreak{}size(),\allowbreak{} i,\allowbreak{} stream + i $\mapsto$ d\_\allowbreak{}cate\allowbreak{}gory\_\allowbreak{}p\allowbreak{}-$>$name()[i]) $\star$ (stream + d\_\allowbreak{}cate\allowbreak{}gory\_\allowbreak{}p\allowbreak{}-$>$name().\allowbreak{}size() $\mapsto$ '.\allowbreak{}') $\star$ SEPFORALL(0,\allowbreak{} d\_\allowbreak{}name\allowbreak{}\_\allowbreak{}p.\allowbreak{}siz\allowbreak{}e(),\allowbreak{} j,\allowbreak{} stream + d\_\allowbreak{}cate\allowbreak{}gory\_\allowbreak{}p\allowbreak{}-$>$name().\allowbreak{}size() + 1 + j $\mapsto$ d\_\allowbreak{}name\_\allowbreak{}p[j]))
    \end{blockFOSL}
    \item \textbf{Filename:} balm\_\allowbreak{}m\allowbreak{}etrici\allowbreak{}d.\allowbreak{}cpp, \textbf{Function:} Metric\allowbreak{}Id::\allowbreak{}pr\allowbreak{}int, \textbf{Logic:} \logicFOSL, \href{https://github.com/bloomberg/bde/blob/main/groups/bal/balm/balm_metricid.cpp#L22}{\textcolor{blue!60!black}{[code]}}
    \begin{blockFOSL}
    \noindent \textbf{requires:} true \\
    \noindent \textbf{ensures:} (d\_\allowbreak{}desc\allowbreak{}riptio\allowbreak{}n\_\allowbreak{}p == 0 ==$>$ SEPFORALL(0,\allowbreak{} strlen("INVAL\allowbreak{}ID\_\allowbreak{}ID"),\allowbreak{} i,\allowbreak{} stream + i $\mapsto$ "INVAL\allowbreak{}ID\_\allowbreak{}ID"[i])) \&\& (d\_\allowbreak{}desc\allowbreak{}riptio\allowbreak{}n\_\allowbreak{}p != 0 ==$>$ SEPFORALL(0,\allowbreak{} strlen(*d\_\allowbreak{}des\allowbreak{}cripti\allowbreak{}on\_\allowbreak{}p),\allowbreak{} i,\allowbreak{} stream + i $\mapsto$ (*d\_\allowbreak{}des\allowbreak{}cripti\allowbreak{}on\_\allowbreak{}p)[i])) \&\& (\_\allowbreak{}\_\allowbreak{}out == stream)
    \end{blockFOSL}
    \item \textbf{Filename:} balm\_\allowbreak{}m\allowbreak{}etrics\allowbreak{}ample.\allowbreak{}\allowbreak{}cpp, \textbf{Function:} Metric\allowbreak{}Sample\allowbreak{}::\allowbreak{}prin\allowbreak{}t, \textbf{Logic:} \logicFOSL, \href{https://github.com/bloomberg/bde/blob/main/groups/bal/balm/balm_metricsample.cpp#L70}{\textcolor{blue!60!black}{[code]}}
    \begin{blockFOSL}
    \noindent \textbf{requires:} stream\allowbreak{}.\allowbreak{}good() \&\& (spaces\allowbreak{}PerLev\allowbreak{}el $>$= 0) \\
    \noindent \textbf{ensures:} \_\allowbreak{}\_\allowbreak{}out == stream \&\& (stream\allowbreak{}.\allowbreak{}bad() $||$ (stream\allowbreak{}.\allowbreak{}good() \&\& SEPFORALL(0,\allowbreak{} strlen(NL),\allowbreak{} i,\allowbreak{} stream + i $\mapsto$ NL[i])))
    \end{blockFOSL}
    \item \textbf{Filename:} balst\_\allowbreak{}\allowbreak{}stackt\allowbreak{}racepr\allowbreak{}inter.\allowbreak{}\allowbreak{}cpp, \textbf{Function:} balst:\allowbreak{}:opera\allowbreak{}tor$<$$<$, \textbf{Logic:} \logicFOSL, \href{https://github.com/bloomberg/bde/blob/main/groups/bal/balst/balst_stacktraceprinter.cpp#L39}{\textcolor{blue!60!black}{[code]}}
    \begin{blockFOSL}
    \noindent \textbf{requires:} true \\
    \noindent \textbf{ensures:} \_\allowbreak{}\_\allowbreak{}out == stream \&\& (stream + 0 $\mapsto$ bsl::\allowbreak{}endl $\star$ SEPFORALL(0,\allowbreak{} balst:\allowbreak{}:Stack\allowbreak{}TraceP\allowbreak{}rintUt\allowbreak{}il::\allowbreak{}pr\allowbreak{}intSta\allowbreak{}ckTrac\allowbreak{}e(stream,\allowbreak{} object\allowbreak{}.\allowbreak{}d\_\allowbreak{}max\allowbreak{}Frames\allowbreak{},\allowbreak{} object\allowbreak{}.\allowbreak{}d\_\allowbreak{}dem\allowbreak{}anglin\allowbreak{}gPrefe\allowbreak{}rredFl\allowbreak{}ag,\allowbreak{} object\allowbreak{}.\allowbreak{}d\_\allowbreak{}add\allowbreak{}itiona\allowbreak{}lIgnor\allowbreak{}eFrame\allowbreak{}s + 1) - 1,\allowbreak{} i,\allowbreak{} stream + i + 1 $\mapsto$ sep\_\allowbreak{}v))
    \end{blockFOSL}
    \item \textbf{Filename:} balst\_\allowbreak{}\allowbreak{}stackt\allowbreak{}racete\allowbreak{}stallo\allowbreak{}cator.\allowbreak{}\allowbreak{}cpp, \textbf{Function:} StackT\allowbreak{}raceTe\allowbreak{}stAllo\allowbreak{}cator:\allowbreak{}:alloc\allowbreak{}ate, \textbf{Logic:} \logicFOSL, \href{https://github.com/bloomberg/bde/blob/main/groups/bal/balst/balst_stacktracetestallocator.cpp#L340}{\textcolor{blue!60!black}{[code]}}
    \begin{blockFOSL}
    \noindent \textbf{requires:} size $>$= 0 \\
    \noindent \textbf{ensures:} (size == 0 \&\& \_\allowbreak{}\_\allowbreak{}out == 0) $||$ (size != 0 \&\& \_\allowbreak{}\_\allowbreak{}out != 0 \&\& (\_\allowbreak{}\_\allowbreak{}out $\mapsto$ \_\allowbreak{} $\star$ (\_\allowbreak{}\_\allowbreak{}out - 1) $\mapsto$ \_\allowbreak{} $\star$ SEPFORALL(0,\allowbreak{} d\_\allowbreak{}trac\allowbreak{}eBuffe\allowbreak{}rLengt\allowbreak{}h,\allowbreak{} i,\allowbreak{} (\_\allowbreak{}\_\allowbreak{}out - 1 - i) $\mapsto$ \_\allowbreak{})))
    \end{blockFOSL}
    \item \textbf{Filename:} balst\_\allowbreak{}\allowbreak{}stackt\allowbreak{}raceut\allowbreak{}il.\allowbreak{}cpp, \textbf{Function:} findBa\allowbreak{}sename, \textbf{Logic:} \logicPL, \href{https://github.com/bloomberg/bde/blob/main/groups/bal/balst/balst_stacktraceutil.cpp#L40}{\textcolor{blue!60!black}{[code]}}
    \begin{blockPL}
    \noindent \textbf{requires:} pathName != nullptr \&\& strlen(pathName) $>$ 0 \\
    \noindent \textbf{ensures:} (\_\allowbreak{}\_\allowbreak{}out $>$= pathName) \&\& (\_\allowbreak{}\_\allowbreak{}out == pathName $||$ (*(\_\allowbreak{}\_\allowbreak{}out - 1) == '/' $||$ *(\_\allowbreak{}\_\allowbreak{}out - 1) == '\textbackslash{}\textbackslash{}'))
    \end{blockPL}
    \item \textbf{Filename:} baltzo\allowbreak{}\_\allowbreak{}zonei\allowbreak{}nfo.\allowbreak{}cp\allowbreak{}p, \textbf{Function:} contai\allowbreak{}nsDesc\allowbreak{}riptor, \textbf{Logic:} \logicFOSL, \href{https://github.com/bloomberg/bde/blob/main/groups/bal/baltzo/baltzo_zoneinfo.cpp#L25}{\textcolor{blue!60!black}{[code]}}
    \begin{blockFOSL}
    \noindent \textbf{requires:} (res\_\allowbreak{}tmp == true ==$>$ SEPEXISTS(0,\allowbreak{} transi\allowbreak{}tions.\allowbreak{}\allowbreak{}size(),\allowbreak{} i,\allowbreak{} transi\allowbreak{}tions[i].\allowbreak{}descr\allowbreak{}iptor() == descri\allowbreak{}ptor)) \&\& (res\_\allowbreak{}tmp == false ==$>$ SEPFORALL(0,\allowbreak{} transi\allowbreak{}tions.\allowbreak{}\allowbreak{}size(),\allowbreak{} i,\allowbreak{} transi\allowbreak{}tions[i].\allowbreak{}descr\allowbreak{}iptor() != descri\allowbreak{}ptor)) \\
    \noindent \textbf{ensures:} (\_\allowbreak{}\_\allowbreak{}out == true ==$>$ SEPEXISTS(0,\allowbreak{} transi\allowbreak{}tions.\allowbreak{}\allowbreak{}size(),\allowbreak{} i,\allowbreak{} transi\allowbreak{}tions[i].\allowbreak{}descr\allowbreak{}iptor() == descri\allowbreak{}ptor)) \&\& (\_\allowbreak{}\_\allowbreak{}out == false ==$>$ SEPFORALL(0,\allowbreak{} transi\allowbreak{}tions.\allowbreak{}\allowbreak{}size(),\allowbreak{} i,\allowbreak{} transi\allowbreak{}tions[i].\allowbreak{}descr\allowbreak{}iptor() != descri\allowbreak{}ptor))
    \end{blockFOSL}
    \item \textbf{Filename:} baltzo\allowbreak{}\_\allowbreak{}zonei\allowbreak{}nfobin\allowbreak{}aryrea\allowbreak{}der.\allowbreak{}cp\allowbreak{}p, \textbf{Function:} areAll\allowbreak{}Printa\allowbreak{}ble, \textbf{Logic:} \logicFOSL, \href{https://github.com/bloomberg/bde/blob/main/groups/bal/baltzo/baltzo_zoneinfobinaryreader.cpp#L133}{\textcolor{blue!60!black}{[code]}}
    \begin{blockFOSL}
    \noindent \textbf{requires:} buffer != 0 \&\& length $>$= 0 \&\& SEPFORALL(0,\allowbreak{} length,\allowbreak{} i,\allowbreak{} (buffer + i $\mapsto$ \_\allowbreak{})) \\
    \noindent \textbf{ensures:} (\_\allowbreak{}\_\allowbreak{}out == true ==$>$ SEPFORALL(0,\allowbreak{} length,\allowbreak{} i,\allowbreak{} (buffer + i $\mapsto$ sep\_\allowbreak{}v) \&\& bdlb::\allowbreak{}\allowbreak{}CharTy\allowbreak{}pe::\allowbreak{}is\allowbreak{}Print(sep\_\allowbreak{}v))) \&\& (\_\allowbreak{}\_\allowbreak{}out == false ==$>$ SEPEXISTS(0,\allowbreak{} length,\allowbreak{} i,\allowbreak{} (buffer + i $\mapsto$ sep\_\allowbreak{}v) \&\& !bdlb:\allowbreak{}:CharT\allowbreak{}ype::\allowbreak{}i\allowbreak{}sPrint(sep\_\allowbreak{}v)))
    \end{blockFOSL}
    \item \textbf{Filename:} balxml\allowbreak{}\_\allowbreak{}prefi\allowbreak{}xstack\allowbreak{}.\allowbreak{}cpp, \textbf{Function:} Prefix\allowbreak{}Stack:\allowbreak{}:looku\allowbreak{}pNames\allowbreak{}paceId, \textbf{Logic:} \logicFOSL, \href{https://github.com/bloomberg/bde/blob/main/groups/bal/balxml/balxml_prefixstack.cpp#L109}{\textcolor{blue!60!black}{[code]}}
    \begin{blockFOSL}
    \noindent \textbf{requires:} true \\
    \noindent \textbf{ensures:} (SEPEXISTS(0,\allowbreak{} d\_\allowbreak{}numP\allowbreak{}refixe\allowbreak{}s,\allowbreak{} i,\allowbreak{} (d\_\allowbreak{}pref\allowbreak{}ixes[i].\allowbreak{}first == prefix) \&\& (\_\allowbreak{}\_\allowbreak{}out == d\_\allowbreak{}pref\allowbreak{}ixes[i].\allowbreak{}second))) $||$ (SEPFORALL(0,\allowbreak{} d\_\allowbreak{}numP\allowbreak{}refixe\allowbreak{}s,\allowbreak{} i,\allowbreak{} (d\_\allowbreak{}pref\allowbreak{}ixes[i].\allowbreak{}first != prefix)) \&\& (\_\allowbreak{}\_\allowbreak{}out == lookup\allowbreak{}Predef\allowbreak{}inedPr\allowbreak{}efix(prefix).\allowbreak{}d\_\allowbreak{}nsid))
    \end{blockFOSL}
    \item \textbf{Filename:} balxml\allowbreak{}\_\allowbreak{}types\allowbreak{}parser\allowbreak{}util.\allowbreak{}c\allowbreak{}pp, \textbf{Function:} parseU\allowbreak{}nsigne\allowbreak{}dInt, \textbf{Logic:} \logicFOSL, \href{https://github.com/bloomberg/bde/blob/main/groups/bal/balxml/balxml_typesparserutil.cpp#L228}{\textcolor{blue!60!black}{[code]}}
    \begin{blockFOSL}
    \noindent \textbf{requires:} inputL\allowbreak{}ength $>$ 0 \&\& SEPFORALL(0,\allowbreak{} inputL\allowbreak{}ength,\allowbreak{} i,\allowbreak{} (input + i $\mapsto$ \_\allowbreak{} \&\& input[i] $>$= '0' \&\& input[i] $<$= '9')) \\
    \noindent \textbf{ensures:} (\_\allowbreak{}\_\allowbreak{}out == BAEXML\allowbreak{}\_\allowbreak{}SUCCE\allowbreak{}SS) $||$ (\_\allowbreak{}\_\allowbreak{}out == BAEXML\allowbreak{}\_\allowbreak{}FAILU\allowbreak{}RE)
    \end{blockFOSL}
    \item \textbf{Filename:} balxml\allowbreak{}\_\allowbreak{}types\allowbreak{}printu\allowbreak{}til.\allowbreak{}cp\allowbreak{}p, \textbf{Function:} printT\allowbreak{}extRep\allowbreak{}lacing\allowbreak{}XMLEsc\allowbreak{}apes, \textbf{Logic:} \logicFOSL, \href{https://github.com/bloomberg/bde/blob/main/groups/bal/balxml/balxml_typesprintutil.cpp#L220}{\textcolor{blue!60!black}{[code]}}
    \begin{blockFOSL}
    \noindent \textbf{requires:} (dataLe\allowbreak{}ngth $>$= 0 \&\& SEPFORALL(0,\allowbreak{} dataLe\allowbreak{}ngth,\allowbreak{} i,\allowbreak{} (data + i) $\mapsto$ \_\allowbreak{})) $||$ (dataLe\allowbreak{}ngth == -1 \&\& SEPEXISTS(0,\allowbreak{} strlen(data),\allowbreak{} i,\allowbreak{} (data + i) $\mapsto$ \_\allowbreak{})) \&\& stream\allowbreak{}.\allowbreak{}good() \\
    \noindent \textbf{ensures:} (\_\allowbreak{}\_\allowbreak{}out == 0 ==$>$ stream\allowbreak{}.\allowbreak{}good()) \&\& (\_\allowbreak{}\_\allowbreak{}out != 0 ==$>$ stream\allowbreak{}.\allowbreak{}bad())
    \end{blockFOSL}
    \item \textbf{Filename:} bbldc\_\allowbreak{}\allowbreak{}daycou\allowbreak{}ntconv\allowbreak{}ention\allowbreak{}.\allowbreak{}cpp, \textbf{Function:} DayCou\allowbreak{}ntConv\allowbreak{}ention\allowbreak{}::\allowbreak{}prin\allowbreak{}t, \textbf{Logic:} \logicFOSL, \href{https://github.com/bloomberg/bde/blob/main/groups/bbl/bbldc/bbldc_daycountconvention.cpp#L19}{\textcolor{blue!60!black}{[code]}}
    \begin{blockFOSL}
    \noindent \textbf{requires:} true \\
    \noindent \textbf{ensures:} \_\allowbreak{}\_\allowbreak{}out == stream \&\& (SEPFORALL(0,\allowbreak{} strlen(toAscii(value)),\allowbreak{} i,\allowbreak{} (stream + i) $\mapsto$ toAscii(value)[i]))
    \end{blockFOSL}
    \item \textbf{Filename:} bdlb\_\allowbreak{}b\allowbreak{}itstri\allowbreak{}ngutil\allowbreak{}.\allowbreak{}cpp, \textbf{Function:} indent, \textbf{Logic:} \logicFOSL, \href{https://github.com/bloomberg/bde/blob/main/groups/bdl/bdlb/bdlb_bitstringutil.cpp#L1025}{\textcolor{blue!60!black}{[code]}}
    \begin{blockFOSL}
    \noindent \textbf{requires:} level $>$= 0 \&\& spaces\allowbreak{}PerLev\allowbreak{}el != 0 \\
    \noindent \textbf{ensures:} \_\allowbreak{}\_\allowbreak{}out == stream \&\& SEPFORALL(0,\allowbreak{} level * spaces\allowbreak{}PerLev\allowbreak{}el,\allowbreak{} i,\allowbreak{} (stream + i $\mapsto$ ' '))
    \end{blockFOSL}
    \item \textbf{Filename:} bdlb\_\allowbreak{}s\allowbreak{}tring.\allowbreak{}\allowbreak{}cpp, \textbf{Function:} String\allowbreak{}::\allowbreak{}areE\allowbreak{}qualCa\allowbreak{}seless, \textbf{Logic:} \logicFOSL, \href{https://github.com/bloomberg/bde/blob/main/groups/bdl/bdlb/bdlb_string.cpp#L18}{\textcolor{blue!60!black}{[code]}}
    \begin{blockFOSL}
    \noindent \textbf{requires:} lhsString != nullptr \&\& rhsString != nullptr \&\& SEPFORALL(0,\allowbreak{} strlen(lhsString),\allowbreak{} i,\allowbreak{} (lhsString + i $\mapsto$ \_\allowbreak{})) \&\& SEPFORALL(0,\allowbreak{} strlen(rhsString),\allowbreak{} i,\allowbreak{} (rhsString + i $\mapsto$ \_\allowbreak{})) \\
    \noindent \textbf{ensures:} (\_\allowbreak{}\_\allowbreak{}out == true ==$>$ SEPFORALL(0,\allowbreak{} strlen(lhsString),\allowbreak{} i,\allowbreak{} (bdlb::\allowbreak{}\allowbreak{}CharTy\allowbreak{}pe::\allowbreak{}to\allowbreak{}Lower(lhsString[i]) == bdlb::\allowbreak{}\allowbreak{}CharTy\allowbreak{}pe::\allowbreak{}to\allowbreak{}Lower(rhsString[i])))) \&\& (\_\allowbreak{}\_\allowbreak{}out == false ==$>$ SEPEXISTS(0,\allowbreak{} strlen(lhsString),\allowbreak{} i,\allowbreak{} (bdlb::\allowbreak{}\allowbreak{}CharTy\allowbreak{}pe::\allowbreak{}to\allowbreak{}Lower(lhsString[i]) != bdlb::\allowbreak{}\allowbreak{}CharTy\allowbreak{}pe::\allowbreak{}to\allowbreak{}Lower(rhsString[i]))) $||$ strlen(lhsString) != strlen(rhsString))
    \end{blockFOSL}
    \item \textbf{Filename:} bdlbb\_\allowbreak{}\allowbreak{}blob.\allowbreak{}c\allowbreak{}pp, \textbf{Function:} BlobBu\allowbreak{}ffer::\allowbreak{}\allowbreak{}print, \textbf{Logic:} \logicFOSL, \href{https://github.com/bloomberg/bde/blob/main/groups/bdl/bdlbb/bdlbb_blob.cpp#L100}{\textcolor{blue!60!black}{[code]}}
    \begin{blockFOSL}
    \noindent \textbf{requires:} stream\allowbreak{}.\allowbreak{}good() \\
    \noindent \textbf{ensures:} \_\allowbreak{}\_\allowbreak{}out == stream \&\& (SEPFORALL(0,\allowbreak{} d\_\allowbreak{}size,\allowbreak{} i,\allowbreak{} stream + i $\mapsto$ d\_\allowbreak{}buff\allowbreak{}er.\allowbreak{}get()[i]) $\star$ stream\allowbreak{}.\allowbreak{}flush\allowbreak{}ed())
    \end{blockFOSL}
    \item \textbf{Filename:} bdlc\_\allowbreak{}i\allowbreak{}ndexcl\allowbreak{}erk.\allowbreak{}cp\allowbreak{}p, \textbf{Function:} IndexC\allowbreak{}lerk::\allowbreak{}\allowbreak{}isInUs\allowbreak{}e, \textbf{Logic:} \logicFOSL, \href{https://github.com/bloomberg/bde/blob/main/groups/bdl/bdlc/bdlc_indexclerk.cpp#L63}{\textcolor{blue!60!black}{[code]}}
    \begin{blockFOSL}
    \noindent \textbf{requires:} (unsigned int)index $<$ (unsigned int)d\_\allowbreak{}next\allowbreak{}NewInd\allowbreak{}ex \\
    \noindent \textbf{ensures:} (\_\allowbreak{}\_\allowbreak{}out == false ==$>$ SEPEXISTS(0,\allowbreak{} d\_\allowbreak{}unus\allowbreak{}edStac\allowbreak{}k.\allowbreak{}size(),\allowbreak{} i,\allowbreak{} d\_\allowbreak{}unus\allowbreak{}edStac\allowbreak{}k[i] == index)) \&\& (\_\allowbreak{}\_\allowbreak{}out == true ==$>$ SEPFORALL(0,\allowbreak{} d\_\allowbreak{}unus\allowbreak{}edStac\allowbreak{}k.\allowbreak{}size(),\allowbreak{} i,\allowbreak{} d\_\allowbreak{}unus\allowbreak{}edStac\allowbreak{}k[i] != index))
    \end{blockFOSL}
    \item \textbf{Filename:} bdld\_\allowbreak{}d\allowbreak{}atum.\allowbreak{}c\allowbreak{}pp, \textbf{Function:} DatumA\allowbreak{}rrayRe\allowbreak{}f::\allowbreak{}pri\allowbreak{}nt, \textbf{Logic:} \logicFOSL, \href{https://github.com/bloomberg/bde/blob/main/groups/bdl/bdld/bdld_datum.cpp#L1358}{\textcolor{blue!60!black}{[code]}}
    \begin{blockFOSL}
    \noindent \textbf{requires:} stream\allowbreak{}.\allowbreak{}bad() $||$ stream\allowbreak{}.\allowbreak{}good() \\
    \noindent \textbf{ensures:} \_\allowbreak{}\_\allowbreak{}out == stream \&\& (stream\allowbreak{}.\allowbreak{}bad() $||$ (stream\allowbreak{}.\allowbreak{}good() \&\& (SEPFORALL(0,\allowbreak{} d\_\allowbreak{}length,\allowbreak{} i,\allowbreak{} stream + i $\mapsto$ d\_\allowbreak{}data\_\allowbreak{}p[i])) \&\& stream\allowbreak{}.\allowbreak{}flush\allowbreak{}ed()))
    \end{blockFOSL}
    \item \textbf{Filename:} bdld\_\allowbreak{}d\allowbreak{}atum.\allowbreak{}c\allowbreak{}pp, \textbf{Function:} DatumM\allowbreak{}apEntr\allowbreak{}y::\allowbreak{}pri\allowbreak{}nt, \textbf{Logic:} \logicFOSL, \href{https://github.com/bloomberg/bde/blob/main/groups/bdl/bdld/bdld_datum.cpp#L1408}{\textcolor{blue!60!black}{[code]}}
    \begin{blockFOSL}
    \noindent \textbf{requires:} !strea\allowbreak{}m.\allowbreak{}bad() \\
    \noindent \textbf{ensures:} \_\allowbreak{}\_\allowbreak{}out == stream \&\& (stream\allowbreak{}.\allowbreak{}bad() $||$ (stream\allowbreak{}.\allowbreak{}good() \&\& (SEPFORALL(0,\allowbreak{} strlen(d\_\allowbreak{}key\_\allowbreak{}p),\allowbreak{} i,\allowbreak{} stream + i $\mapsto$ d\_\allowbreak{}key\_\allowbreak{}p[i]) $\star$ SEPFORALL(0,\allowbreak{} strlen(d\_\allowbreak{}value),\allowbreak{} j,\allowbreak{} stream + strlen(d\_\allowbreak{}key\_\allowbreak{}p) + 1 + j $\mapsto$ d\_\allowbreak{}value[j]))))
    \end{blockFOSL}
    \item \textbf{Filename:} bdld\_\allowbreak{}d\allowbreak{}atumbi\allowbreak{}naryre\allowbreak{}f.\allowbreak{}cpp, \textbf{Function:} DatumB\allowbreak{}inaryR\allowbreak{}ef::\allowbreak{}pr\allowbreak{}int, \textbf{Logic:} \logicFOSL, \href{https://github.com/bloomberg/bde/blob/main/groups/bdl/bdld/bdld_datumbinaryref.cpp#L19}{\textcolor{blue!60!black}{[code]}}
    \begin{blockFOSL}
    \noindent \textbf{requires:} stream\allowbreak{}.\allowbreak{}good() \\
    \noindent \textbf{ensures:} (stream\allowbreak{}.\allowbreak{}bad() ==$>$ \_\allowbreak{}\_\allowbreak{}out == stream) \&\& (stream\allowbreak{}.\allowbreak{}good() ==$>$ (\_\allowbreak{}\_\allowbreak{}out == (stream $<$$<$ bsl::\allowbreak{}f\allowbreak{}lush) \&\& (SEPFORALL(0,\allowbreak{} d\_\allowbreak{}data\allowbreak{}\_\allowbreak{}p.\allowbreak{}siz\allowbreak{}e(),\allowbreak{} i,\allowbreak{} stream + i $\mapsto$ d\_\allowbreak{}data\_\allowbreak{}p[i]) $\star$ (stream + d\_\allowbreak{}data\allowbreak{}\_\allowbreak{}p.\allowbreak{}siz\allowbreak{}e() $\mapsto$ ' ') $\star$ SEPFORALL(0,\allowbreak{} d\_\allowbreak{}size\allowbreak{}.\allowbreak{}size(),\allowbreak{} j,\allowbreak{} stream + d\_\allowbreak{}data\allowbreak{}\_\allowbreak{}p.\allowbreak{}siz\allowbreak{}e() + 1 + j $\mapsto$ d\_\allowbreak{}size[j]))))
    \end{blockFOSL}
    \item \textbf{Filename:} bdld\_\allowbreak{}d\allowbreak{}atumma\allowbreak{}ker.\allowbreak{}cp\allowbreak{}p, \textbf{Function:} DatumM\allowbreak{}aker::\allowbreak{}\allowbreak{}operat\allowbreak{}or(), \textbf{Logic:} \logicFOSL, \href{https://github.com/bloomberg/bde/blob/main/groups/bdl/bdld/bdld_datummaker.cpp#L32}{\textcolor{blue!60!black}{[code]}}
    \begin{blockFOSL}
    \noindent \textbf{requires:} SEPFORALL(0,\allowbreak{} size,\allowbreak{} i,\allowbreak{} elements + i $\mapsto$ \_\allowbreak{}) \&\& size $>$= 0 \\
    \noindent \textbf{ensures:} \_\allowbreak{}\_\allowbreak{}out == bdld::\allowbreak{}\allowbreak{}Datum:\allowbreak{}:adopt\allowbreak{}Map(map) \&\& (SEPFORALL(0,\allowbreak{} size,\allowbreak{} i,\allowbreak{} map.\allowbreak{}data()[i] == elements[i]) $\star$ (*map.\allowbreak{}size() == size) $\star$ (*map.\allowbreak{}s\allowbreak{}orted() == sorted))
    \end{blockFOSL}
    \item \textbf{Filename:} bdlde\_\allowbreak{}\allowbreak{}base64\allowbreak{}alphab\allowbreak{}et.\allowbreak{}cpp, \textbf{Function:} Base64\allowbreak{}Alphab\allowbreak{}et::\allowbreak{}pr\allowbreak{}int, \textbf{Logic:} \logicFOSL, \href{https://github.com/bloomberg/bde/blob/main/groups/bdl/bdlde/bdlde_base64alphabet.cpp#L19}{\textcolor{blue!60!black}{[code]}}
    \begin{blockFOSL}
    \noindent \textbf{requires:} !strea\allowbreak{}m.\allowbreak{}bad() \&\& (value $>$= Base64\allowbreak{}Alphab\allowbreak{}et::\allowbreak{}En\allowbreak{}um::\allowbreak{}MI\allowbreak{}N \&\& value $<$= Base64\allowbreak{}Alphab\allowbreak{}et::\allowbreak{}En\allowbreak{}um::\allowbreak{}MA\allowbreak{}X) \\
    \noindent \textbf{ensures:} \_\allowbreak{}\_\allowbreak{}out == stream \&\& (stream\allowbreak{}.\allowbreak{}bad() $||$ (stream\allowbreak{}.\allowbreak{}good() \&\& (SEPFORALL(0,\allowbreak{} strlen(Base64\allowbreak{}Alphab\allowbreak{}et::\allowbreak{}to\allowbreak{}Ascii(value)),\allowbreak{} i,\allowbreak{} (stream + i) $\mapsto$ Base64\allowbreak{}Alphab\allowbreak{}et::\allowbreak{}to\allowbreak{}Ascii(value)[i]))))
    \end{blockFOSL}
    \item \textbf{Filename:} bdlde\_\allowbreak{}\allowbreak{}base64\allowbreak{}decode\allowbreak{}roptio\allowbreak{}ns.\allowbreak{}cpp, \textbf{Function:} Base64\allowbreak{}Decode\allowbreak{}rOptio\allowbreak{}ns::\allowbreak{}pr\allowbreak{}int, \textbf{Logic:} \logicFOSL, \href{https://github.com/bloomberg/bde/blob/main/groups/bdl/bdlde/bdlde_base64decoderoptions.cpp#L19}{\textcolor{blue!60!black}{[code]}}
    \begin{blockFOSL}
    \noindent \textbf{requires:} stream\allowbreak{}.\allowbreak{}good() \\
    \noindent \textbf{ensures:} \_\allowbreak{}\_\allowbreak{}out == stream \&\& (SEPFORALL(0,\allowbreak{} 3,\allowbreak{} i,\allowbreak{} (\_\allowbreak{}\_\allowbreak{}out + i $\mapsto$ sep\_\allowbreak{}v \&\& sep\_\allowbreak{}v == printe\allowbreak{}r\_\allowbreak{}attr\allowbreak{}ibute[i])))
    \end{blockFOSL}
    \item \textbf{Filename:} bdlde\_\allowbreak{}\allowbreak{}base64\allowbreak{}ignore\allowbreak{}mode.\allowbreak{}c\allowbreak{}pp, \textbf{Function:} Base64\allowbreak{}Ignore\allowbreak{}Mode::\allowbreak{}\allowbreak{}print, \textbf{Logic:} \logicFOSL, \href{https://github.com/bloomberg/bde/blob/main/groups/bdl/bdlde/bdlde_base64ignoremode.cpp#L19}{\textcolor{blue!60!black}{[code]}}
    \begin{blockFOSL}
    \noindent \textbf{requires:} stream\allowbreak{}.\allowbreak{}good() \\
    \noindent \textbf{ensures:} \_\allowbreak{}\_\allowbreak{}out == stream \&\& (stream\allowbreak{}.\allowbreak{}bad() $||$ (stream\allowbreak{}.\allowbreak{}good() \&\& SEPFORALL(0,\allowbreak{} strlen(Base64\allowbreak{}Ignore\allowbreak{}Mode::\allowbreak{}\allowbreak{}toAsci\allowbreak{}i(value)),\allowbreak{} i,\allowbreak{} (stream + i $\mapsto$ Base64\allowbreak{}Ignore\allowbreak{}Mode::\allowbreak{}\allowbreak{}toAsci\allowbreak{}i(value)[i]))))
    \end{blockFOSL}
    \item \textbf{Filename:} bdlde\_\allowbreak{}\allowbreak{}charco\allowbreak{}nverts\allowbreak{}tatus.\allowbreak{}\allowbreak{}cpp, \textbf{Function:} CharCo\allowbreak{}nvertS\allowbreak{}tatus:\allowbreak{}:print, \textbf{Logic:} \logicFOSL, \href{https://github.com/bloomberg/bde/blob/main/groups/bdl/bdlde/bdlde_charconvertstatus.cpp#L19}{\textcolor{blue!60!black}{[code]}}
    \begin{blockFOSL}
    \noindent \textbf{requires:} !strea\allowbreak{}m.\allowbreak{}bad() \\
    \noindent \textbf{ensures:} \_\allowbreak{}\_\allowbreak{}out == stream \&\& (stream\allowbreak{}.\allowbreak{}bad() $||$ (stream\allowbreak{}.\allowbreak{}good() \&\& (SEPFORALL(0,\allowbreak{} strlen(CharCo\allowbreak{}nvertS\allowbreak{}tatus:\allowbreak{}:toAsc\allowbreak{}ii(value)),\allowbreak{} i,\allowbreak{} stream + i $\mapsto$ CharCo\allowbreak{}nvertS\allowbreak{}tatus:\allowbreak{}:toAsc\allowbreak{}ii(value)[i]))))
    \end{blockFOSL}
    \item \textbf{Filename:} bdlde\_\allowbreak{}\allowbreak{}charco\allowbreak{}nvertu\allowbreak{}tf32.\allowbreak{}c\allowbreak{}pp, \textbf{Function:} decode\allowbreak{}TwoOct\allowbreak{}ets, \textbf{Logic:} \logicFOSL, \href{https://github.com/bloomberg/bde/blob/main/groups/bdl/bdlde/bdlde_charconvertutf32.cpp#L694}{\textcolor{blue!60!black}{[code]}}
    \begin{blockFOSL}
    \noindent \textbf{requires:} SEPFORALL(0,\allowbreak{} 2,\allowbreak{} i,\allowbreak{} octBuf + i $\mapsto$ \_\allowbreak{}) \\
    \noindent \textbf{ensures:} \_\allowbreak{}\_\allowbreak{}out == ((octBuf[1] \& \textasciitilde{}k\_\allowbreak{}CON\allowbreak{}TINUE\_\allowbreak{}\allowbreak{}MASK) | ((octBuf[0] \& \textasciitilde{}k\_\allowbreak{}TWO\allowbreak{}\_\allowbreak{}OCTET\allowbreak{}\_\allowbreak{}MASK) $<$$<$ k\_\allowbreak{}CONT\allowbreak{}INUE\_\allowbreak{}C\allowbreak{}ONT\_\allowbreak{}WI\allowbreak{}D))
    \end{blockFOSL}
    \item \textbf{Filename:} balm\_\allowbreak{}m\allowbreak{}etricf\allowbreak{}ormat.\allowbreak{}\allowbreak{}cpp, \textbf{Function:} Metric\allowbreak{}Format\allowbreak{}Spec::\allowbreak{}\allowbreak{}format\allowbreak{}Value, \textbf{Logic:} \logicFOSL, \href{https://github.com/bloomberg/bde/blob/main/groups/bal/balm/balm_metricformat.cpp#L32}{\textcolor{blue!60!black}{[code]}}
    \begin{blockFOSL}
    \noindent \textbf{requires:} true \\
    \noindent \textbf{ensures:} \_\allowbreak{}\_\allowbreak{}out == stream \&\& (SEPFORALL(0,\allowbreak{} strlen(buffer),\allowbreak{} i,\allowbreak{} stream + i $\mapsto$ buffer[i]) $||$ SEPFORALL(0,\allowbreak{} newBuf\allowbreak{}fer.\allowbreak{}si\allowbreak{}ze(),\allowbreak{} i,\allowbreak{} stream + i $\mapsto$ newBuf\allowbreak{}fer.\allowbreak{}da\allowbreak{}ta()[i]))
    \end{blockFOSL}
    \item \textbf{Filename:} bdljsn\allowbreak{}\_\allowbreak{}error\allowbreak{}.\allowbreak{}cpp, \textbf{Function:} bdljsn\allowbreak{}::\allowbreak{}oper\allowbreak{}ator$<$$<$, \textbf{Logic:} \logicFOSL, \href{https://github.com/bloomberg/bde/blob/main/groups/bdl/bdljsn/bdljsn_error.cpp#L46}{\textcolor{blue!60!black}{[code]}}
    \begin{blockFOSL}
    \noindent \textbf{requires:} true \\
    \noindent \textbf{ensures:} \_\allowbreak{}\_\allowbreak{}out == stream \&\& (SEPFORALL(0,\allowbreak{} object\allowbreak{}.\allowbreak{}locat\allowbreak{}ion().\allowbreak{}size(),\allowbreak{} i,\allowbreak{} stream + i $\mapsto$ object\allowbreak{}.\allowbreak{}locat\allowbreak{}ion().\allowbreak{}data()[i]) $\star$ SEPFORALL(0,\allowbreak{} object\allowbreak{}.\allowbreak{}messa\allowbreak{}ge().\allowbreak{}size(),\allowbreak{} j,\allowbreak{} stream + object\allowbreak{}.\allowbreak{}locat\allowbreak{}ion().\allowbreak{}size() + j $\mapsto$ object\allowbreak{}.\allowbreak{}messa\allowbreak{}ge().\allowbreak{}data()[j]))
    \end{blockFOSL}
    \item \textbf{Filename:} bdljsn\allowbreak{}\_\allowbreak{}jsont\allowbreak{}estsui\allowbreak{}teutil\allowbreak{}.\allowbreak{}cpp, \textbf{Function:} getLef\allowbreak{}tBrack\allowbreak{}ets100\allowbreak{}000, \textbf{Logic:} \logicFOSL, \href{https://github.com/bloomberg/bde/blob/main/groups/bdl/bdljsn/bdljsn_jsontestsuiteutil.cpp#L31}{\textcolor{blue!60!black}{[code]}}
    \begin{blockFOSL}
    \noindent \textbf{requires:} true \\
    \noindent \textbf{ensures:} \_\allowbreak{}\_\allowbreak{}out != 0 \&\& SEPFORALL(0,\allowbreak{} 100000,\allowbreak{} i,\allowbreak{} \_\allowbreak{}\_\allowbreak{}out + i $\mapsto$ '[')
    \end{blockFOSL}
    \item \textbf{Filename:} baltzo\allowbreak{}\_\allowbreak{}dataf\allowbreak{}ileloa\allowbreak{}der.\allowbreak{}cp\allowbreak{}p, \textbf{Function:} valida\allowbreak{}teTime\allowbreak{}ZoneId, \textbf{Logic:} \logicFOSL, \href{https://github.com/bloomberg/bde/blob/main/groups/bal/baltzo/baltzo_datafileloader.cpp#L97}{\textcolor{blue!60!black}{[code]}}
    \begin{blockFOSL}
    \noindent \textbf{requires:} timeZo\allowbreak{}neId != 0 \&\& (timeZo\allowbreak{}neId[0] == '/' ==$>$ res\_\allowbreak{}tmp == -1) \&\& (SEPEXISTS(0,\allowbreak{} strlen(timeZo\allowbreak{}neId),\allowbreak{} i,\allowbreak{} (timeZo\allowbreak{}neId + i $\mapsto$ sep\_\allowbreak{}v) \&\& !bsl::\allowbreak{}\allowbreak{}strchr("ABCDE\allowbreak{}FGHIJK\allowbreak{}LMNOPQ\allowbreak{}RSTUVW\allowbreak{}XYZabc\allowbreak{}defghi\allowbreak{}jklmno\allowbreak{}pqrstu\allowbreak{}vwxyz1\allowbreak{}234567\allowbreak{}890/\_\allowbreak{}+\allowbreak{}-",\allowbreak{} sep\_\allowbreak{}v)) ==$>$ res\_\allowbreak{}tmp == -2) \&\& (SEPFORALL(0,\allowbreak{} strlen(timeZo\allowbreak{}neId),\allowbreak{} i,\allowbreak{} (timeZo\allowbreak{}neId + i $\mapsto$ sep\_\allowbreak{}v) \&\& bsl::\allowbreak{}s\allowbreak{}trchr("ABCDE\allowbreak{}FGHIJK\allowbreak{}LMNOPQ\allowbreak{}RSTUVW\allowbreak{}XYZabc\allowbreak{}defghi\allowbreak{}jklmno\allowbreak{}pqrstu\allowbreak{}vwxyz1\allowbreak{}234567\allowbreak{}890/\_\allowbreak{}+\allowbreak{}-",\allowbreak{} sep\_\allowbreak{}v)) ==$>$ res\_\allowbreak{}tmp == 0) \\
    \noindent \textbf{ensures:} (\_\allowbreak{}\_\allowbreak{}out == -1 ==$>$ timeZo\allowbreak{}neId[0] == '/') \&\& (\_\allowbreak{}\_\allowbreak{}out == -2 ==$>$ SEPEXISTS(0,\allowbreak{} strlen(timeZo\allowbreak{}neId),\allowbreak{} i,\allowbreak{} (timeZo\allowbreak{}neId + i $\mapsto$ sep\_\allowbreak{}v) \&\& !bsl::\allowbreak{}\allowbreak{}strchr("ABCDE\allowbreak{}FGHIJK\allowbreak{}LMNOPQ\allowbreak{}RSTUVW\allowbreak{}XYZabc\allowbreak{}defghi\allowbreak{}jklmno\allowbreak{}pqrstu\allowbreak{}vwxyz1\allowbreak{}234567\allowbreak{}890/\_\allowbreak{}+\allowbreak{}-",\allowbreak{} sep\_\allowbreak{}v))) \&\& (\_\allowbreak{}\_\allowbreak{}out == 0 ==$>$ SEPFORALL(0,\allowbreak{} strlen(timeZo\allowbreak{}neId),\allowbreak{} i,\allowbreak{} (timeZo\allowbreak{}neId + i $\mapsto$ sep\_\allowbreak{}v) \&\& bsl::\allowbreak{}s\allowbreak{}trchr("ABCDE\allowbreak{}FGHIJK\allowbreak{}LMNOPQ\allowbreak{}RSTUVW\allowbreak{}XYZabc\allowbreak{}defghi\allowbreak{}jklmno\allowbreak{}pqrstu\allowbreak{}vwxyz1\allowbreak{}234567\allowbreak{}890/\_\allowbreak{}+\allowbreak{}-",\allowbreak{} sep\_\allowbreak{}v)))
    \end{blockFOSL}
    \item \textbf{Filename:} baltzo\allowbreak{}\_\allowbreak{}local\allowbreak{}dateti\allowbreak{}me.\allowbreak{}cpp, \textbf{Function:} baltzo\allowbreak{}::\allowbreak{}oper\allowbreak{}ator$<$$<$, \textbf{Logic:} \logicFOSL, \href{https://github.com/bloomberg/bde/blob/main/groups/bal/baltzo/baltzo_localdatetime.cpp#L41}{\textcolor{blue!60!black}{[code]}}
    \begin{blockFOSL}
    \noindent \textbf{requires:} true \\
    \noindent \textbf{ensures:} \_\allowbreak{}\_\allowbreak{}out == stream \&\& (stream\allowbreak{}.\allowbreak{}bad() $||$ (stream\allowbreak{}.\allowbreak{}good() \&\& (SEPFORALL(0,\allowbreak{} object\allowbreak{}.\allowbreak{}datet\allowbreak{}imeTz().\allowbreak{}size(),\allowbreak{} j,\allowbreak{} stream + j $\mapsto$ object\allowbreak{}.\allowbreak{}datet\allowbreak{}imeTz().\allowbreak{}data()[j]) $\star$ SEPFORALL(0,\allowbreak{} object\allowbreak{}.\allowbreak{}timeZ\allowbreak{}oneId().\allowbreak{}size(),\allowbreak{} i,\allowbreak{} stream + object\allowbreak{}.\allowbreak{}datet\allowbreak{}imeTz().\allowbreak{}size() + i $\mapsto$ object\allowbreak{}.\allowbreak{}timeZ\allowbreak{}oneId().\allowbreak{}data()[i]))))
    \end{blockFOSL}
    \item \textbf{Filename:} baltzo\allowbreak{}\_\allowbreak{}local\allowbreak{}timede\allowbreak{}script\allowbreak{}or.\allowbreak{}cpp, \textbf{Function:} baltzo\allowbreak{}::\allowbreak{}oper\allowbreak{}ator$<$$<$, \textbf{Logic:} \logicFOSL, \href{https://github.com/bloomberg/bde/blob/main/groups/bal/baltzo/baltzo_localtimedescriptor.cpp#L40}{\textcolor{blue!60!black}{[code]}}
    \begin{blockFOSL}
    \noindent \textbf{requires:} true \\
    \noindent \textbf{ensures:} \_\allowbreak{}\_\allowbreak{}out == stream \&\& (SEPFORALL(0,\allowbreak{} object\allowbreak{}.\allowbreak{}utcOf\allowbreak{}fsetIn\allowbreak{}Second\allowbreak{}s().\allowbreak{}size(),\allowbreak{} i,\allowbreak{} stream + i $\mapsto$ object\allowbreak{}.\allowbreak{}utcOf\allowbreak{}fsetIn\allowbreak{}Second\allowbreak{}s().\allowbreak{}data()[i]) $\star$ SEPFORALL(0,\allowbreak{} object\allowbreak{}.\allowbreak{}dstIn\allowbreak{}Effect\allowbreak{}Flag().\allowbreak{}size(),\allowbreak{} j,\allowbreak{} stream + object\allowbreak{}.\allowbreak{}utcOf\allowbreak{}fsetIn\allowbreak{}Second\allowbreak{}s().\allowbreak{}size() + j $\mapsto$ object\allowbreak{}.\allowbreak{}dstIn\allowbreak{}Effect\allowbreak{}Flag().\allowbreak{}data()[j]) $\star$ SEPFORALL(0,\allowbreak{} object\allowbreak{}.\allowbreak{}descr\allowbreak{}iption().\allowbreak{}size(),\allowbreak{} k,\allowbreak{} stream + object\allowbreak{}.\allowbreak{}utcOf\allowbreak{}fsetIn\allowbreak{}Second\allowbreak{}s().\allowbreak{}size() + object\allowbreak{}.\allowbreak{}dstIn\allowbreak{}Effect\allowbreak{}Flag().\allowbreak{}size() + k $\mapsto$ object\allowbreak{}.\allowbreak{}descr\allowbreak{}iption().\allowbreak{}c\_\allowbreak{}str()[k]))
    \end{blockFOSL}
    \item \textbf{Filename:} bdlbb\_\allowbreak{}\allowbreak{}blobut\allowbreak{}il.\allowbreak{}cpp, \textbf{Function:} asciiD\allowbreak{}umpFro\allowbreak{}mBuffe\allowbreak{}rStart, \textbf{Logic:} \logicFOSL, \href{https://github.com/bloomberg/bde/blob/main/groups/bdl/bdlbb/bdlbb_blobutil.cpp#L63}{\textcolor{blue!60!black}{[code]}}
    \begin{blockFOSL}
    \noindent \textbf{requires:} 0 $<$= buffer\allowbreak{}Index \&\& buffer\allowbreak{}Index $<$ source\allowbreak{}.\allowbreak{}numDa\allowbreak{}taBuff\allowbreak{}ers() \&\& 0 $<$= numBytes \&\& numBytes $<$= source\allowbreak{}.\allowbreak{}total\allowbreak{}Size() - source\allowbreak{}.\allowbreak{}cumul\allowbreak{}ativeS\allowbreak{}ize(buffer\allowbreak{}Index) \\
    \noindent \textbf{ensures:} \_\allowbreak{}\_\allowbreak{}out == stream \&\& (stream\allowbreak{}.\allowbreak{}bad() $||$ (stream\allowbreak{}.\allowbreak{}good() \&\& SEPFORALL(0,\allowbreak{} numBytes,\allowbreak{} i,\allowbreak{} stream + i $\mapsto$ source\allowbreak{}.\allowbreak{}buffe\allowbreak{}r(buffer\allowbreak{}Index + i / source\allowbreak{}.\allowbreak{}buffe\allowbreak{}r(buffer\allowbreak{}Index).\allowbreak{}size()).\allowbreak{}data()[i \% source\allowbreak{}.\allowbreak{}buffe\allowbreak{}r(buffer\allowbreak{}Index).\allowbreak{}size()])))
    \end{blockFOSL}
    \item \textbf{Filename:} bdlde\_\allowbreak{}\allowbreak{}charco\allowbreak{}nvertu\allowbreak{}tf32.\allowbreak{}c\allowbreak{}pp, \textbf{Function:} decode\allowbreak{}ThreeO\allowbreak{}ctets, \textbf{Logic:} \logicFOSL, \href{https://github.com/bloomberg/bde/blob/main/groups/bdl/bdlde/bdlde_charconvertutf32.cpp#L703}{\textcolor{blue!60!black}{[code]}}
    \begin{blockFOSL}
    \noindent \textbf{requires:} SEPFORALL(0,\allowbreak{} 3,\allowbreak{} i,\allowbreak{} octBuf + i $\mapsto$ \_\allowbreak{}) \\
    \noindent \textbf{ensures:} \_\allowbreak{}\_\allowbreak{}out == ((octBuf[2] \& \textasciitilde{}k\_\allowbreak{}CON\allowbreak{}TINUE\_\allowbreak{}\allowbreak{}MASK) | ((octBuf[1] \& \textasciitilde{}k\_\allowbreak{}CON\allowbreak{}TINUE\_\allowbreak{}\allowbreak{}MASK) $<$$<$ k\_\allowbreak{}CONT\allowbreak{}INUE\_\allowbreak{}C\allowbreak{}ONT\_\allowbreak{}WI\allowbreak{}D) | ((octBuf[0] \& \textasciitilde{}k\_\allowbreak{}THR\allowbreak{}EE\_\allowbreak{}OCT\allowbreak{}ET\_\allowbreak{}MAS\allowbreak{}K) $<$$<$ (2 * k\_\allowbreak{}CONT\allowbreak{}INUE\_\allowbreak{}C\allowbreak{}ONT\_\allowbreak{}WI\allowbreak{}D)))
    \end{blockFOSL}
    \item \textbf{Filename:} bdlde\_\allowbreak{}\allowbreak{}charco\allowbreak{}nvertu\allowbreak{}tf32.\allowbreak{}c\allowbreak{}pp, \textbf{Function:} decode\allowbreak{}FourOc\allowbreak{}tets, \textbf{Logic:} \logicFOSL, \href{https://github.com/bloomberg/bde/blob/main/groups/bdl/bdlde/bdlde_charconvertutf32.cpp#L713}{\textcolor{blue!60!black}{[code]}}
    \begin{blockFOSL}
    \noindent \textbf{requires:} SEPFORALL(0,\allowbreak{} 4,\allowbreak{} i,\allowbreak{} octBuf + i $\mapsto$ \_\allowbreak{}) \\
    \noindent \textbf{ensures:} \_\allowbreak{}\_\allowbreak{}out == ((octBuf[3] \& \textasciitilde{}k\_\allowbreak{}CON\allowbreak{}TINUE\_\allowbreak{}\allowbreak{}MASK) | ((octBuf[2] \& \textasciitilde{}k\_\allowbreak{}CON\allowbreak{}TINUE\_\allowbreak{}\allowbreak{}MASK) $<$$<$ k\_\allowbreak{}CONT\allowbreak{}INUE\_\allowbreak{}C\allowbreak{}ONT\_\allowbreak{}WI\allowbreak{}D) | ((octBuf[1] \& \textasciitilde{}k\_\allowbreak{}CON\allowbreak{}TINUE\_\allowbreak{}\allowbreak{}MASK) $<$$<$ 2 * k\_\allowbreak{}CONT\allowbreak{}INUE\_\allowbreak{}C\allowbreak{}ONT\_\allowbreak{}WI\allowbreak{}D) | ((octBuf[0] \& \textasciitilde{}k\_\allowbreak{}FOU\allowbreak{}R\_\allowbreak{}OCTE\allowbreak{}T\_\allowbreak{}MASK) $<$$<$ 3 * k\_\allowbreak{}CONT\allowbreak{}INUE\_\allowbreak{}C\allowbreak{}ONT\_\allowbreak{}WI\allowbreak{}D))
    \end{blockFOSL}
    \item \textbf{Filename:} bdlde\_\allowbreak{}\allowbreak{}charco\allowbreak{}nvertu\allowbreak{}tf32.\allowbreak{}c\allowbreak{}pp, \textbf{Function:} lookah\allowbreak{}eadCon\allowbreak{}tinuat\allowbreak{}ions, \textbf{Logic:} \logicFOSL, \href{https://github.com/bloomberg/bde/blob/main/groups/bdl/bdlde/bdlde_charconvertutf32.cpp#L725}{\textcolor{blue!60!black}{[code]}}
    \begin{blockFOSL}
    \noindent \textbf{requires:} SEPFORALL(0,\allowbreak{} n,\allowbreak{} i,\allowbreak{} (octBuf + i $\mapsto$ \_\allowbreak{})) \\
    \noindent \textbf{ensures:} SEPFORALL(0,\allowbreak{} \_\allowbreak{}\_\allowbreak{}out,\allowbreak{} i,\allowbreak{} (octBuf + i $\mapsto$ sep\_\allowbreak{}v \&\& isCont\allowbreak{}inuati\allowbreak{}on(sep\_\allowbreak{}v))) \&\& (\_\allowbreak{}\_\allowbreak{}out == n $||$ !isCon\allowbreak{}tinuat\allowbreak{}ion(*(octBuf + \_\allowbreak{}\_\allowbreak{}out)))
    \end{blockFOSL}
    \item \textbf{Filename:} bdlde\_\allowbreak{}\allowbreak{}crc32c\allowbreak{}.\allowbreak{}cpp, \textbf{Function:} calcul\allowbreak{}ateCrc\allowbreak{}32c, \textbf{Logic:} \logicFOSL, \href{https://github.com/bloomberg/bde/blob/main/groups/bdl/bdlde/bdlde_crc32c.cpp#L719}{\textcolor{blue!60!black}{[code]}}
    \begin{blockFOSL}
    \noindent \textbf{requires:} (length == 0) $||$ (data != 0 \&\& SEPFORALL(0,\allowbreak{} length,\allowbreak{} i,\allowbreak{} data + i $\mapsto$ \_\allowbreak{})) \\
    \noindent \textbf{ensures:} \_\allowbreak{}\_\allowbreak{}out == crc
    \end{blockFOSL}
    \item \textbf{Filename:} bdlde\_\allowbreak{}\allowbreak{}sha2.\allowbreak{}c\allowbreak{}pp, \textbf{Function:} bdlde:\allowbreak{}:opera\allowbreak{}tor==, \textbf{Logic:} \logicFOSL, \href{https://github.com/bloomberg/bde/blob/main/groups/bdl/bdlde/bdlde_sha2.cpp#L656}{\textcolor{blue!60!black}{[code]}}
    \begin{blockFOSL}
    \noindent \textbf{requires:} lhs.\allowbreak{}d\_\allowbreak{}\allowbreak{}buffer\allowbreak{}Size $>$= 0 \&\& rhs.\allowbreak{}d\_\allowbreak{}\allowbreak{}buffer\allowbreak{}Size $>$= 0 \&\& SEPFORALL(0,\allowbreak{} lhs.\allowbreak{}d\_\allowbreak{}\allowbreak{}buffer\allowbreak{}Size,\allowbreak{} i,\allowbreak{} lhs.\allowbreak{}d\_\allowbreak{}\allowbreak{}buffer + i $\mapsto$ \_\allowbreak{}) \&\& SEPFORALL(0,\allowbreak{} rhs.\allowbreak{}d\_\allowbreak{}\allowbreak{}buffer\allowbreak{}Size,\allowbreak{} i,\allowbreak{} rhs.\allowbreak{}d\_\allowbreak{}\allowbreak{}buffer + i $\mapsto$ \_\allowbreak{}) \&\& SEPFORALL(0,\allowbreak{} 8,\allowbreak{} i,\allowbreak{} lhs.\allowbreak{}d\_\allowbreak{}\allowbreak{}state + i $\mapsto$ \_\allowbreak{}) \&\& SEPFORALL(0,\allowbreak{} 8,\allowbreak{} i,\allowbreak{} rhs.\allowbreak{}d\_\allowbreak{}\allowbreak{}state + i $\mapsto$ \_\allowbreak{}) \\
    \noindent \textbf{ensures:} (\_\allowbreak{}\_\allowbreak{}out == true ==$>$ (lhs.\allowbreak{}d\_\allowbreak{}\allowbreak{}totalS\allowbreak{}ize == rhs.\allowbreak{}d\_\allowbreak{}\allowbreak{}totalS\allowbreak{}ize \&\& lhs.\allowbreak{}d\_\allowbreak{}\allowbreak{}buffer\allowbreak{}Size == rhs.\allowbreak{}d\_\allowbreak{}\allowbreak{}buffer\allowbreak{}Size \&\& SEPFORALL(0,\allowbreak{} lhs.\allowbreak{}d\_\allowbreak{}\allowbreak{}buffer\allowbreak{}Size,\allowbreak{} i,\allowbreak{} lhs.\allowbreak{}d\_\allowbreak{}\allowbreak{}buffer[i] $\mapsto$ rhs.\allowbreak{}d\_\allowbreak{}\allowbreak{}buffer[i]) \&\& SEPFORALL(0,\allowbreak{} 8,\allowbreak{} i,\allowbreak{} lhs.\allowbreak{}d\_\allowbreak{}\allowbreak{}state[i] $\mapsto$ rhs.\allowbreak{}d\_\allowbreak{}\allowbreak{}state[i]))) \&\& (\_\allowbreak{}\_\allowbreak{}out == false ==$>$ !(lhs.\allowbreak{}d\_\allowbreak{}\allowbreak{}totalS\allowbreak{}ize == rhs.\allowbreak{}d\_\allowbreak{}\allowbreak{}totalS\allowbreak{}ize \&\& lhs.\allowbreak{}d\_\allowbreak{}\allowbreak{}buffer\allowbreak{}Size == rhs.\allowbreak{}d\_\allowbreak{}\allowbreak{}buffer\allowbreak{}Size \&\& SEPFORALL(0,\allowbreak{} lhs.\allowbreak{}d\_\allowbreak{}\allowbreak{}buffer\allowbreak{}Size,\allowbreak{} i,\allowbreak{} lhs.\allowbreak{}d\_\allowbreak{}\allowbreak{}buffer[i] $\mapsto$ rhs.\allowbreak{}d\_\allowbreak{}\allowbreak{}buffer[i]) \&\& SEPFORALL(0,\allowbreak{} 8,\allowbreak{} i,\allowbreak{} lhs.\allowbreak{}d\_\allowbreak{}\allowbreak{}state[i] $\mapsto$ rhs.\allowbreak{}d\_\allowbreak{}\allowbreak{}state[i])))
    \end{blockFOSL}
    \item \textbf{Filename:} bdlde\_\allowbreak{}\allowbreak{}utf8ut\allowbreak{}il.\allowbreak{}cpp, \textbf{Function:} get4By\allowbreak{}teValu\allowbreak{}e, \textbf{Logic:} \logicFOSL, \href{https://github.com/bloomberg/bde/blob/main/groups/bdl/bdlde/bdlde_utf8util.cpp#L209}{\textcolor{blue!60!black}{[code]}}
    \begin{blockFOSL}
    \noindent \textbf{requires:} SEPFORALL(0,\allowbreak{} 4,\allowbreak{} i,\allowbreak{} pc + i $\mapsto$ \_\allowbreak{}) \\
    \noindent \textbf{ensures:} \_\allowbreak{}\_\allowbreak{}out == ((*pc \& 0x7) $<$$<$ 18) | ((pc[1] \& k\_\allowbreak{}CONT\allowbreak{}\_\allowbreak{}VALUE\allowbreak{}\_\allowbreak{}MASK) $<$$<$ 12) | ((pc[2] \& k\_\allowbreak{}CONT\allowbreak{}\_\allowbreak{}VALUE\allowbreak{}\_\allowbreak{}MASK) $<$$<$ 6) | (pc[3] \& k\_\allowbreak{}CONT\allowbreak{}\_\allowbreak{}VALUE\allowbreak{}\_\allowbreak{}MASK)
    \end{blockFOSL}
    \item \textbf{Filename:} bdlb\_\allowbreak{}b\allowbreak{}itstri\allowbreak{}ngutil\allowbreak{}.\allowbreak{}cpp, \textbf{Function:} lt64Raw, \textbf{Logic:} \logicPL, \href{https://github.com/bloomberg/bde/blob/main/groups/bdl/bdlb/bdlb_bitstringutil.cpp#L892}{\textcolor{blue!60!black}{[code]}}
    \begin{blockPL}
    \noindent \textbf{requires:} 0 $<$= numBits \&\& numBits $<$ 64 \\
    \noindent \textbf{ensures:} \_\allowbreak{}\_\allowbreak{}out == ((1ULL $<$$<$ numBits) - 1)
    \end{blockPL}
    \item \textbf{Filename:} bdlb\_\allowbreak{}b\allowbreak{}itstri\allowbreak{}ngutil\allowbreak{}.\allowbreak{}cpp, \textbf{Function:} ge64Raw, \textbf{Logic:} \logicPL, \href{https://github.com/bloomberg/bde/blob/main/groups/bdl/bdlb/bdlb_bitstringutil.cpp#L906}{\textcolor{blue!60!black}{[code]}}
    \begin{blockPL}
    \noindent \textbf{requires:} 0 $<$= numBits \&\& numBits $<$ 64 \\
    \noindent \textbf{ensures:} \_\allowbreak{}\_\allowbreak{}out == (\textasciitilde{}0ULL $<$$<$ numBits)
    \end{blockPL}
    \item \textbf{Filename:} bdlde\_\allowbreak{}\allowbreak{}utf8ut\allowbreak{}il.\allowbreak{}cpp, \textbf{Function:} get2By\allowbreak{}teValu\allowbreak{}e, \textbf{Logic:} \logicPSL, \href{https://github.com/bloomberg/bde/blob/main/groups/bdl/bdlde/bdlde_utf8util.cpp#L188}{\textcolor{blue!60!black}{[code]}}
    \begin{blockPSL}
    \noindent \textbf{requires:} (pc $\mapsto$ \_\allowbreak{}) $\star$ (pc + 1 $\mapsto$ \_\allowbreak{}) \\
    \noindent \textbf{ensures:} \_\allowbreak{}\_\allowbreak{}out == ((*pc \& 0x1f) $<$$<$ 6) | (pc[1] \& k\_\allowbreak{}CONT\allowbreak{}\_\allowbreak{}VALUE\allowbreak{}\_\allowbreak{}MASK)
    \end{blockPSL}
    \item \textbf{Filename:} bdlde\_\allowbreak{}\allowbreak{}utf8ut\allowbreak{}il.\allowbreak{}cpp, \textbf{Function:} get3By\allowbreak{}teValu\allowbreak{}e, \textbf{Logic:} \logicPSL, \href{https://github.com/bloomberg/bde/blob/main/groups/bdl/bdlde/bdlde_utf8util.cpp#L198}{\textcolor{blue!60!black}{[code]}}
    \begin{blockPSL}
    \noindent \textbf{requires:} pc != nullptr \&\& (pc $\mapsto$ \_\allowbreak{} $\star$ (pc + 1) $\mapsto$ \_\allowbreak{} $\star$ (pc + 2) $\mapsto$ \_\allowbreak{}) \\
    \noindent \textbf{ensures:} \_\allowbreak{}\_\allowbreak{}out == ((*pc \& 0xf) $<$$<$ 12) | ((pc[1] \& k\_\allowbreak{}CONT\allowbreak{}\_\allowbreak{}VALUE\allowbreak{}\_\allowbreak{}MASK) $<$$<$ 6) | (pc[2] \& k\_\allowbreak{}CONT\allowbreak{}\_\allowbreak{}VALUE\allowbreak{}\_\allowbreak{}MASK)
    \end{blockPSL}
    \item \textbf{Filename:} ball\_\allowbreak{}a\allowbreak{}ttribu\allowbreak{}tecont\allowbreak{}ainerl\allowbreak{}ist.\allowbreak{}cp\allowbreak{}p, \textbf{Function:} Attrib\allowbreak{}uteCon\allowbreak{}tainer\allowbreak{}List::\allowbreak{}\allowbreak{}hasVal\allowbreak{}ue, \textbf{Logic:} \logicPL, \href{https://github.com/bloomberg/bde/blob/main/groups/bal/ball/ball_attributecontainerlist.cpp#L157}{\textcolor{blue!60!black}{[code]}}
    \begin{blockPL}
    \noindent \textbf{requires:} true \\
    \noindent \textbf{ensures:} (\_\allowbreak{}\_\allowbreak{}out == true ==$>$ d\_\allowbreak{}head\_\allowbreak{}p != 0) \&\& (\_\allowbreak{}\_\allowbreak{}out == false ==$>$ d\_\allowbreak{}head\_\allowbreak{}p == 0 $||$ !d\_\allowbreak{}hea\allowbreak{}d\_\allowbreak{}p-$>$d\allowbreak{}\_\allowbreak{}value\allowbreak{}\_\allowbreak{}p-$>$ha\allowbreak{}sValue(value))
    \end{blockPL}
    \item \textbf{Filename:} baltzo\allowbreak{}\_\allowbreak{}zonei\allowbreak{}nfobin\allowbreak{}aryrea\allowbreak{}der.\allowbreak{}cp\allowbreak{}p, \textbf{Function:} decode32, \textbf{Logic:} \logicPL, \href{https://github.com/bloomberg/bde/blob/main/groups/bal/baltzo/baltzo_zoneinfobinaryreader.cpp#L232}{\textcolor{blue!60!black}{[code]}}
    \begin{blockPL}
    \noindent \textbf{requires:} address != nullptr \&\& strlen(address) $>$= 4 \\
    \noindent \textbf{ensures:} \_\allowbreak{}\_\allowbreak{}out == BSLS\_\allowbreak{}B\allowbreak{}YTEORD\allowbreak{}ER\_\allowbreak{}BE\_\allowbreak{}\allowbreak{}U32\_\allowbreak{}TO\allowbreak{}\_\allowbreak{}HOST(*reint\allowbreak{}erpret\allowbreak{}\_\allowbreak{}cast$<$\allowbreak{}const int*$>$(address))
    \end{blockPL}
    \item \textbf{Filename:} bdlbb\_\allowbreak{}\allowbreak{}blob.\allowbreak{}c\allowbreak{}pp, \textbf{Function:} bdlbb:\allowbreak{}:opera\allowbreak{}tor==, \textbf{Logic:} \logicPL, \href{https://github.com/bloomberg/bde/blob/main/groups/bdl/bdlbb/bdlbb_blob.cpp#L764}{\textcolor{blue!60!black}{[code]}}
    \begin{blockPL}
    \noindent \textbf{requires:} true \\
    \noindent \textbf{ensures:} (\_\allowbreak{}\_\allowbreak{}out == true ==$>$ (lhs.\allowbreak{}d\_\allowbreak{}\allowbreak{}buffer\allowbreak{}s == rhs.\allowbreak{}d\_\allowbreak{}\allowbreak{}buffer\allowbreak{}s \&\& lhs.\allowbreak{}d\_\allowbreak{}\allowbreak{}totalS\allowbreak{}ize == rhs.\allowbreak{}d\_\allowbreak{}\allowbreak{}totalS\allowbreak{}ize \&\& lhs.\allowbreak{}d\_\allowbreak{}\allowbreak{}dataLe\allowbreak{}ngth == rhs.\allowbreak{}d\_\allowbreak{}\allowbreak{}dataLe\allowbreak{}ngth \&\& lhs.\allowbreak{}d\_\allowbreak{}\allowbreak{}dataIn\allowbreak{}dex == rhs.\allowbreak{}d\_\allowbreak{}\allowbreak{}dataIn\allowbreak{}dex \&\& lhs.\allowbreak{}d\_\allowbreak{}\allowbreak{}preDat\allowbreak{}aIndex\allowbreak{}Length == rhs.\allowbreak{}d\_\allowbreak{}\allowbreak{}preDat\allowbreak{}aIndex\allowbreak{}Length)) \&\& (\_\allowbreak{}\_\allowbreak{}out == false ==$>$ !(lhs.\allowbreak{}d\_\allowbreak{}\allowbreak{}buffer\allowbreak{}s == rhs.\allowbreak{}d\_\allowbreak{}\allowbreak{}buffer\allowbreak{}s \&\& lhs.\allowbreak{}d\_\allowbreak{}\allowbreak{}totalS\allowbreak{}ize == rhs.\allowbreak{}d\_\allowbreak{}\allowbreak{}totalS\allowbreak{}ize \&\& lhs.\allowbreak{}d\_\allowbreak{}\allowbreak{}dataLe\allowbreak{}ngth == rhs.\allowbreak{}d\_\allowbreak{}\allowbreak{}dataLe\allowbreak{}ngth \&\& lhs.\allowbreak{}d\_\allowbreak{}\allowbreak{}dataIn\allowbreak{}dex == rhs.\allowbreak{}d\_\allowbreak{}\allowbreak{}dataIn\allowbreak{}dex \&\& lhs.\allowbreak{}d\_\allowbreak{}\allowbreak{}preDat\allowbreak{}aIndex\allowbreak{}Length == rhs.\allowbreak{}d\_\allowbreak{}\allowbreak{}preDat\allowbreak{}aIndex\allowbreak{}Length))
    \end{blockPL}
    \item \textbf{Filename:} bmqa\_\allowbreak{}q\allowbreak{}ueueid\allowbreak{}.\allowbreak{}cpp, \textbf{Function:} QueueI\allowbreak{}d::\allowbreak{}pri\allowbreak{}nt, \textbf{Logic:} \logicFOSL, \href{https://github.com/bloomberg/blazingmq/blob/main/src/groups/bmq/bmqa/bmqa_queueid.cpp#L137}{\textcolor{blue!60!black}{[code]}}
    \begin{blockFOSL}
    \noindent \textbf{requires:} stream\allowbreak{}.\allowbreak{}good() \&\& uri().\allowbreak{}size() $>$= 0 \&\& correl\allowbreak{}ationI\allowbreak{}d().\allowbreak{}size() $>$= 0 \\
    \noindent \textbf{ensures:} (stream\allowbreak{}.\allowbreak{}bad() ==$>$ \_\allowbreak{}\_\allowbreak{}out == stream) \&\& (stream\allowbreak{}.\allowbreak{}good() ==$>$ (\_\allowbreak{}\_\allowbreak{}out == stream \&\& (SEPFORALL(0,\allowbreak{} uri().\allowbreak{}size(),\allowbreak{} j,\allowbreak{} stream + j $\mapsto$ uri().\allowbreak{}data()[j]) $\star$ SEPFORALL(0,\allowbreak{} correl\allowbreak{}ationI\allowbreak{}d().\allowbreak{}size(),\allowbreak{} i,\allowbreak{} stream + uri().\allowbreak{}size() + 1 + i $\mapsto$ correl\allowbreak{}ationI\allowbreak{}d().\allowbreak{}data()[i]))))
    \end{blockFOSL}
    \item \textbf{Filename:} bmqp\_\allowbreak{}p\allowbreak{}rotoco\allowbreak{}lutil.\allowbreak{}\allowbreak{}cpp, \textbf{Function:} Protoc\allowbreak{}olUtil\allowbreak{}::\allowbreak{}veri\allowbreak{}fy, \textbf{Logic:} \logicPL, \href{https://github.com/bloomberg/blazingmq/blob/main/src/groups/bmq/bmqp/bmqp_protocolutil.cpp#L666}{\textcolor{blue!60!black}{[code]}}
    \begin{blockPL}
    \noindent \textbf{requires:} true \\
    \noindent \textbf{ensures:} (ci.\allowbreak{}con\allowbreak{}sumerP\allowbreak{}riorit\allowbreak{}y() == bmqp::\allowbreak{}\allowbreak{}Protoc\allowbreak{}ol::\allowbreak{}k\_\allowbreak{}\allowbreak{}CONSUM\allowbreak{}ER\_\allowbreak{}PRI\allowbreak{}ORITY\_\allowbreak{}\allowbreak{}INVALI\allowbreak{}D ==$>$ \_\allowbreak{}\_\allowbreak{}out == (ci.\allowbreak{}con\allowbreak{}sumerP\allowbreak{}riorit\allowbreak{}yCount() == 0)) \&\& (ci.\allowbreak{}con\allowbreak{}sumerP\allowbreak{}riorit\allowbreak{}y() != bmqp::\allowbreak{}\allowbreak{}Protoc\allowbreak{}ol::\allowbreak{}k\_\allowbreak{}\allowbreak{}CONSUM\allowbreak{}ER\_\allowbreak{}PRI\allowbreak{}ORITY\_\allowbreak{}\allowbreak{}INVALI\allowbreak{}D ==$>$ \_\allowbreak{}\_\allowbreak{}out == (ci.\allowbreak{}con\allowbreak{}sumerP\allowbreak{}riorit\allowbreak{}yCount() $>$ 0))
    \end{blockPL}
    \item \textbf{Filename:} bmqst\_\allowbreak{}\allowbreak{}statco\allowbreak{}ntext.\allowbreak{}\allowbreak{}cpp, \textbf{Function:} StatCo\allowbreak{}ntext:\allowbreak{}:value\allowbreak{}Index, \textbf{Logic:} \logicFOSL, \href{https://github.com/bloomberg/blazingmq/blob/main/src/groups/bmq/bmqst/bmqst_statcontext.cpp#L727}{\textcolor{blue!60!black}{[code]}}
    \begin{blockFOSL}
    \noindent \textbf{requires:} true \\
    \noindent \textbf{ensures:} (\_\allowbreak{}\_\allowbreak{}out != -1 ==$>$ SEPEXISTS(0,\allowbreak{} d\_\allowbreak{}valu\allowbreak{}eDefs\_\allowbreak{}\allowbreak{}p-$>$siz\allowbreak{}e(),\allowbreak{} i,\allowbreak{} (*d\_\allowbreak{}val\allowbreak{}ueDefs\allowbreak{}\_\allowbreak{}p)[i].\allowbreak{}d\_\allowbreak{}name == name)) \&\& (\_\allowbreak{}\_\allowbreak{}out == -1 ==$>$ FORALL(0,\allowbreak{} d\_\allowbreak{}valu\allowbreak{}eDefs\_\allowbreak{}\allowbreak{}p-$>$siz\allowbreak{}e(),\allowbreak{} i,\allowbreak{} (*d\_\allowbreak{}val\allowbreak{}ueDefs\allowbreak{}\_\allowbreak{}p)[i].\allowbreak{}d\_\allowbreak{}name != name))
    \end{blockFOSL}
    \item \textbf{Filename:} bmqp\_\allowbreak{}e\allowbreak{}vent.\allowbreak{}c\allowbreak{}pp, \textbf{Function:} Event:\allowbreak{}:print, \textbf{Logic:} \logicFOSL, \href{https://github.com/bloomberg/blazingmq/blob/main/src/groups/bmq/bmqp/bmqp_event.cpp#L30}{\textcolor{blue!60!black}{[code]}}
    \begin{blockFOSL}
    \noindent \textbf{requires:} !strea\allowbreak{}m.\allowbreak{}bad() \\
    \noindent \textbf{ensures:} (stream\allowbreak{}.\allowbreak{}bad() ==$>$ \_\allowbreak{}\_\allowbreak{}out == stream) \&\& (stream\allowbreak{}.\allowbreak{}good() ==$>$ (\_\allowbreak{}\_\allowbreak{}out == stream \&\& (SEPFORALL(0,\allowbreak{} strlen("type"),\allowbreak{} i,\allowbreak{} stream + i $\mapsto$ "type"[i]) $\star$ (stream + strlen("type") $\mapsto$ ':') $\star$ (stream + strlen("type") + 1 $\mapsto$ ' ') $\star$ SEPFORALL(0,\allowbreak{} strlen(type()),\allowbreak{} j,\allowbreak{} stream + strlen("type") + 2 + j $\mapsto$ type()[j]))))
    \end{blockFOSL}
    \item \textbf{Filename:} bmqp\_\allowbreak{}e\allowbreak{}ventut\allowbreak{}il.\allowbreak{}cpp, \textbf{Function:} Flatte\allowbreak{}ner::\allowbreak{}c\allowbreak{}loneAn\allowbreak{}dPackE\allowbreak{}achSub\allowbreak{}QId, \textbf{Logic:} \logicFOSL, \href{https://github.com/bloomberg/blazingmq/blob/main/src/groups/bmq/bmqp/bmqp_eventutil.cpp#L220}{\textcolor{blue!60!black}{[code]}}
    \begin{blockFOSL}
    \noindent \textbf{requires:} SEPFORALL(0,\allowbreak{} subQIn\allowbreak{}fos.\allowbreak{}si\allowbreak{}ze(),\allowbreak{} i,\allowbreak{} (subQInfos[i] $\mapsto$ sep\_\allowbreak{}v) \&\& (sep\_\allowbreak{}v != 0)) \\
    \noindent \textbf{ensures:} (\_\allowbreak{}\_\allowbreak{}out == rc\_\allowbreak{}SUC\allowbreak{}CESS) $||$ (\_\allowbreak{}\_\allowbreak{}out != rc\_\allowbreak{}SUC\allowbreak{}CESS ==$>$ (SEPFORALL(0,\allowbreak{} subQIn\allowbreak{}fos.\allowbreak{}si\allowbreak{}ze(),\allowbreak{} i,\allowbreak{} (subQInfos[i] $\mapsto$ sep\_\allowbreak{}v) \&\& (sep\_\allowbreak{}v != 0))))
    \end{blockFOSL}
    \item \textbf{Filename:} bmqp\_\allowbreak{}p\allowbreak{}rotoco\allowbreak{}l.\allowbreak{}cpp, \textbf{Function:} PutHea\allowbreak{}derFla\allowbreak{}gUtil:\allowbreak{}:isVal\allowbreak{}id, \textbf{Logic:} \logicPL, \href{https://github.com/bloomberg/blazingmq/blob/main/src/groups/bmq/bmqp/bmqp_protocol.cpp#L483}{\textcolor{blue!60!black}{[code]}}
    \begin{blockPL}
    \noindent \textbf{requires:} true \\
    \noindent \textbf{ensures:} (\_\allowbreak{}\_\allowbreak{}out == false ==$>$ (isSet(flags,\allowbreak{} PutHea\allowbreak{}derFla\allowbreak{}gs::\allowbreak{}e\_\allowbreak{}\allowbreak{}UNUSED\allowbreak{}3) $||$ isSet(flags,\allowbreak{} PutHea\allowbreak{}derFla\allowbreak{}gs::\allowbreak{}e\_\allowbreak{}\allowbreak{}UNUSED\allowbreak{}4))) \&\& (\_\allowbreak{}\_\allowbreak{}out == true ==$>$ !(isSet(flags,\allowbreak{} PutHea\allowbreak{}derFla\allowbreak{}gs::\allowbreak{}e\_\allowbreak{}\allowbreak{}UNUSED\allowbreak{}3) $||$ isSet(flags,\allowbreak{} PutHea\allowbreak{}derFla\allowbreak{}gs::\allowbreak{}e\_\allowbreak{}\allowbreak{}UNUSED\allowbreak{}4)))
    \end{blockPL}
    \item \textbf{Filename:} bmqp\_\allowbreak{}p\allowbreak{}rotoco\allowbreak{}l.\allowbreak{}cpp, \textbf{Function:} PushHe\allowbreak{}aderFl\allowbreak{}agUtil\allowbreak{}::\allowbreak{}isVa\allowbreak{}lid, \textbf{Logic:} \logicPL, \href{https://github.com/bloomberg/blazingmq/blob/main/src/groups/bmq/bmqp/bmqp_protocol.cpp#L657}{\textcolor{blue!60!black}{[code]}}
    \begin{blockPL}
    \noindent \textbf{requires:} true \\
    \noindent \textbf{ensures:} (\_\allowbreak{}\_\allowbreak{}out == false ==$>$ isSet(flags,\allowbreak{} PushHe\allowbreak{}aderFl\allowbreak{}ags::\allowbreak{}e\allowbreak{}\_\allowbreak{}UNUSE\allowbreak{}D4)) \&\& (\_\allowbreak{}\_\allowbreak{}out == true ==$>$ !isSet(flags,\allowbreak{} PushHe\allowbreak{}aderFl\allowbreak{}ags::\allowbreak{}e\allowbreak{}\_\allowbreak{}UNUSE\allowbreak{}D4))
    \end{blockPL}
    \item \textbf{Filename:} bmqp\_\allowbreak{}p\allowbreak{}rotoco\allowbreak{}l.\allowbreak{}cpp, \textbf{Function:} Storag\allowbreak{}eHeade\allowbreak{}rFlagU\allowbreak{}til::\allowbreak{}f\allowbreak{}romStr\allowbreak{}ing, \textbf{Logic:} \logicFOSL, \href{https://github.com/bloomberg/blazingmq/blob/main/src/groups/bmq/bmqp/bmqp_protocol.cpp#L896}{\textcolor{blue!60!black}{[code]}}
    \begin{blockFOSL}
    \noindent \textbf{requires:} true \\
    \noindent \textbf{ensures:} (\_\allowbreak{}\_\allowbreak{}out == 0 ==$>$ (out $\mapsto$ sep\_\allowbreak{}v \&\& SEPFORALL(0,\allowbreak{} bdlb::\allowbreak{}\allowbreak{}Tokeni\allowbreak{}zer(str,\allowbreak{} ",\allowbreak{}").\allowbreak{}size(),\allowbreak{} i,\allowbreak{} (bdlb::\allowbreak{}\allowbreak{}Tokeni\allowbreak{}zer(str,\allowbreak{} ",\allowbreak{}").\allowbreak{}begin() + i) != bdlb::\allowbreak{}\allowbreak{}Tokeni\allowbreak{}zer(str,\allowbreak{} ",\allowbreak{}").\allowbreak{}end() \&\& Storag\allowbreak{}eHeade\allowbreak{}rFlags\allowbreak{}::\allowbreak{}from\allowbreak{}Ascii(\&sep\_\allowbreak{}v,\allowbreak{} *(bdlb::\allowbreak{}\allowbreak{}Tokeni\allowbreak{}zer(str,\allowbreak{} ",\allowbreak{}").\allowbreak{}begin() + i)) == true \&\& sep\_\allowbreak{}v == (sep\_\allowbreak{}v | *out)))) \&\& (\_\allowbreak{}\_\allowbreak{}out == -1 ==$>$ EXISTS(0,\allowbreak{} bdlb::\allowbreak{}\allowbreak{}Tokeni\allowbreak{}zer(str,\allowbreak{} ",\allowbreak{}").\allowbreak{}size(),\allowbreak{} i,\allowbreak{} (bdlb::\allowbreak{}\allowbreak{}Tokeni\allowbreak{}zer(str,\allowbreak{} ",\allowbreak{}").\allowbreak{}begin() + i) != bdlb::\allowbreak{}\allowbreak{}Tokeni\allowbreak{}zer(str,\allowbreak{} ",\allowbreak{}").\allowbreak{}end() \&\& Storag\allowbreak{}eHeade\allowbreak{}rFlags\allowbreak{}::\allowbreak{}from\allowbreak{}Ascii(\&sep\_\allowbreak{}v,\allowbreak{} *(bdlb::\allowbreak{}\allowbreak{}Tokeni\allowbreak{}zer(str,\allowbreak{} ",\allowbreak{}").\allowbreak{}begin() + i)) == false \&\& errorD\allowbreak{}escrip\allowbreak{}tion.\allowbreak{}s\allowbreak{}tr().\allowbreak{}find(*(bdlb::\allowbreak{}\allowbreak{}Tokeni\allowbreak{}zer(str,\allowbreak{} ",\allowbreak{}").\allowbreak{}begin() + i)) != std::\allowbreak{}s\allowbreak{}tring:\allowbreak{}:npos))
    \end{blockFOSL}
    \item \textbf{Filename:} bmqp\_\allowbreak{}q\allowbreak{}ueueid\allowbreak{}.\allowbreak{}cpp, \textbf{Function:} QueueI\allowbreak{}d::\allowbreak{}pri\allowbreak{}nt, \textbf{Logic:} \logicFOSL, \href{https://github.com/bloomberg/blazingmq/blob/main/src/groups/bmq/bmqp/bmqp_queueid.cpp#L61}{\textcolor{blue!60!black}{[code]}}
    \begin{blockFOSL}
    \noindent \textbf{requires:} stream\allowbreak{}.\allowbreak{}good() \&\& d\_\allowbreak{}id[strlen(d\_\allowbreak{}id)] == '\textbackslash{}0' \&\& d\_\allowbreak{}subId[strlen(d\_\allowbreak{}subId)] == '\textbackslash{}0' \\
    \noindent \textbf{ensures:} \_\allowbreak{}\_\allowbreak{}out == stream \&\& (stream\allowbreak{}.\allowbreak{}bad() $||$ (stream\allowbreak{}.\allowbreak{}good() \&\& (SEPFORALL(0,\allowbreak{} strlen(d\_\allowbreak{}id),\allowbreak{} i,\allowbreak{} stream + i $\mapsto$ d\_\allowbreak{}id[i]) $\star$ SEPFORALL(0,\allowbreak{} strlen(d\_\allowbreak{}subId),\allowbreak{} j,\allowbreak{} stream + strlen(d\_\allowbreak{}id) + j $\mapsto$ d\_\allowbreak{}subId[j]))))
    \end{blockFOSL}
    \item \textbf{Filename:} bmqst\_\allowbreak{}\allowbreak{}printu\allowbreak{}til.\allowbreak{}cp\allowbreak{}p, \textbf{Function:} PrintU\allowbreak{}til::\allowbreak{}p\allowbreak{}rintVa\allowbreak{}lueWit\allowbreak{}hSepar\allowbreak{}ator, \textbf{Logic:} \logicFOSL, \href{https://github.com/bloomberg/blazingmq/blob/main/src/groups/bmq/bmqst/bmqst_printutil.cpp#L237}{\textcolor{blue!60!black}{[code]}}
    \begin{blockFOSL}
    \noindent \textbf{requires:} stream\allowbreak{}.\allowbreak{}good() \&\& groupSize $>$= 0 \\
    \noindent \textbf{ensures:} \_\allowbreak{}\_\allowbreak{}out == stream \&\& (stream\allowbreak{}.\allowbreak{}bad() $||$ (stream\allowbreak{}.\allowbreak{}good() \&\& (SEPFORALL(0,\allowbreak{} 64,\allowbreak{} i,\allowbreak{} (buf + i) $\mapsto$ sep\_\allowbreak{}v) $\star$ printV\allowbreak{}alueWi\allowbreak{}thSepa\allowbreak{}ratorI\allowbreak{}mp(buf + 63,\allowbreak{} value,\allowbreak{} groupS\allowbreak{}ize,\allowbreak{} separator))))
    \end{blockFOSL}
    \item \textbf{Filename:} bmqt\_\allowbreak{}c\allowbreak{}ompres\allowbreak{}sional\allowbreak{}gorith\allowbreak{}mtype.\allowbreak{}\allowbreak{}cpp, \textbf{Function:} Compre\allowbreak{}ssionA\allowbreak{}lgorit\allowbreak{}hmType\allowbreak{}::\allowbreak{}isVa\allowbreak{}lid, \textbf{Logic:} \logicFOL, \href{https://github.com/bloomberg/blazingmq/blob/main/src/groups/bmq/bmqt/bmqt_compressionalgorithmtype.cpp#L82}{\textcolor{blue!60!black}{[code]}}
    \begin{blockFOL}
    \noindent \textbf{requires:} str != 0 \\
    \noindent \textbf{ensures:} (\_\allowbreak{}\_\allowbreak{}out == true ==$>$ EXISTS(0,\allowbreak{} 1,\allowbreak{} dummy,\allowbreak{} fromAscii(str) == true)) \&\& (\_\allowbreak{}\_\allowbreak{}out == false ==$>$ EXISTS(0,\allowbreak{} 1,\allowbreak{} dummy,\allowbreak{} fromAscii(str) == false))
    \end{blockFOL}
    \item \textbf{Filename:} bmqt\_\allowbreak{}c\allowbreak{}orrela\allowbreak{}tionid\allowbreak{}.\allowbreak{}cpp, \textbf{Function:} Correl\allowbreak{}ationI\allowbreak{}d::\allowbreak{}pri\allowbreak{}nt, \textbf{Logic:} \logicFOSL, \href{https://github.com/bloomberg/blazingmq/blob/main/src/groups/bmq/bmqt/bmqt_correlationid.cpp#L48}{\textcolor{blue!60!black}{[code]}}
    \begin{blockFOSL}
    \noindent \textbf{requires:} !strea\allowbreak{}m.\allowbreak{}bad() \\
    \noindent \textbf{ensures:} (stream\allowbreak{}.\allowbreak{}bad() ==$>$ \_\allowbreak{}\_\allowbreak{}out == stream) \&\& (stream\allowbreak{}.\allowbreak{}good() ==$>$ (\_\allowbreak{}\_\allowbreak{}out == stream \&\& (SEPFORALL(0,\allowbreak{} printe\allowbreak{}r.\allowbreak{}outp\allowbreak{}utSize(),\allowbreak{} i,\allowbreak{} stream + i $\mapsto$ printe\allowbreak{}r.\allowbreak{}outp\allowbreak{}ut()[i]))))
    \end{blockFOSL}
    \item \textbf{Filename:} bmqt\_\allowbreak{}e\allowbreak{}ncodin\allowbreak{}gtype.\allowbreak{}\allowbreak{}cpp, \textbf{Function:} Encodi\allowbreak{}ngType\allowbreak{}::\allowbreak{}isVa\allowbreak{}lid, \textbf{Logic:} \logicFOL, \href{https://github.com/bloomberg/blazingmq/blob/main/src/groups/bmq/bmqt/bmqt_encodingtype.cpp#L92}{\textcolor{blue!60!black}{[code]}}
    \begin{blockFOL}
    \noindent \textbf{requires:} string != 0 \\
    \noindent \textbf{ensures:} (\_\allowbreak{}\_\allowbreak{}out == true ==$>$ EXISTS(0,\allowbreak{} 1,\allowbreak{} dummy,\allowbreak{} fromAscii(string) == true)) \&\& (\_\allowbreak{}\_\allowbreak{}out == false ==$>$ EXISTS(0,\allowbreak{} 1,\allowbreak{} dummy,\allowbreak{} fromAscii(string) == false))
    \end{blockFOL}
    \item \textbf{Filename:} bmqt\_\allowbreak{}q\allowbreak{}ueuefl\allowbreak{}ags.\allowbreak{}cp\allowbreak{}p, \textbf{Function:} QueueF\allowbreak{}lagsUt\allowbreak{}il::\allowbreak{}is\allowbreak{}Valid, \textbf{Logic:} \logicFOSL, \href{https://github.com/bloomberg/blazingmq/blob/main/src/groups/bmq/bmqt/bmqt_queueflags.cpp#L33}{\textcolor{blue!60!black}{[code]}}
    \begin{blockFOSL}
    \noindent \textbf{requires:} (!isSet(flags,\allowbreak{} QueueF\allowbreak{}lags::\allowbreak{}\allowbreak{}e\_\allowbreak{}ADMI\allowbreak{}N) $||$ /* valid for BlazingMQ admin tasks */) \&\& (isSet(flags,\allowbreak{} QueueF\allowbreak{}lags::\allowbreak{}\allowbreak{}e\_\allowbreak{}READ) $||$ isSet(flags,\allowbreak{} QueueF\allowbreak{}lags::\allowbreak{}\allowbreak{}e\_\allowbreak{}WRIT\allowbreak{}E)) \\
    \noindent \textbf{ensures:} (\_\allowbreak{}\_\allowbreak{}out == false ==$>$ SEPEXISTS(0,\allowbreak{} errorD\allowbreak{}escrip\allowbreak{}tion.\allowbreak{}t\allowbreak{}ellp(),\allowbreak{} i,\allowbreak{} errorD\allowbreak{}escrip\allowbreak{}tion + i $\mapsto$ \_\allowbreak{})) \&\& (\_\allowbreak{}\_\allowbreak{}out == true ==$>$ SEPFORALL(0,\allowbreak{} errorD\allowbreak{}escrip\allowbreak{}tion.\allowbreak{}t\allowbreak{}ellp(),\allowbreak{} i,\allowbreak{} errorD\allowbreak{}escrip\allowbreak{}tion + i $\mapsto$ old\_\allowbreak{}er\allowbreak{}rorDes\allowbreak{}cripti\allowbreak{}on[i]))
    \end{blockFOSL}
    \item \textbf{Filename:} bmqt\_\allowbreak{}q\allowbreak{}ueueop\allowbreak{}tions.\allowbreak{}\allowbreak{}cpp, \textbf{Function:} QueueO\allowbreak{}ptions\allowbreak{}::\allowbreak{}prin\allowbreak{}t, \textbf{Logic:} \logicFOSL, \href{https://github.com/bloomberg/blazingmq/blob/main/src/groups/bmq/bmqt/bmqt_queueoptions.cpp#L91}{\textcolor{blue!60!black}{[code]}}
    \begin{blockFOSL}
    \noindent \textbf{requires:} stream\allowbreak{}.\allowbreak{}good() \\
    \noindent \textbf{ensures:} \_\allowbreak{}\_\allowbreak{}out == stream \&\& (stream\allowbreak{}.\allowbreak{}bad() $||$ (stream\allowbreak{}.\allowbreak{}good() \&\& (SEPFORALL(0,\allowbreak{} strlen("maxUn\allowbreak{}confir\allowbreak{}medMes\allowbreak{}sages"),\allowbreak{} i,\allowbreak{} stream + i $\mapsto$ "maxUn\allowbreak{}confir\allowbreak{}medMes\allowbreak{}sages"[i]) $\star$ SEPFORALL(0,\allowbreak{} strlen("maxUn\allowbreak{}confir\allowbreak{}medByt\allowbreak{}es"),\allowbreak{} i,\allowbreak{} stream + strlen("maxUn\allowbreak{}confir\allowbreak{}medMes\allowbreak{}sages") + 1 + i $\mapsto$ "maxUn\allowbreak{}confir\allowbreak{}medByt\allowbreak{}es"[i]) $\star$ SEPFORALL(0,\allowbreak{} strlen("consu\allowbreak{}merPri\allowbreak{}ority"),\allowbreak{} i,\allowbreak{} stream + strlen("maxUn\allowbreak{}confir\allowbreak{}medMes\allowbreak{}sages") + strlen("maxUn\allowbreak{}confir\allowbreak{}medByt\allowbreak{}es") + 2 + i $\mapsto$ "consu\allowbreak{}merPri\allowbreak{}ority"[i]) $\star$ SEPFORALL(0,\allowbreak{} strlen("suspe\allowbreak{}ndsOnB\allowbreak{}adHost\allowbreak{}Health\allowbreak{}"),\allowbreak{} i,\allowbreak{} stream + strlen("maxUn\allowbreak{}confir\allowbreak{}medMes\allowbreak{}sages") + strlen("maxUn\allowbreak{}confir\allowbreak{}medByt\allowbreak{}es") + strlen("consu\allowbreak{}merPri\allowbreak{}ority") + 3 + i $\mapsto$ "suspe\allowbreak{}ndsOnB\allowbreak{}adHost\allowbreak{}Health\allowbreak{}"[i]) $\star$ (SEPEXISTS(0,\allowbreak{} d\_\allowbreak{}subs\allowbreak{}cripti\allowbreak{}ons.\allowbreak{}si\allowbreak{}ze(),\allowbreak{} j,\allowbreak{} stream + offset + j $\mapsto$ "Subsc\allowbreak{}riptio\allowbreak{}ns:"[j])))))
    \end{blockFOSL}
    \item \textbf{Filename:} bmqt\_\allowbreak{}v\allowbreak{}ersion\allowbreak{}.\allowbreak{}cpp, \textbf{Function:} Versio\allowbreak{}n::\allowbreak{}pri\allowbreak{}nt, \textbf{Logic:} \logicPL, \href{https://github.com/bloomberg/blazingmq/blob/main/src/groups/bmq/bmqt/bmqt_version.cpp#L30}{\textcolor{blue!60!black}{[code]}}
    \begin{blockPL}
    \noindent \textbf{requires:} true \\
    \noindent \textbf{ensures:} \_\allowbreak{}\_\allowbreak{}out == stream
    \end{blockPL}
    \item \textbf{Filename:} bmqu\_\allowbreak{}b\allowbreak{}lob.\allowbreak{}cp\allowbreak{}p, \textbf{Function:} BlobPo\allowbreak{}sition\allowbreak{}::\allowbreak{}prin\allowbreak{}t, \textbf{Logic:} \logicFOSL, \href{https://github.com/bloomberg/blazingmq/blob/main/src/groups/bmq/bmqu/bmqu_blob.cpp#L33}{\textcolor{blue!60!black}{[code]}}
    \begin{blockFOSL}
    \noindent \textbf{requires:} stream\allowbreak{}.\allowbreak{}good() \\
    \noindent \textbf{ensures:} (stream\allowbreak{}.\allowbreak{}bad() ==$>$ \_\allowbreak{}\_\allowbreak{}out == stream) \&\& (stream\allowbreak{}.\allowbreak{}good() ==$>$ (\_\allowbreak{}\_\allowbreak{}out == stream \&\& (SEPFORALL(0,\allowbreak{} d\_\allowbreak{}buff\allowbreak{}er.\allowbreak{}siz\allowbreak{}e(),\allowbreak{} i,\allowbreak{} stream + i $\mapsto$ d\_\allowbreak{}buff\allowbreak{}er.\allowbreak{}dat\allowbreak{}a()[i]) $\star$ SEPFORALL(0,\allowbreak{} d\_\allowbreak{}byte\allowbreak{}.\allowbreak{}size(),\allowbreak{} j,\allowbreak{} stream + d\_\allowbreak{}buff\allowbreak{}er.\allowbreak{}siz\allowbreak{}e() + j $\mapsto$ d\_\allowbreak{}byte\allowbreak{}.\allowbreak{}data()[j]))))
    \end{blockFOSL}
    \item \textbf{Filename:} bmqu\_\allowbreak{}b\allowbreak{}lob.\allowbreak{}cp\allowbreak{}p, \textbf{Function:} BlobUt\allowbreak{}il::\allowbreak{}is\allowbreak{}ValidP\allowbreak{}os, \textbf{Logic:} \logicPL, \href{https://github.com/bloomberg/blazingmq/blob/main/src/groups/bmq/bmqu/bmqu_blob.cpp#L72}{\textcolor{blue!60!black}{[code]}}
    \begin{blockPL}
    \noindent \textbf{requires:} pos.\allowbreak{}bu\allowbreak{}ffer() $>$= 0 \&\& pos.\allowbreak{}byte() $>$= 0 \\
    \noindent \textbf{ensures:} (pos.\allowbreak{}bu\allowbreak{}ffer() $>$ blob.\allowbreak{}n\allowbreak{}umData\allowbreak{}Buffer\allowbreak{}s() ==$>$ \_\allowbreak{}\_\allowbreak{}out == false) \&\& (pos.\allowbreak{}bu\allowbreak{}ffer() == blob.\allowbreak{}n\allowbreak{}umData\allowbreak{}Buffer\allowbreak{}s() \&\& pos.\allowbreak{}byte() != 0 ==$>$ \_\allowbreak{}\_\allowbreak{}out == false) \&\& (pos.\allowbreak{}bu\allowbreak{}ffer() $<$ blob.\allowbreak{}n\allowbreak{}umData\allowbreak{}Buffer\allowbreak{}s() \&\& pos.\allowbreak{}byte() $>$= 0 \&\& pos.\allowbreak{}byte() $<$ buffer\allowbreak{}Size(blob,\allowbreak{} pos.\allowbreak{}bu\allowbreak{}ffer()) ==$>$ \_\allowbreak{}\_\allowbreak{}out == true)
    \end{blockPL}
    \item \textbf{Filename:} bmqu\_\allowbreak{}b\allowbreak{}lob.\allowbreak{}cp\allowbreak{}p, \textbf{Function:} BlobUt\allowbreak{}il::\allowbreak{}wr\allowbreak{}iteByt\allowbreak{}es, \textbf{Logic:} \logicFOSL, \href{https://github.com/bloomberg/blazingmq/blob/main/src/groups/bmq/bmqu/bmqu_blob.cpp#L217}{\textcolor{blue!60!black}{[code]}}
    \begin{blockFOSL}
    \noindent \textbf{requires:} pos.\allowbreak{}byte() + length $<$= buffer\allowbreak{}Size(*blob,\allowbreak{} pos.\allowbreak{}bu\allowbreak{}ffer()) \\
    \noindent \textbf{ensures:} (\_\allowbreak{}\_\allowbreak{}out != 0) $||$ (\_\allowbreak{}\_\allowbreak{}out == 0 \&\& SEPFORALL(0,\allowbreak{} length,\allowbreak{} i,\allowbreak{} (blob-$>$\allowbreak{}buffer((pos.\allowbreak{}bu\allowbreak{}ffer() + i / buffer\allowbreak{}Size(*blob,\allowbreak{} pos.\allowbreak{}bu\allowbreak{}ffer()))).\allowbreak{}data() + (pos.\allowbreak{}byte() + i \% buffer\allowbreak{}Size(*blob,\allowbreak{} pos.\allowbreak{}bu\allowbreak{}ffer()))) $\mapsto$ buf[i]))
    \end{blockFOSL}
    \item \textbf{Filename:} bmqu\_\allowbreak{}p\allowbreak{}rintut\allowbreak{}il.\allowbreak{}cpp, \textbf{Function:} pretty\allowbreak{}Number\allowbreak{}Imp, \textbf{Logic:} \logicFOSL, \href{https://github.com/bloomberg/blazingmq/blob/main/src/groups/bmq/bmqu/bmqu_printutil.cpp#L43}{\textcolor{blue!60!black}{[code]}}
    \begin{blockFOSL}
    \noindent \textbf{requires:} buf != nullptr \&\& groupSize $>$ 0 \&\& (separator $>$= 0 \&\& separator $<$= 127) \\
    \noindent \textbf{ensures:} (\_\allowbreak{}\_\allowbreak{}out == buf) \&\& SEPFORALL(0,\allowbreak{} strlen(buf),\allowbreak{} i,\allowbreak{} buf + i $\mapsto$ buf[i])
    \end{blockFOSL}
    \item \textbf{Filename:} bmqu\_\allowbreak{}p\allowbreak{}rintut\allowbreak{}il.\allowbreak{}cpp, \textbf{Function:} pretty\allowbreak{}Bytes, \textbf{Logic:} \logicFOSL, \href{https://github.com/bloomberg/blazingmq/blob/main/src/groups/bmq/bmqu/bmqu_printutil.cpp#L145}{\textcolor{blue!60!black}{[code]}}
    \begin{blockFOSL}
    \noindent \textbf{requires:} true \\
    \noindent \textbf{ensures:} \_\allowbreak{}\_\allowbreak{}out == stream \&\& (SEPFORALL(0,\allowbreak{} temp.\allowbreak{}str().\allowbreak{}size(),\allowbreak{} i,\allowbreak{} stream + i $\mapsto$ temp.\allowbreak{}str().\allowbreak{}data()[i]))
    \end{blockFOSL}
    \item \textbf{Filename:} bmqu\_\allowbreak{}p\allowbreak{}rintut\allowbreak{}il.\allowbreak{}cpp, \textbf{Function:} pretty\allowbreak{}TimeIn\allowbreak{}terval, \textbf{Logic:} \logicFOSL, \href{https://github.com/bloomberg/blazingmq/blob/main/src/groups/bmq/bmqu/bmqu_printutil.cpp#L231}{\textcolor{blue!60!black}{[code]}}
    \begin{blockFOSL}
    \noindent \textbf{requires:} stream\allowbreak{}.\allowbreak{}good() \&\& precision $>$= 0 \\
    \noindent \textbf{ensures:} \_\allowbreak{}\_\allowbreak{}out == stream \&\& stream\allowbreak{}.\allowbreak{}good() \&\& (SEPFORALL(0,\allowbreak{} temp.\allowbreak{}str().\allowbreak{}size(),\allowbreak{} i,\allowbreak{} stream + i $\mapsto$ temp.\allowbreak{}str()[i]))
    \end{blockFOSL}
    \item \textbf{Filename:} bmqu\_\allowbreak{}s\allowbreak{}tringu\allowbreak{}til.\allowbreak{}cp\allowbreak{}p, \textbf{Function:} remove\allowbreak{}IfPrec\allowbreak{}ededBy\allowbreak{}Same, \textbf{Logic:} \logicFOSL, \href{https://github.com/bloomberg/blazingmq/blob/main/src/groups/bmq/bmqu/bmqu_stringutil.cpp#L37}{\textcolor{blue!60!black}{[code]}}
    \begin{blockFOSL}
    \noindent \textbf{requires:} begin $<$= end \&\& allowl\allowbreak{}istBeg\allowbreak{}in $<$= allowl\allowbreak{}istEnd \&\& SEPFORALL(0,\allowbreak{} end - begin,\allowbreak{} i,\allowbreak{} (begin + i $\mapsto$ \_\allowbreak{})) \&\& SEPFORALL(0,\allowbreak{} allowl\allowbreak{}istEnd - allowl\allowbreak{}istBeg\allowbreak{}in,\allowbreak{} j,\allowbreak{} (allowl\allowbreak{}istBeg\allowbreak{}in + j $\mapsto$ \_\allowbreak{})) \\
    \noindent \textbf{ensures:} (\_\allowbreak{}\_\allowbreak{}out $>$= begin \&\& \_\allowbreak{}\_\allowbreak{}out $<$= end) \&\& SEPFORALL(0,\allowbreak{} \_\allowbreak{}\_\allowbreak{}out - begin,\allowbreak{} i,\allowbreak{} (begin + i $\mapsto$ sep\_\allowbreak{}v) \&\& (!SEPEX\allowbreak{}ISTS(allowl\allowbreak{}istBeg\allowbreak{}in,\allowbreak{} allowl\allowbreak{}istEnd\allowbreak{},\allowbreak{} j,\allowbreak{} allowl\allowbreak{}istBeg\allowbreak{}in + j $\mapsto$ sep\_\allowbreak{}v) $||$ (i == 0 $||$ *(begin + i - 1) != sep\_\allowbreak{}v)))
    \end{blockFOSL}
    \item \textbf{Filename:} bmqu\_\allowbreak{}s\allowbreak{}tringu\allowbreak{}til.\allowbreak{}cp\allowbreak{}p, \textbf{Function:} String\allowbreak{}Util::\allowbreak{}\allowbreak{}starts\allowbreak{}With, \textbf{Logic:} \logicFOSL, \href{https://github.com/bloomberg/blazingmq/blob/main/src/groups/bmq/bmqu/bmqu_stringutil.cpp#L79}{\textcolor{blue!60!black}{[code]}}
    \begin{blockFOSL}
    \noindent \textbf{requires:} (offset $<$= str.\allowbreak{}le\allowbreak{}ngth()) \&\& SEPFORALL(0,\allowbreak{} str.\allowbreak{}le\allowbreak{}ngth(),\allowbreak{} i,\allowbreak{} str.\allowbreak{}data() + i $\mapsto$ \_\allowbreak{}) \&\& SEPFORALL(0,\allowbreak{} prefix\allowbreak{}.\allowbreak{}lengt\allowbreak{}h(),\allowbreak{} i,\allowbreak{} prefix\allowbreak{}.\allowbreak{}data() + i $\mapsto$ \_\allowbreak{}) \\
    \noindent \textbf{ensures:} (\_\allowbreak{}\_\allowbreak{}out == true ==$>$ SEPFORALL(0,\allowbreak{} prefix\allowbreak{}.\allowbreak{}lengt\allowbreak{}h(),\allowbreak{} i,\allowbreak{} str.\allowbreak{}data() + offset + i $\mapsto$ prefix\allowbreak{}.\allowbreak{}data()[i])) \&\& (\_\allowbreak{}\_\allowbreak{}out == false ==$>$ ((offset $>$ str.\allowbreak{}le\allowbreak{}ngth()) $||$ ((str.\allowbreak{}le\allowbreak{}ngth() - offset) $<$ prefix\allowbreak{}.\allowbreak{}lengt\allowbreak{}h()) $||$ SEPEXISTS(0,\allowbreak{} prefix\allowbreak{}.\allowbreak{}lengt\allowbreak{}h(),\allowbreak{} i,\allowbreak{} str.\allowbreak{}data() + offset + i $\mapsto$ sep\_\allowbreak{}v \&\& sep\_\allowbreak{}v != prefix\allowbreak{}.\allowbreak{}data()[i])))
    \end{blockFOSL}
    \item \textbf{Filename:} bmqu\_\allowbreak{}s\allowbreak{}tringu\allowbreak{}til.\allowbreak{}cp\allowbreak{}p, \textbf{Function:} String\allowbreak{}Util::\allowbreak{}\allowbreak{}endsWi\allowbreak{}th, \textbf{Logic:} \logicFOSL, \href{https://github.com/bloomberg/blazingmq/blob/main/src/groups/bmq/bmqu/bmqu_stringutil.cpp#L100}{\textcolor{blue!60!black}{[code]}}
    \begin{blockFOSL}
    \noindent \textbf{requires:} true \\
    \noindent \textbf{ensures:} (\_\allowbreak{}\_\allowbreak{}out == true ==$>$ (str.\allowbreak{}le\allowbreak{}ngth() $>$= suffix\allowbreak{}.\allowbreak{}lengt\allowbreak{}h() \&\& SEPFORALL(0,\allowbreak{} suffix\allowbreak{}.\allowbreak{}lengt\allowbreak{}h(),\allowbreak{} i,\allowbreak{} str[str.\allowbreak{}le\allowbreak{}ngth() - 1 - i] == suffix[suffix\allowbreak{}.\allowbreak{}lengt\allowbreak{}h() - 1 - i]))) \&\& (\_\allowbreak{}\_\allowbreak{}out == false ==$>$ (str.\allowbreak{}le\allowbreak{}ngth() $<$ suffix\allowbreak{}.\allowbreak{}lengt\allowbreak{}h() $||$ SEPEXISTS(0,\allowbreak{} suffix\allowbreak{}.\allowbreak{}lengt\allowbreak{}h(),\allowbreak{} i,\allowbreak{} str[str.\allowbreak{}le\allowbreak{}ngth() - 1 - i] != suffix[suffix\allowbreak{}.\allowbreak{}lengt\allowbreak{}h() - 1 - i])))
    \end{blockFOSL}
    \item \textbf{Filename:} bmqu\_\allowbreak{}s\allowbreak{}tringu\allowbreak{}til.\allowbreak{}cp\allowbreak{}p, \textbf{Function:} String\allowbreak{}Util::\allowbreak{}\allowbreak{}ltrim, \textbf{Logic:} \logicFOSL, \href{https://github.com/bloomberg/blazingmq/blob/main/src/groups/bmq/bmqu/bmqu_stringutil.cpp#L125}{\textcolor{blue!60!black}{[code]}}
    \begin{blockFOSL}
    \noindent \textbf{requires:} str != 0 \&\& SEPFORALL(0,\allowbreak{} str-$>$size(),\allowbreak{} i,\allowbreak{} ((str-$>$b\allowbreak{}egin() + i) $\mapsto$ \_\allowbreak{})) \\
    \noindent \textbf{ensures:} \_\allowbreak{}\_\allowbreak{}out == *str \&\& SEPFORALL(0,\allowbreak{} \_\allowbreak{}\_\allowbreak{}out.\allowbreak{}\allowbreak{}size(),\allowbreak{} i,\allowbreak{} (\_\allowbreak{}\_\allowbreak{}out[i] $\mapsto$ sep\_\allowbreak{}v \&\& (sep\_\allowbreak{}v != ' ' $||$ i $>$ 0)))
    \end{blockFOSL}
    \item \textbf{Filename:} bmqp\_\allowbreak{}p\allowbreak{}rotoco\allowbreak{}l.\allowbreak{}cpp, \textbf{Function:} Storag\allowbreak{}eHeade\allowbreak{}rFlagU\allowbreak{}til::\allowbreak{}i\allowbreak{}sValid, \textbf{Logic:} \logicPL, \href{https://github.com/bloomberg/blazingmq/blob/main/src/groups/bmq/bmqp/bmqp_protocol.cpp#L861}{\textcolor{blue!60!black}{[code]}}
    \begin{blockPL}
    \noindent \textbf{requires:} true \\
    \noindent \textbf{ensures:} (\_\allowbreak{}\_\allowbreak{}out == false ==$>$ (isSet(flags,\allowbreak{} Storag\allowbreak{}eHeade\allowbreak{}rFlags\allowbreak{}::\allowbreak{}e\_\allowbreak{}UN\allowbreak{}USED2) $||$ isSet(flags,\allowbreak{} Storag\allowbreak{}eHeade\allowbreak{}rFlags\allowbreak{}::\allowbreak{}e\_\allowbreak{}UN\allowbreak{}USED3) $||$ isSet(flags,\allowbreak{} Storag\allowbreak{}eHeade\allowbreak{}rFlags\allowbreak{}::\allowbreak{}e\_\allowbreak{}UN\allowbreak{}USED4))) \&\& (\_\allowbreak{}\_\allowbreak{}out == true ==$>$ !(isSet(flags,\allowbreak{} Storag\allowbreak{}eHeade\allowbreak{}rFlags\allowbreak{}::\allowbreak{}e\_\allowbreak{}UN\allowbreak{}USED2) $||$ isSet(flags,\allowbreak{} Storag\allowbreak{}eHeade\allowbreak{}rFlags\allowbreak{}::\allowbreak{}e\_\allowbreak{}UN\allowbreak{}USED3) $||$ isSet(flags,\allowbreak{} Storag\allowbreak{}eHeade\allowbreak{}rFlags\allowbreak{}::\allowbreak{}e\_\allowbreak{}UN\allowbreak{}USED4)))
    \end{blockPL}
    \item \textbf{Filename:} bmqu\_\allowbreak{}s\allowbreak{}tringu\allowbreak{}til.\allowbreak{}cp\allowbreak{}p, \textbf{Function:} String\allowbreak{}Util::\allowbreak{}\allowbreak{}squeez\allowbreak{}e, \textbf{Logic:} \logicFOSL, \href{https://github.com/bloomberg/blazingmq/blob/main/src/groups/bmq/bmqu/bmqu_stringutil.cpp#L241}{\textcolor{blue!60!black}{[code]}}
    \begin{blockFOSL}
    \noindent \textbf{requires:} str != nullptr \&\& (SEPFORALL(0,\allowbreak{} str-$>$size(),\allowbreak{} i,\allowbreak{} (*str)[i] $\mapsto$ \_\allowbreak{})) \&\& (charac\allowbreak{}ters.\allowbreak{}s\allowbreak{}ize() $>$= 0) \\
    \noindent \textbf{ensures:} \_\allowbreak{}\_\allowbreak{}out == *str \&\& !(SEPEXISTS(0,\allowbreak{} \_\allowbreak{}\_\allowbreak{}out.\allowbreak{}\allowbreak{}size() - 1,\allowbreak{} i,\allowbreak{} (\_\allowbreak{}\_\allowbreak{}out[i] == \_\allowbreak{}\_\allowbreak{}out[i + 1]) \&\& (charac\allowbreak{}ters.\allowbreak{}f\allowbreak{}ind(\_\allowbreak{}\_\allowbreak{}out[i]) != bslstl\allowbreak{}::\allowbreak{}Stri\allowbreak{}ngRef:\allowbreak{}:npos)))
    \end{blockFOSL}
    \item \textbf{Filename:} bmqimp\allowbreak{}\_\allowbreak{}messa\allowbreak{}gecorr\allowbreak{}elatio\allowbreak{}nidcon\allowbreak{}tainer\allowbreak{}.\allowbreak{}cpp, \textbf{Function:} Messag\allowbreak{}eCorre\allowbreak{}lation\allowbreak{}IdCont\allowbreak{}ainer:\allowbreak{}:itera\allowbreak{}teAndI\allowbreak{}nvoke, \textbf{Logic:} \logicPL, \href{https://github.com/bloomberg/blazingmq/blob/main/src/groups/bmq/bmqimp/bmqimp_messagecorrelationidcontainer.cpp#L234}{\textcolor{blue!60!black}{[code]}}
    \begin{blockPL}
    \noindent \textbf{requires:} true \\
    \noindent \textbf{ensures:} (\_\allowbreak{}\_\allowbreak{}out == true ==$>$ true) \&\& (\_\allowbreak{}\_\allowbreak{}out == false ==$>$ true)
    \end{blockPL}
    \item \textbf{Filename:} bmqt\_\allowbreak{}m\allowbreak{}essage\allowbreak{}guid.\allowbreak{}c\allowbreak{}pp, \textbf{Function:} Messag\allowbreak{}eGUID:\allowbreak{}:isVal\allowbreak{}idHexR\allowbreak{}eprese\allowbreak{}ntatio\allowbreak{}n, \textbf{Logic:} \logicFOL, \href{https://github.com/bloomberg/blazingmq/blob/main/src/groups/bmq/bmqt/bmqt_messageguid.cpp#L77}{\textcolor{blue!60!black}{[code]}}
    \begin{blockFOL}
    \noindent \textbf{requires:} true \\
    \noindent \textbf{ensures:} \_\allowbreak{}\_\allowbreak{}out == FORALL(0,\allowbreak{} Messag\allowbreak{}eGUID:\allowbreak{}:e\_\allowbreak{}SIZ\allowbreak{}E\_\allowbreak{}HEX,\allowbreak{} i,\allowbreak{} (buffer[i] $>$= '0' \&\& buffer[i] $<$= '9') $||$ (buffer[i] $>$= 'A' \&\& buffer[i] $<$= 'F'))
    \end{blockFOL}
    \item \textbf{Filename:} bmqt\_\allowbreak{}m\allowbreak{}essage\allowbreak{}guid.\allowbreak{}c\allowbreak{}pp, \textbf{Function:} Messag\allowbreak{}eGUID:\allowbreak{}:fromH\allowbreak{}ex, \textbf{Logic:} \logicFOL, \href{https://github.com/bloomberg/blazingmq/blob/main/src/groups/bmq/bmqt/bmqt_messageguid.cpp#L96}{\textcolor{blue!60!black}{[code]}}
    \begin{blockFOL}
    \noindent \textbf{requires:} strlen(buffer) $>$= 2 * Messag\allowbreak{}eGUID:\allowbreak{}:e\_\allowbreak{}SIZ\allowbreak{}E\_\allowbreak{}BINA\allowbreak{}RY \&\& FORALL(0,\allowbreak{} 2 * Messag\allowbreak{}eGUID:\allowbreak{}:e\_\allowbreak{}SIZ\allowbreak{}E\_\allowbreak{}BINA\allowbreak{}RY,\allowbreak{} i,\allowbreak{} (buffer[i] $>$= '0' \&\& buffer[i] $<$= '9') $||$ (buffer[i] $>$= 'A' \&\& buffer[i] $<$= 'F') $||$ (buffer[i] $>$= 'a' \&\& buffer[i] $<$= 'f')) \\
    \noindent \textbf{ensures:} EXISTS(0,\allowbreak{} Messag\allowbreak{}eGUID:\allowbreak{}:e\_\allowbreak{}SIZ\allowbreak{}E\_\allowbreak{}BINA\allowbreak{}RY,\allowbreak{} i,\allowbreak{} \_\allowbreak{}\_\allowbreak{}out.\allowbreak{}\allowbreak{}d\_\allowbreak{}buff\allowbreak{}er[i] == ((k\_\allowbreak{}HEX\_\allowbreak{}\allowbreak{}INT\_\allowbreak{}TA\allowbreak{}BLE[buffer[2 * i] - '0'] $<$$<$ 4) | (k\_\allowbreak{}HEX\_\allowbreak{}\allowbreak{}INT\_\allowbreak{}TA\allowbreak{}BLE[buffer[2 * i + 1] - '0'])))
    \end{blockFOL}
    \item \textbf{Filename:} bmqio\_\allowbreak{}\allowbreak{}channe\allowbreak{}lutil.\allowbreak{}\allowbreak{}cpp, \textbf{Function:} Channe\allowbreak{}lUtil:\allowbreak{}:handl\allowbreak{}eRead, \textbf{Logic:} \logicPL, \href{https://github.com/bloomberg/blazingmq/blob/main/src/groups/bmq/bmqio/bmqio_channelutil.cpp#L101}{\textcolor{blue!60!black}{[code]}}
    \begin{blockPL}
    \noindent \textbf{requires:} outPac\allowbreak{}ket-$>$l\allowbreak{}ength() == 0 \\
    \noindent \textbf{ensures:} (\_\allowbreak{}\_\allowbreak{}out == 0 $||$ \_\allowbreak{}\_\allowbreak{}out == -1) \&\& (\_\allowbreak{}\_\allowbreak{}out == -1 ==$>$ outPac\allowbreak{}ket-$>$l\allowbreak{}ength() == 0)
    \end{blockPL}
    \item \textbf{Filename:} bmqtst\allowbreak{}\_\allowbreak{}blobt\allowbreak{}estuti\allowbreak{}l.\allowbreak{}cpp, \textbf{Function:} BlobTe\allowbreak{}stUtil\allowbreak{}::\allowbreak{}toSt\allowbreak{}ring, \textbf{Logic:} \logicFOL, \href{https://github.com/bloomberg/blazingmq/blob/main/src/groups/bmq/bmqtst/bmqtst_blobtestutil.cpp#L82}{\textcolor{blue!60!black}{[code]}}
    \begin{blockFOL}
    \noindent \textbf{requires:} str != NULL \&\& blob.\allowbreak{}l\allowbreak{}ength() $>$= 0 \&\& blob.\allowbreak{}t\allowbreak{}otalSi\allowbreak{}ze() $>$= 0 \\
    \noindent \textbf{ensures:} (\_\allowbreak{}\_\allowbreak{}out == *str) \&\& (\_\allowbreak{}\_\allowbreak{}out.\allowbreak{}\allowbreak{}length() == blob.\allowbreak{}t\allowbreak{}otalSi\allowbreak{}ze()) \&\& (blob.\allowbreak{}l\allowbreak{}ength() $<$ blob.\allowbreak{}t\allowbreak{}otalSi\allowbreak{}ze() ==$>$ EXISTS(\_\allowbreak{}\_\allowbreak{}out.\allowbreak{}\allowbreak{}length() - (blob.\allowbreak{}t\allowbreak{}otalSi\allowbreak{}ze() - blob.\allowbreak{}l\allowbreak{}ength()),\allowbreak{} \_\allowbreak{}\_\allowbreak{}out.\allowbreak{}\allowbreak{}length(),\allowbreak{} i,\allowbreak{} \_\allowbreak{}\_\allowbreak{}out[i] == 'X'))
    \end{blockFOL}
    \item \textbf{Filename:} bmqvt\_\allowbreak{}\allowbreak{}proper\allowbreak{}tybag.\allowbreak{}\allowbreak{}cpp, \textbf{Function:} Proper\allowbreak{}tyBag:\allowbreak{}:impor\allowbreak{}t, \textbf{Logic:} \logicPL, \href{https://github.com/bloomberg/blazingmq/blob/main/src/groups/bmq/bmqvt/bmqvt_propertybag.cpp#L131}{\textcolor{blue!60!black}{[code]}}
    \begin{blockPL}
    \noindent \textbf{requires:} true \\
    \noindent \textbf{ensures:} \&\_\allowbreak{}\_\allowbreak{}out == this
    \end{blockPL}
    \item \textbf{Filename:} bmqa\_\allowbreak{}a\allowbreak{}bstrac\allowbreak{}tsessi\allowbreak{}on.\allowbreak{}cpp, \textbf{Function:} Abstra\allowbreak{}ctSess\allowbreak{}ion::\allowbreak{}o\allowbreak{}penQue\allowbreak{}ueSync, \textbf{Logic:} \logicPL, \href{https://github.com/bloomberg/blazingmq/blob/main/src/groups/bmq/bmqa/bmqa_abstractsession.cpp#L130}{\textcolor{blue!60!black}{[code]}}
    \begin{blockPL}
    \noindent \textbf{requires:} true \\
    \noindent \textbf{ensures:} \_\allowbreak{}\_\allowbreak{}out.\allowbreak{}\allowbreak{}queueI\allowbreak{}d() == bmqa::\allowbreak{}\allowbreak{}QueueI\allowbreak{}d() \&\& \_\allowbreak{}\_\allowbreak{}out.\allowbreak{}\allowbreak{}result() == bmqt::\allowbreak{}\allowbreak{}OpenQu\allowbreak{}eueRes\allowbreak{}ult::\allowbreak{}e\allowbreak{}\_\allowbreak{}NOT\_\allowbreak{}S\allowbreak{}UPPORT\allowbreak{}ED
    \end{blockPL}
    \item \textbf{Filename:} bmqa\_\allowbreak{}a\allowbreak{}bstrac\allowbreak{}tsessi\allowbreak{}on.\allowbreak{}cpp, \textbf{Function:} Abstra\allowbreak{}ctSess\allowbreak{}ion::\allowbreak{}c\allowbreak{}onfigu\allowbreak{}reQueu\allowbreak{}eSync, \textbf{Logic:} \logicPL, \href{https://github.com/bloomberg/blazingmq/blob/main/src/groups/bmq/bmqa/bmqa_abstractsession.cpp#L181}{\textcolor{blue!60!black}{[code]}}
    \begin{blockPL}
    \noindent \textbf{requires:} true \\
    \noindent \textbf{ensures:} \_\allowbreak{}\_\allowbreak{}out.\allowbreak{}\allowbreak{}result() == bmqt::\allowbreak{}\allowbreak{}Config\allowbreak{}ureQue\allowbreak{}ueResu\allowbreak{}lt::\allowbreak{}e\_\allowbreak{}\allowbreak{}NOT\_\allowbreak{}SU\allowbreak{}PPORTE\allowbreak{}D
    \end{blockPL}
    \item \textbf{Filename:} bmqa\_\allowbreak{}a\allowbreak{}bstrac\allowbreak{}tsessi\allowbreak{}on.\allowbreak{}cpp, \textbf{Function:} Abstra\allowbreak{}ctSess\allowbreak{}ion::\allowbreak{}c\allowbreak{}loseQu\allowbreak{}eueSyn\allowbreak{}c, \textbf{Logic:} \logicPL, \href{https://github.com/bloomberg/blazingmq/blob/main/src/groups/bmq/bmqa/bmqa_abstractsession.cpp#L227}{\textcolor{blue!60!black}{[code]}}
    \begin{blockPL}
    \noindent \textbf{requires:} true \\
    \noindent \textbf{ensures:} \_\allowbreak{}\_\allowbreak{}out.\allowbreak{}\allowbreak{}queueI\allowbreak{}d() == bmqa::\allowbreak{}\allowbreak{}QueueI\allowbreak{}d() \&\& \_\allowbreak{}\_\allowbreak{}out.\allowbreak{}\allowbreak{}result() == bmqt::\allowbreak{}\allowbreak{}CloseQ\allowbreak{}ueueRe\allowbreak{}sult::\allowbreak{}\allowbreak{}e\_\allowbreak{}NOT\_\allowbreak{}\allowbreak{}SUPPOR\allowbreak{}TED \&\& \_\allowbreak{}\_\allowbreak{}out.\allowbreak{}\allowbreak{}messag\allowbreak{}e() == "Method is undefined in base protocol"
    \end{blockPL}
    \item \textbf{Filename:} bmqa\_\allowbreak{}e\allowbreak{}vent.\allowbreak{}c\allowbreak{}pp, \textbf{Function:} Event:\allowbreak{}:isSes\allowbreak{}sionEv\allowbreak{}ent, \textbf{Logic:} \logicPL, \href{https://github.com/bloomberg/blazingmq/blob/main/src/groups/bmq/bmqa/bmqa_event.cpp#L74}{\textcolor{blue!60!black}{[code]}}
    \begin{blockPL}
    \noindent \textbf{requires:} (d\_\allowbreak{}impl\_\allowbreak{}sp == nullptr) $||$ (d\_\allowbreak{}impl\_\allowbreak{}sp != nullptr \&\& d\_\allowbreak{}impl\allowbreak{}\_\allowbreak{}sp-$>$t\allowbreak{}ype() == d\_\allowbreak{}impl\allowbreak{}\_\allowbreak{}sp-$>$t\allowbreak{}ype()) \\
    \noindent \textbf{ensures:} (\_\allowbreak{}\_\allowbreak{}out == true ==$>$ (d\_\allowbreak{}impl\_\allowbreak{}sp != nullptr \&\& d\_\allowbreak{}impl\allowbreak{}\_\allowbreak{}sp-$>$t\allowbreak{}ype() == bmqimp\allowbreak{}::\allowbreak{}Even\allowbreak{}t::\allowbreak{}Eve\allowbreak{}ntType\allowbreak{}::\allowbreak{}e\_\allowbreak{}SE\allowbreak{}SSION)) \&\& (\_\allowbreak{}\_\allowbreak{}out == false ==$>$ (d\_\allowbreak{}impl\_\allowbreak{}sp == nullptr $||$ d\_\allowbreak{}impl\allowbreak{}\_\allowbreak{}sp-$>$t\allowbreak{}ype() != bmqimp\allowbreak{}::\allowbreak{}Even\allowbreak{}t::\allowbreak{}Eve\allowbreak{}ntType\allowbreak{}::\allowbreak{}e\_\allowbreak{}SE\allowbreak{}SSION))
    \end{blockPL}
    \item \textbf{Filename:} bmqa\_\allowbreak{}e\allowbreak{}vent.\allowbreak{}c\allowbreak{}pp, \textbf{Function:} Event:\allowbreak{}:isMes\allowbreak{}sageEv\allowbreak{}ent, \textbf{Logic:} \logicPL, \href{https://github.com/bloomberg/blazingmq/blob/main/src/groups/bmq/bmqa/bmqa_event.cpp#L80}{\textcolor{blue!60!black}{[code]}}
    \begin{blockPL}
    \noindent \textbf{requires:} (d\_\allowbreak{}impl\_\allowbreak{}sp == nullptr) $||$ (d\_\allowbreak{}impl\_\allowbreak{}sp != nullptr \&\& d\_\allowbreak{}impl\allowbreak{}\_\allowbreak{}sp-$>$t\allowbreak{}ype() == bmqimp\allowbreak{}::\allowbreak{}Even\allowbreak{}t::\allowbreak{}Eve\allowbreak{}ntType\allowbreak{}::\allowbreak{}e\_\allowbreak{}ME\allowbreak{}SSAGE $||$ d\_\allowbreak{}impl\allowbreak{}\_\allowbreak{}sp-$>$t\allowbreak{}ype() != bmqimp\allowbreak{}::\allowbreak{}Even\allowbreak{}t::\allowbreak{}Eve\allowbreak{}ntType\allowbreak{}::\allowbreak{}e\_\allowbreak{}ME\allowbreak{}SSAGE) \\
    \noindent \textbf{ensures:} (\_\allowbreak{}\_\allowbreak{}out == true ==$>$ (d\_\allowbreak{}impl\_\allowbreak{}sp != nullptr \&\& d\_\allowbreak{}impl\allowbreak{}\_\allowbreak{}sp-$>$t\allowbreak{}ype() == bmqimp\allowbreak{}::\allowbreak{}Even\allowbreak{}t::\allowbreak{}Eve\allowbreak{}ntType\allowbreak{}::\allowbreak{}e\_\allowbreak{}ME\allowbreak{}SSAGE)) \&\& (\_\allowbreak{}\_\allowbreak{}out == false ==$>$ (d\_\allowbreak{}impl\_\allowbreak{}sp == nullptr $||$ d\_\allowbreak{}impl\allowbreak{}\_\allowbreak{}sp-$>$t\allowbreak{}ype() != bmqimp\allowbreak{}::\allowbreak{}Even\allowbreak{}t::\allowbreak{}Eve\allowbreak{}ntType\allowbreak{}::\allowbreak{}e\_\allowbreak{}ME\allowbreak{}SSAGE))
    \end{blockPL}
    \item \textbf{Filename:} bmqa\_\allowbreak{}m\allowbreak{}essage\allowbreak{}.\allowbreak{}cpp, \textbf{Function:} Messag\allowbreak{}e::\allowbreak{}com\allowbreak{}pressi\allowbreak{}onAlgo\allowbreak{}rithmT\allowbreak{}ype, \textbf{Logic:} \logicPL, \href{https://github.com/bloomberg/blazingmq/blob/main/src/groups/bmq/bmqa/bmqa_message.cpp#L290}{\textcolor{blue!60!black}{[code]}}
    \begin{blockPL}
    \noindent \textbf{requires:} isInit\allowbreak{}ialize\allowbreak{}d() \\
    \noindent \textbf{ensures:} (rawEve\allowbreak{}nt.\allowbreak{}isP\allowbreak{}ushEve\allowbreak{}nt() ==$>$ \_\allowbreak{}\_\allowbreak{}out == d\_\allowbreak{}impl\allowbreak{}.\allowbreak{}d\_\allowbreak{}eve\allowbreak{}nt\_\allowbreak{}p-$>$\allowbreak{}pushMe\allowbreak{}ssageI\allowbreak{}terato\allowbreak{}r()-$>$header().\allowbreak{}compr\allowbreak{}ession\allowbreak{}Algori\allowbreak{}thmTyp\allowbreak{}e()) \&\& (rawEve\allowbreak{}nt.\allowbreak{}isP\allowbreak{}utEven\allowbreak{}t() ==$>$ \_\allowbreak{}\_\allowbreak{}out == d\_\allowbreak{}impl\allowbreak{}.\allowbreak{}d\_\allowbreak{}eve\allowbreak{}nt\_\allowbreak{}p-$>$\allowbreak{}putMes\allowbreak{}sageIt\allowbreak{}erator()-$>$header().\allowbreak{}compr\allowbreak{}ession\allowbreak{}Algori\allowbreak{}thmTyp\allowbreak{}e()) \&\& (!(rawEve\allowbreak{}nt.\allowbreak{}isP\allowbreak{}ushEve\allowbreak{}nt() $||$ rawEve\allowbreak{}nt.\allowbreak{}isP\allowbreak{}utEven\allowbreak{}t()) ==$>$ \_\allowbreak{}\_\allowbreak{}out == bmqt::\allowbreak{}\allowbreak{}Compre\allowbreak{}ssionA\allowbreak{}lgorit\allowbreak{}hmType\allowbreak{}::\allowbreak{}e\_\allowbreak{}NO\allowbreak{}NE)
    \end{blockPL}
    \item \textbf{Filename:} bmqa\_\allowbreak{}m\allowbreak{}essage\allowbreak{}.\allowbreak{}cpp, \textbf{Function:} Messag\allowbreak{}e::\allowbreak{}ack\allowbreak{}Status, \textbf{Logic:} \logicPL, \href{https://github.com/bloomberg/blazingmq/blob/main/src/groups/bmq/bmqa/bmqa_message.cpp#L389}{\textcolor{blue!60!black}{[code]}}
    \begin{blockPL}
    \noindent \textbf{requires:} isInit\allowbreak{}ialize\allowbreak{}d() \&\& d\_\allowbreak{}impl\allowbreak{}.\allowbreak{}d\_\allowbreak{}eve\allowbreak{}nt\_\allowbreak{}p-$>$\allowbreak{}rawEve\allowbreak{}nt().\allowbreak{}isAck\allowbreak{}Event() \\
    \noindent \textbf{ensures:} \_\allowbreak{}\_\allowbreak{}out == bmqp::\allowbreak{}\allowbreak{}Protoc\allowbreak{}olUtil\allowbreak{}::\allowbreak{}ackR\allowbreak{}esultF\allowbreak{}romCod\allowbreak{}e(d\_\allowbreak{}impl\allowbreak{}.\allowbreak{}d\_\allowbreak{}eve\allowbreak{}nt\_\allowbreak{}p-$>$\allowbreak{}ackMes\allowbreak{}sageIt\allowbreak{}erator()-$>$message().\allowbreak{}status())
    \end{blockPL}
    \item \textbf{Filename:} bmqa\_\allowbreak{}m\allowbreak{}essage\allowbreak{}.\allowbreak{}cpp, \textbf{Function:} Messag\allowbreak{}e::\allowbreak{}get\allowbreak{}Data, \textbf{Logic:} \logicPL, \href{https://github.com/bloomberg/blazingmq/blob/main/src/groups/bmq/bmqa/bmqa_message.cpp#L404}{\textcolor{blue!60!black}{[code]}}
    \begin{blockPL}
    \noindent \textbf{requires:} blob != nullptr \\
    \noindent \textbf{ensures:} (rawEve\allowbreak{}nt.\allowbreak{}isP\allowbreak{}ushEve\allowbreak{}nt() ==$>$ \_\allowbreak{}\_\allowbreak{}out == d\_\allowbreak{}impl\allowbreak{}.\allowbreak{}d\_\allowbreak{}eve\allowbreak{}nt\_\allowbreak{}p-$>$\allowbreak{}pushMe\allowbreak{}ssageI\allowbreak{}terato\allowbreak{}r()-$>$load\allowbreak{}Messag\allowbreak{}ePaylo\allowbreak{}ad(blob)) \&\& (rawEve\allowbreak{}nt.\allowbreak{}isP\allowbreak{}utEven\allowbreak{}t() ==$>$ \_\allowbreak{}\_\allowbreak{}out == d\_\allowbreak{}impl\allowbreak{}.\allowbreak{}d\_\allowbreak{}eve\allowbreak{}nt\_\allowbreak{}p-$>$\allowbreak{}putMes\allowbreak{}sageIt\allowbreak{}erator()-$>$load\allowbreak{}Messag\allowbreak{}ePaylo\allowbreak{}ad(blob)) \&\& (!(rawEve\allowbreak{}nt.\allowbreak{}isP\allowbreak{}ushEve\allowbreak{}nt() $||$ rawEve\allowbreak{}nt.\allowbreak{}isP\allowbreak{}utEven\allowbreak{}t()) ==$>$ \_\allowbreak{}\_\allowbreak{}out == -1)
    \end{blockPL}
    \item \textbf{Filename:} bmqeva\allowbreak{}l\_\allowbreak{}simp\allowbreak{}leeval\allowbreak{}uator.\allowbreak{}\allowbreak{}cpp, \textbf{Function:} Simple\allowbreak{}Evalua\allowbreak{}tor::\allowbreak{}P\allowbreak{}ropert\allowbreak{}y::\allowbreak{}eva\allowbreak{}luate, \textbf{Logic:} \logicPL, \href{https://github.com/bloomberg/blazingmq/blob/main/src/groups/bmq/bmqeval/bmqeval_simpleevaluator.cpp#L175}{\textcolor{blue!60!black}{[code]}}
    \begin{blockPL}
    \noindent \textbf{requires:} true \\
    \noindent \textbf{ensures:} (\_\allowbreak{}\_\allowbreak{}out.\allowbreak{}\allowbreak{}isErro\allowbreak{}r() ==$>$ contex\allowbreak{}t.\allowbreak{}d\_\allowbreak{}st\allowbreak{}op == true)
    \end{blockPL}
    \item \textbf{Filename:} bmqimp\allowbreak{}\_\allowbreak{}event\allowbreak{}.\allowbreak{}cpp, \textbf{Function:} Event:\allowbreak{}:confi\allowbreak{}gureAs\allowbreak{}Messag\allowbreak{}eEvent, \textbf{Logic:} \logicPL, \href{https://github.com/bloomberg/blazingmq/blob/main/src/groups/bmq/bmqimp/bmqimp_event.cpp#L258}{\textcolor{blue!60!black}{[code]}}
    \begin{blockPL}
    \noindent \textbf{requires:} rawEve\allowbreak{}nt.\allowbreak{}isP\allowbreak{}ushEve\allowbreak{}nt() $||$ rawEve\allowbreak{}nt.\allowbreak{}isA\allowbreak{}ckEven\allowbreak{}t() $||$ rawEve\allowbreak{}nt.\allowbreak{}isP\allowbreak{}utEven\allowbreak{}t() \\
    \noindent \textbf{ensures:} (\_\allowbreak{}\_\allowbreak{}out.\allowbreak{}\allowbreak{}type() == EventT\allowbreak{}ype::\allowbreak{}e\allowbreak{}\_\allowbreak{}MESSA\allowbreak{}GE) \&\& (\_\allowbreak{}\_\allowbreak{}out.\allowbreak{}\allowbreak{}d\_\allowbreak{}msgE\allowbreak{}ventMo\allowbreak{}de == Messag\allowbreak{}eEvent\allowbreak{}Mode::\allowbreak{}\allowbreak{}e\_\allowbreak{}READ)
    \end{blockPL}
    \item \textbf{Filename:} bmqio\_\allowbreak{}\allowbreak{}resolv\allowbreak{}ingcha\allowbreak{}nnelfa\allowbreak{}ctory.\allowbreak{}\allowbreak{}cpp, \textbf{Function:} Resolv\allowbreak{}ingCha\allowbreak{}nnelFa\allowbreak{}ctory\_\allowbreak{}\allowbreak{}Channe\allowbreak{}l::\allowbreak{}pee\allowbreak{}rUri, \textbf{Logic:} \logicPSL, \href{https://github.com/bloomberg/blazingmq/blob/main/src/groups/bmq/bmqio/bmqio_resolvingchannelfactory.cpp#L120}{\textcolor{blue!60!black}{[code]}}
    \begin{blockPSL}
    \noindent \textbf{requires:} d\_\allowbreak{}peerUri != nullptr \&\& d\_\allowbreak{}peerUri $\mapsto$ \_\allowbreak{} \\
    \noindent \textbf{ensures:} \_\allowbreak{}\_\allowbreak{}out == *d\_\allowbreak{}pee\allowbreak{}rUri
    \end{blockPSL}
    \item \textbf{Filename:} bmqma\_\allowbreak{}\allowbreak{}counti\allowbreak{}ngallo\allowbreak{}catoru\allowbreak{}til.\allowbreak{}cp\allowbreak{}p, \textbf{Function:} Counti\allowbreak{}ngAllo\allowbreak{}catorU\allowbreak{}til::\allowbreak{}g\allowbreak{}lobalS\allowbreak{}tatCon\allowbreak{}text, \textbf{Logic:} \logicPL, \href{https://github.com/bloomberg/blazingmq/blob/main/src/groups/bmq/bmqma/bmqma_countingallocatorutil.cpp#L102}{\textcolor{blue!60!black}{[code]}}
    \begin{blockPL}
    \noindent \textbf{requires:} g\_\allowbreak{}init\allowbreak{}ialize\allowbreak{}d \\
    \noindent \textbf{ensures:} \_\allowbreak{}\_\allowbreak{}out != 0 \&\& (\_\allowbreak{}\_\allowbreak{}out == \&g\_\allowbreak{}sta\allowbreak{}tConte\allowbreak{}xt.\allowbreak{}obj\allowbreak{}ect())
    \end{blockPL}
    \item \textbf{Filename:} bmqp\_\allowbreak{}a\allowbreak{}ckmess\allowbreak{}ageite\allowbreak{}rator.\allowbreak{}\allowbreak{}cpp, \textbf{Function:} AckMes\allowbreak{}sageIt\allowbreak{}erator\allowbreak{}::\allowbreak{}rese\allowbreak{}t, \textbf{Logic:} \logicPL, \href{https://github.com/bloomberg/blazingmq/blob/main/src/groups/bmq/bmqp/bmqp_ackmessageiterator.cpp#L105}{\textcolor{blue!60!black}{[code]}}
    \begin{blockPL}
    \noindent \textbf{requires:} blob != 0 \&\& blob-$>$\allowbreak{}length() $>$= eventH\allowbreak{}eader.\allowbreak{}\allowbreak{}header\allowbreak{}Words() * Protoc\allowbreak{}ol::\allowbreak{}k\_\allowbreak{}\allowbreak{}WORD\_\allowbreak{}S\allowbreak{}IZE + AckHea\allowbreak{}der::\allowbreak{}k\allowbreak{}\_\allowbreak{}MIN\_\allowbreak{}H\allowbreak{}EADER\_\allowbreak{}\allowbreak{}SIZE \\
    \noindent \textbf{ensures:} (\_\allowbreak{}\_\allowbreak{}out == rc\_\allowbreak{}SUC\allowbreak{}CESS) $||$ (\_\allowbreak{}\_\allowbreak{}out == rc\_\allowbreak{}INV\allowbreak{}ALID\_\allowbreak{}E\allowbreak{}VENTHE\allowbreak{}ADER) $||$ (\_\allowbreak{}\_\allowbreak{}out == rc\_\allowbreak{}INV\allowbreak{}ALID\_\allowbreak{}A\allowbreak{}CKHEAD\allowbreak{}ER) $||$ (\_\allowbreak{}\_\allowbreak{}out == rc\_\allowbreak{}NOT\allowbreak{}\_\allowbreak{}ENOUG\allowbreak{}H\_\allowbreak{}BYTE\allowbreak{}S)
    \end{blockPL}
    \item \textbf{Filename:} bmqp\_\allowbreak{}c\allowbreak{}ompres\allowbreak{}sion.\allowbreak{}c\allowbreak{}pp, \textbf{Function:} ZLib::\allowbreak{}\allowbreak{}writeO\allowbreak{}utput, \textbf{Logic:} \logicPL, \href{https://github.com/bloomberg/blazingmq/blob/main/src/groups/bmq/bmqp/bmqp_compression.cpp#L223}{\textcolor{blue!60!black}{[code]}}
    \begin{blockPL}
    \noindent \textbf{requires:} output != NULL \&\& factory != NULL \&\& stream != NULL \&\& errorS\allowbreak{}tream != NULL \&\& \&input != NULL \\
    \noindent \textbf{ensures:} (\_\allowbreak{}\_\allowbreak{}out == rc\_\allowbreak{}SUC\allowbreak{}CESS) $||$ (\_\allowbreak{}\_\allowbreak{}out == rc\_\allowbreak{}STR\allowbreak{}EAM\_\allowbreak{}IN\allowbreak{}IT\_\allowbreak{}FAI\allowbreak{}LURE) $||$ (\_\allowbreak{}\_\allowbreak{}out == rc\_\allowbreak{}STR\allowbreak{}EAM\_\allowbreak{}PR\allowbreak{}OCESS\_\allowbreak{}\allowbreak{}FAILUR\allowbreak{}E) $||$ (\_\allowbreak{}\_\allowbreak{}out == rc\_\allowbreak{}STR\allowbreak{}EAM\_\allowbreak{}EN\allowbreak{}D\_\allowbreak{}FAIL\allowbreak{}URE)
    \end{blockPL}
    \item \textbf{Filename:} bmqst\_\allowbreak{}\allowbreak{}statco\allowbreak{}ntext.\allowbreak{}\allowbreak{}cpp, \textbf{Function:} conver\allowbreak{}tFromE\allowbreak{}poch, \textbf{Logic:} \logicPL, \href{https://github.com/bloomberg/blazingmq/blob/main/src/groups/bmq/bmqst/bmqst_statcontext.cpp#L155}{\textcolor{blue!60!black}{[code]}}
    \begin{blockPL}
    \noindent \textbf{requires:} true \\
    \noindent \textbf{ensures:} \_\allowbreak{}\_\allowbreak{}out == epochTime - epochO\allowbreak{}ffset()
    \end{blockPL}
    \item \textbf{Filename:} bmqp\_\allowbreak{}p\allowbreak{}rotoco\allowbreak{}lutil.\allowbreak{}\allowbreak{}cpp, \textbf{Function:} Protoc\allowbreak{}olUtil\allowbreak{}::\allowbreak{}ackR\allowbreak{}esultT\allowbreak{}oCode, \textbf{Logic:} \logicPL, \href{https://github.com/bloomberg/blazingmq/blob/main/src/groups/bmq/bmqp/bmqp_protocolutil.cpp#L160}{\textcolor{blue!60!black}{[code]}}
    \begin{blockPL}
    \noindent \textbf{requires:} true \\
    \noindent \textbf{ensures:} \_\allowbreak{}\_\allowbreak{}out == 0 $||$ \_\allowbreak{}\_\allowbreak{}out == 1 $||$ \_\allowbreak{}\_\allowbreak{}out == 2 $||$ \_\allowbreak{}\_\allowbreak{}out == 5 $||$ \_\allowbreak{}\_\allowbreak{}out == 6 $||$ \_\allowbreak{}\_\allowbreak{}out == 7
    \end{blockPL}
\end{enumerate}


\end{document}